\documentclass[physics,hypothesis,accept,moreauthors,LaTeX]{Definitions/mdpi} 

\firstpage{1}
\pubvolume{2}
\issuenum{1}
\articlenumber{213--276}
\pubyear{2020}
\copyrightyear{2020}
\history{Received: 29 February 2020; Version: 30 May 2020; Accepted: 2 June 2020} 

\usepackage[english]{babel}
\usepackage{xcolor}

\Title{Hypothesis about Enrichment of Solar System} 

\Author{E. P. Tito $^{1}$
\orcidA{},
 V. I. Pavlov $^{1,2}$
\orcidB{}
}

\AuthorNames{E. P. Tito and V. I. Pavlov}

\address{%
$^{1}$ \quad Scientific Advisory Group, Pasadena, CA 91125, USA\\
$^{2}$ \quad Universit\'{e} de Lille, Facult\'{e} des Sciences et Technologies, 
F-59000 Lille, 
France
}

\abstract{
Despite significant progress in the understanding of {\em galactic nucleosynthesis} and its influence on the solar system neighborhood, challenges remain in the understanding of enrichment of the {\em solar system} itself. Based on the detailed review of multi-disciplinary literature, we propose a scenario that an event of nucleogenesis---not nucleosynthesis (from lower nucleon numbers $A$ to higher $A$) but nuclear-{\em fission} (from higher $A$ to lower $A$)---occurred in the inner part of the solar system at one of the stages of its evolution. We propose a feasible mechanism of implementation of such event.  The occurrence of such event could help explain the puzzles in yet-unresolved isotopic abundances, certain meteoritic anomalies, as well as peculiarities in the solar system's composition and planetary structure. We also discuss experimental data and available results from existing models (in several relevant sub-fields) that provide support and/or appear consistent with the hypothesis.
}


\begin{document}
 
\newcommand{\araa}{Annu.~Rev.~Astron.~Astrophys. }
\newcommand{\aaa}{Astron.~and~Astrophys. }
\newcommand{\aap}{Astrophys.~Astron. }
\newcommand{\aj}{Astronom.~J } 
\newcommand{\apj}{Astrophys.~J}
\newcommand{\aplett}{Astrophys.~Lett.} 
\newcommand{\aphjl}{Astrophys.~J.~Lett. }
\newcommand{\apjl}{ApJ Lett. }
\newcommand{\apspr}{Astrophys.~Space~Phys.~Res. }%
\newcommand{\apss}{Astrophys.~Space~Sci. }%
\newcommand{\gal}{Galaxies }
\newcommand{\gca}{Geochimica et Cosmochimica Acta} 
\newcommand{\icarus}{Icarus }
\newcommand{\jcap}{J.~Cosmology and Astroparticle Phys.} 
\newcommand{\mnras}{Mon.~Not.~R.~Astron.~Soc. }
\newcommand{\nar}{New~Astron.~Rev.} 
\newcommand{\nat}{Nature }
\newcommand{\nphysa}{Nucl.~Phys.~A } 
\newcommand{\prl}{Phys.~Rev.~Lett. }
\newcommand{\prc}{Phys.~Rev.~C }
\newcommand{\prd}{Phys.~Rev.~D }
\newcommand{\pasp}{PASP}
\newcommand{\sovast}{Soviet Ast.-AJ }
\newcommand{\areps}{Annu.~Rev.~Earth~and~Planet.~Sci.}
\newcommand{\arnps}{Annu.~Rev.~Nucl.~Part.~Sci.}
\newcommand{\epsl}{Earth~and~Planet.~Sci.~Lett.}
\newcommand{\jgr}{J.~Geophys.~Res.}
\newcommand{\jpgnpp}{J.~Phys.~G:~Nucl.~Part.~Phys.}
\newcommand{\mps}{Meteoritics~and~Planet.~Sci.}
\newcommand{\pl}{Phys.~Lett.}
\newcommand{\rmp}{Rev.~Modern~Phys.}

\renewcommand{\thesubparagraph}{(\arabic{subparagraph})}
\setcounter{secnumdepth}{5}
\setcounter{tocdepth}{4}
\tableofcontents

\noindent\rule{\textwidth}{0.5pt}

\section{Introduction}
\label{s:1}

Significant progress in the understanding of {\em galactic nucleosynthesis} and its influence on the solar system neighborhood has been achieved over recent decades, but challenges remain in the understanding of 
enrichment of the {\em solar system} itself. 
Despite the widely-held belief that the solar system evolution is by now well-understood, 
numerous unsolved puzzles persist, among them are:  
the "excess" of $p$-elements, 
the bi-modal planetary structure of the solar system, 
the "solar modeling problem", various meteoritic anomalies, and more. 
These puzzles  (referenced throughout the article) are not just minor discrepancies---
they reflect the fundamentally critical 
gaps in the current state of understanding of the processes that affected the solar system 
over the course of its evolution.   
Experts in various subfields have been struggling to resolve these puzzles for years.  

But stepping away from the conventional path of addressing each problem in isolation, 
taking an "elevated" perspective,    
and considering at once the vast breadth of 
experimental data and theoretical findings from multiple sub-fields, 
helps realize that many of these puzzles may perhaps be better solved 
in the framework of one encompassing scenario instead of many unconnected models. 
Indeed, if the question is asked whether all the puzzles could be the consequences of just {\em one} event  
and what kind of event might it be, one answer is: 
such event had to be  an event of nucleogenesis 
---however not nucleosynthesis (from lower nucleon numbers $A$ to higher $A$) 
 but rather nuclear-{\em fission} (from higher $A$ to lower $A$)---
and the event had to be {\em local} (i.e., in the inner part of the solar system), 
in contrast to the conventionally presumed distant nuclei-producing cataclysms modeled as events of nucleosynthesis. 
Naturally, 
to result in the current solar system's macro-composition and structure,  
 such event had to be "non-destructively impactful".
Hence, in this paper we propose a scenario of {\em how} a nuclear-{\em fission} event 
could occur in the inner part of the solar system 
at one of the stages of its evolution. 
The occurrence of such event could indeed help explain the yet-unresolved puzzles in isotopic abundances and certain meteoritic anomalies 
(because the event's nucleogenetic signature is unique and distinct from all other cataclysms), 
as well as the peculiarities in the solar system's composition and planetary structure 
(because the event is "local").  

We also propose and discuss a feasible mechanism of implementation of such event.   
The key nucleogenetic phenomenon in consideration is  (not {\em nucleosynthesis} 
but) {\em nuclear fission} of super-dense neutron matter (of galactic origin).  
This idea may seem bizarre to those unfamiliar with recent advances in the fields dealing with super-dense matter and super-heavy nuclei,  
but the idea is grounded in 
well-established multi-disciplinary facts, both observed and experimental,  
which are extensively cited throughout the article. 
The mechanism of "delivery" of the super-dense neutron matter is discussed in Sec.~\ref{s:3-1};  
the object was  perhaps a stellar "clump" born in a galactic cataclysm (discussed in Sec.~\ref{s:3-1-1}). 
The key mechanism of the implementation scenario is {\em instability} of super-dense nuclear matter (in the vicinity of its instability threshold).  
 Such instability can lead to {\em fragmentation within the state of nuclear-fog}, 
 with subsequent fission of "mega"-nuclei (nuclear "droplets") along stochastic chains of nuclear transformations. 
Notably, only in recent decades,  experimental studies on fission of super-heavy nuclei and 
expansion of comprehension of properties (stability and instability) of super-dense nuclear matter 
have produced fundamental insights sufficient to 
offer a feasible concept of fission-driven nucleogenesis 
(with appropriate scale for the solar system) which could help expand the paradigm of solar system formation and enrichment. 

Indeed, if---as we propose---the solar system 
(1) formed originally, {\em pre-}4.5~Gyrs ago, containing {\em only} gaseous objects,  
and 
(2) encountered, $\sim 4.5$~Gyrs ago, a fission-capable  appropriately-sized stellar object
which "exploded" in the inner part of the solar system in nucleogenetic cascades 
eventually leading to  formation of terrestrial planets,     
then the puzzles of exotic isotopic abundances (such as of $p$-nuclides, actinides, extinct radionuclides), difficult-to-explain meteoritic anomalies, bi-modal planetary structure, compositional enrichment of the Sun and gaseous giants, and many others, could be explained by such event. 
(Within the framework of such scenario, the radionuclides in meteorites---the products of the event---which have been used as chronometers presumably dating the "age" of the solar system are in fact pointing at the time of the event; this implies that  the actual age of the gaseous solar system is therefore greater than the estimated 4.56~Gyrs.)

The abundant multi-disciplinary references cited in this article offer substantial support to each 
discussed aspect of the proposed hypothesis. 
Once the entirety of the presented arguments and data is fully comprehended and weighed with an open mind, 
the hypothesis should not seem as "wild and extravagant" as it might appear to some at the first glance,   
especially if key parts of the paper are skipped during reading. 
The hypothesis integrates physical phenomena that are well-established 
-- the $T(\rho)$-evolution of super-dense nuclear-matter, the nuclear-fog, the fragmentation and fission of super-heavy nuclei, and so on---and the events similar to the one proposed in this article may broadly occur in the galaxy and the universe  
even if the mankind has not yet considered their occurrence. 
Furthermore, historically the mankind has revised conceptions of the solar system structure and formation a number of times;   
the solar system is merely an ordinary stellar system---one among many---
hence contemplating a possibility of a structure-altering cataclysm in it should not be a taboo 
(even if our innate human egocentrism makes cataclysms seem much more palatable when they happen elsewhere). 

In order to avoid any potential misunderstandings, several comments are worth mentioning from the start: 

  Although our paper presents a hypothesis about enrichment of the {\em solar system}, we want to explicitly state  that it does not {\bf not dismiss} any conventional models concerning {\bf galactic} nucleosynthesis. 
The entire understanding of the galactic chemical evolution remains an integral part of understanding of the evolution of the solar system (defining its composition during the first, pre-event, stage). 
Our focus is on the enrichment (not creation) 
of the solar system (not the entire galaxy) 
with exotic post-$Fe$ elements (not all elements),  
namely $s$-, $r$-, $p$-isotopes, and actinides,  
which form the tail of the element abundance profile and which are "tiny" in comparison with the bulk composition (in particular those which are "excessive" and difficult to explain with current models or which are mixed in meteorites in combinations that cannot be understood at present).  

Furthermore, 
   the nucleogenetic aspect of the hypothesis 
represents {\bf not an "additional model" but a "modified framework"} – the difference is essential. Indeed, existing factual (experimental) data may be explained via distinct "frameworks" (elevated perspectives onto the totality of observed phenomena) which may comprise various internal "models" for individual aspects. 
   To fully grasp the idea, 
a thoughtful evaluation of the proposal necessarily requires 
a mental transition away from the conventional framework 
    (with its {\em presumed} multitude of distant nuclei-producing cataclysms) 
and 
{\em reinterpretation of the fit of  all existing (multi-disciplinary) data within the new framework}. 
The hypothesis should not be regarded
as merely a 
suggestion of yet another "model" to be added to the "existing framework".

Finally, many problems are the so-called "direct problems": the problem is "set up" (via the system of equations, regulating parameters, initial conditions, etc.) and then calculations produce some "result". 
  The presented hypothesis 
is an example of the so-called "inverse problem." A suitable metaphor may
be an example of a police-detective investigating a crime: there is a corpse, there is some evidence – the question is who is the murderer. 
  In other words, the best {\bf mindset} for comprehension of the following sections is the mindset of a "detective". 
  As new evidence becomes available in the future, and as studies in various sub-fields offer their refinements to the outlined propositions and data interpretations, the clarity of understanding of "what actually happened" will certainly increase.  Such pursuit may take a while though---just like any investigation. 

The structure of the paper is as follows:  
In Section~\ref{s:2}, the problem and the proposed solution (hypothesis and implementation scenario) are defined. 
In Section~\ref{s:3}, key physical processes are discussed. 
Section~\ref{s:4} presents the model and results.
Section~\ref{s:5} discusses supporting evidence, possible solutions to the outstanding puzzles of the solar system, and implied refinements to the existing models that may be offered by the proposed expanded paradigm of the solar system evolution. 
Section~\ref{s:6} concludes with final remarks.

\section{Problem Definition and Proposed Solution} 
\label{s:2}

In the context of understanding of {\em galactic nucleogenesis} in general, 
 a number of production-sites (for each type of nuclei) have been proposed and studied.
However, to appear in the solar system, the nuclei produced in distant stellar cataclysms 
had to arrive to the solar system, and therefore, 
various practical constraints must have been satisfied---such as mass and direction of cataclysm ejecta 
(so enrichment could be reaching but not demolishing), 
cataclysm radiation (so scorching evaporation could be survived), 
relative timing of certain events 
(so short-lived isotopes with separate origins could co-mix in meteoritic grains as detected), and so on. 
 These practical issues cannot be ignored when considering {\em nucleogenesis of the solar system nuclei}. 
 However, the co-presence and co-mixing in the solar system of non-native nuclei
 from all of the presumed multiple distant cataclysms 
 needed to match the solar system abundances, 
 is rather perplexing (details of the data are discussed in Sec.~\ref{s:5-3-3-2}).  
 Critically important is the fact that the existing models of $p$-nuclei production cannot currently match the actually measured abundances in the solar system---
 thus implying that an additional physical mechanism of $p$-nuclei production is needed to explain their presence in the samples   
  (more details are below, in Sec.~\ref{s:2-1} and in Sec.~\ref{s:5-3-3-3}).

\subsection{Galactic Nucleogenesis}
\label{s:2-1}

Galactic enrichment is often summarized in ways similar to this conceptual sketch \cite{Jacobson_2014}:
\begin{figure}[h!]
\centering
\includegraphics[width=0.65\columnwidth]{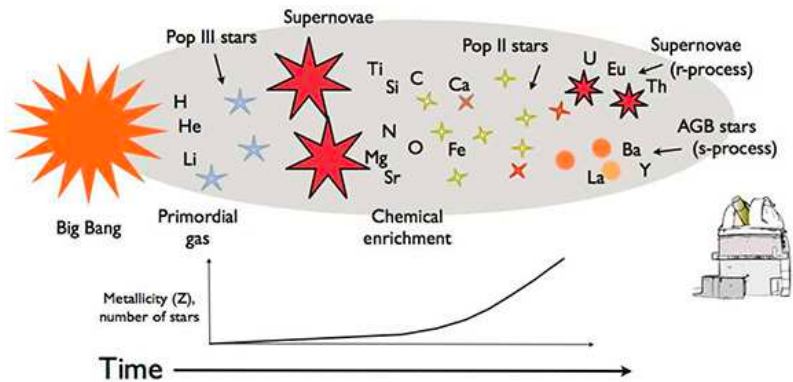}
\caption
[Conceptual illustration of chemical enrichment of the universe]
{Conceptual illustration of chemical enrichment of the universe:
cycles of star formations and cataclysmic deaths enrich the universe over time \cite{Jacobson_2014}.
}
\label{Fig:1}
\end{figure}

\noindent
More specifically,  the following {\em production sites} have been proposed for exotic nuclei detected in the solar system:

The stellar sites of  $s$-process (slow neutron-capture) nuclei are believed to be 
thermally pulsing asymptotic giant branch stars (AGB) that evolved from low- and intermediate-mass stars
\cite{Gallino_1998}, \cite{Arlandini_1999}, \cite{Busso_1999}. 

The (rapid) $r$-process captures occur at much higher temperatures and neutron densities \cite{Burbidge_1957}, \cite{osti_4709881}.  
Predicting the primary stellar site of the $r$-process has been difficult \cite{Arnould_2007}. 
Historically, the proposed scenarios have been divided in two main categories: the high and low entropy scenario \cite{Arnould_2007}. 
The former one includes core-collapse supernova, where the freshly born protoneutron star cools by emitting a large amount of neutrinos which heat the material in the surface of the neutron star and create an outflow of baryonic matter 
\cite{Woosley_1994}, 
\cite{2012ARNPS..62..407J}. 
The low entropy scenario \cite{1974ApJ...192L.145L}  is based on the idea that at high neutron densities the material is neutronized due to continuous electron capture. 
It has been suggested that when this material undergoes a sudden decompression, the existing nuclei will start to capture neutrons producing neutron-rich elements. If the ratio of neutron to seed nuclei is large enough, the seed nuclei are converted by successive neutron captures and beta decays to heavier and heavier elements until the point when they become unstable against fission. The astrophysical plausibility of this process remained unanswered for several years, until the first numerical simulations of the decompression of material during the merger of two neutron stars and a NS with a black hole predicted the ejecta of $r$-process material \cite{1974ApJ...192L.145L}, \cite{1976ApJ...210..549L}, \cite{1982ApL....22..143S}.
Overall, a variety of explosive cataclysms have been proposed as potential $r$-nuclei production-sites%
\footnote{
See Reference \cite{Cowan_2019} for an extensive review of the following processes: 
possible $r$-process sites related to massive stars: 
neutrino winds from core-collapse supernovae, 
electron-capture supernovae, 
neutrino-induced $r$-process in the $He$-shell, 
quark deconfinement supernovae, 
magneto-rotational supernovae with jets, 
collapsars, hypernovae, long-duration gamma-ray bursts; 
and 
neutron-star and neutron-star / black hole mergers:  
dynamic ejecta, 
neutrino winds and the effect of neutrinos, 
accretion disks outflows.
}, 
among them are supernovae (of various types)  \cite{Sneden_2003},  \cite{Cameron_2003}, \cite{Winteler_2012}, \cite{Nishimura_2015}, \cite{M_sta_2018}, 
neutron star (NS-NS) mergers \cite{Fern_ndez_2013}, \cite{Bauswein_2013}, 
neutron star and black hole (NS-BH) mergers 
\cite{Korobkin_2012}, \cite{Foucart_2014}, \cite{Mennekens_2014}, \cite{Kyutoku_2015}, \cite{Roberts_2016}, \cite{Rosswog_2013}, \cite{Perego_2014}, \cite{Martin_2015}, \cite{Sekiguchi_2016}, \cite{Radice_2016}, \cite{Wu_2016}, 
strange-quark-star cataclysms and 
strange-stars mergers \cite{Paulucci_2017}, 
collapsars  \cite{Pruet_2004},  \cite{Siegel_2019},
dark matter-induced neutron star implosions \cite{Bramante_2016}, \cite{Fuller_2017}.

Certain proton-rich nuclides cannot be synthesized through sequences of only neutron-captures ($s$- or $r$-processes) and $\beta$-decays (see, for example, Reference \cite{Meyer_1994} and references therein).  
Since $p$-nuclei can be synthesized 
(1) by successively adding protons to a nuclide ($p$-captures) or 
(2) by removing neutrons from pre-existing $s$- or $r$-nuclides (seeds) through sequences of photodisintegrations (plus/minus $\gamma$-captures),  
the term \emph{$p$-process} is used to generally describe any process synthesizing $p$-nuclei, even when no proton-captures are involved. 
In situ temperatures of the order of several $10^9$~K are required for $p$-captures and $\gamma$-processes. 
 Massive stars are thought to  produce $p$-nuclei through photodisintegration of pre-existing intermediate and heavy nuclei. This so-called $\gamma$-process requires high stellar plasma temperatures and occurs mainly in explosive $O/Ne$ burning during a core-collapse supernova. 
 Although models of  the $\gamma$-process in massive stars have been successful in producing a large range of $p$-nuclei, significant deficiencies remain \cite{Rauscher_2013}.

\subsection{Challenges in Understanding of Origins of Solar System Isotopes}
\label{s:2-2}

{\bf Detected "Excess" (Model Underproduction) of $p$-Nuclides:} 
In terrestrial and meteoritic samples  over thirty $p$-nuclides---with $^{74}Se$  being the lightest and $^{196}Hg$ the heaviest---have been identified (see, for example, Reference \cite{Rauscher_2013}).   
Their isotopic abundances are 1-2 orders of magnitude lower than for the respective $r$- and $s$-nuclei in the same mass region (thus they typically attract less attention in general discussions of the solar abundances). 
So far it seems to be impossible to reproduce the solar abundances of the $p$-isotopes by 
any combination of all of the so-far considered processes---\guillemotleft the mystery of the origin of the $p$-nuclides is still with us\guillemotright{} 
 \cite{Rauscher_2013}. 
In the unresolvable gaps, the measured abundances {\em exceed} the estimates produced by the numerical models. 
This fact is extraordinarily important because it indicates that the solar system contains nuclides which 
cannot be explained by the best-fit superposition of all current models for the so-far considered production 
mechanisms---hence {\em another nucleogenetic mechanisms is fundamentally needed}. 

{\bf Presence of Short-Lived Nuclides:} 
Presence in the early solar system of a number of short-lived nuclides  has been inferred from their daughter nuclide abundances: 
$^{7}Be$~($t_{1/2}$=~53~days),  
$^{10}Be$~($t_{1/2}$=~1.5~Myr),  
$^{26}Al$~($t_{1/2}$=~0.74~Myr), 
$^{36}Cl$~($t_{1/2}$=~0.3~Myr), 
$^{41}Ca$~($t_{1/2}$=~0.1~Myr), 
$^{53}Mn$~($t_{1/2}$=~3.7~Myr), 
$^{60}Fe$~($t_{1/2}$=~2.6~Myr), 
$^{92}Nb$~($t_{1/2}$=~35~Myr), 
$^{99}Tc$~($t_{1/2}$=~0.21~Myr), 
$^{107}Pd$~($t_{1/2}$=~6.5~Myr), 
$^{129}I$~($t_{1/2}$=~16~Myr), 
$^{146}Sm$~($t_{1/2}$=~68 or 103~Myr), 
$^{182}Hf$~($t_{1/2}$=~9~Myr), 
$^{205}Pb$~($t_{1/2}$=~15~Myr),  
$^{244}Pu$~($t_{1/2}$=~81~Myr),    
$^{247}Cm$~($t_{1/2}$=~15.6~Myr). 
Various explanations for their origins have been proposed and studied 
-- multiple contributing events are presumed in order to explain the full inventory 
(see, for example, reviews \cite{Dauphas_2016}, \cite{Qin_2016}, \cite{Pignatari_2016}). 
However, even for stable nuclides, the multitude of cataclysmic events which is 
required to produce them is a challenge because each  explosive 
stellar cataclysm must have been located sufficiently close (to provide adequate enrichment), 
but far enough to not scorch or demolish the presolar nebula \cite{Adams_2010}. 
The more cataclysms are required to explain 
the presence of cataclysm-produced $s$-, $r$-, and $p$-nuclides, 
the lower are the odds for such combination. 
(More details are in Sec.~\ref{s:5-3-3-2}.) 
For the short-lived nuclides, 
the challenge is even greater because of the {\em limited time-window} 
(estimated to be only about 20,000 years  \cite{Gritschneder_2011})
within which all the cataclysms had to occur,  
and occur (what is also remarkable) in the quiet, not famous for star-explosions  neighborhood where the solar system resides.  
Notably, among the extinct radionuclides on the list, $^{92}Nb$ and $^{146}Sm$ are $p$-nuclides. 
    The astrophysical production site of $^{92}Nb$  is still unclear  \cite{Pignatari_2016},   \cite{Travaglio_2014}, 
and while  $^{146}Sm$ may be produced in Type Ia supernova, such scenario does not reconcile well 
with the signature of another extinct radionuclide $^{53}Mn$ \cite{Pignatari_2016},  \cite{Lugaro_2016}. 
Furthermore, the detection of "excessive" $^{7}Li$ in the solar system \cite{Chaussidon_2001}, \cite{Chaussidon_2002} 
necessarily 
points at its  "local" production-site because  $^{7}Li$ is produced by decay of $^{7}Be$ whose half-life is only 53 days.   
It has been suggested that the nuclei were produced by spallation within the solar system as it was forming, but  
such explanation is not fully self-consistent 
\cite{Gounelle_2001},  \cite{Goswami_2001},  \cite{Leya_2003}, \cite{2006GeCoA..70..224C}, \cite{Dauphas_2011}. 

\vspace{9pt}

The mentioned challenges are not just minor discrepancies---
they reflect the fundamentally critical gaps in the current state of understanding of the processes that enriched the solar system.  
Besides the above-mentioned issues, there remain other unresolved questions about the origins of exotic nuclei found in the solar system (the production site of actinides is, for example, one such challenge). 
Specific issues are too numerous and too nuanced to continue their mentioning here, but experts in the specialized fields are well-aware of them 
(see, among others, References \cite{Rauscher_2013}, \cite{Dauphas_2016}, \cite{Qin_2016}, \cite{Pignatari_2016}, \cite{Zinner_2014}, \cite{Tissot_2016}).     
For example, as stated with respect to the $r$-process nuclides: 
\guillemotleft 
 The limit of our understanding of the $r$-process is illustrated by the fact that there have been as many $r$-processes proposed as short-lived $r$-nuclides investigated\guillemotright  {} 
\cite{Tissot_2016}. 

By taking the perspective that these puzzles may be the consequences of just one event 
rather than numerous events,  
in view of the presence of short-lived nuclides in the solar system  
and, most importantly, in view of the 
{\em apparently existing need to find yet another mechanism of $p$-nuclei production}, 
we see a possible solution in the suggestion that a distinctly different (from the so-far considered)  
nucleogenetic "event" enriched the solar system with the difficult-to-explain nuclides. 
As elaborated below in Sec.~\ref{s:3-3-3},  
a nuclear-{\it fission} production event would indeed be capable of producing all of the 
nuclides commonly presumed to originate in $s$- and $r$-  nucleo{\it synthesis} and $p$-processes, 
and such event would have a distinct signature from all other production-mechanisms  
(so the "anomalies" of nucleosynthesis-events would be simply the "signatures" of the fission-event). 
And we propose that this event  
occurred "locally"---in the inner part of the solar system---thus accounting for the multitude of 
the time- and location-specific features (discussed in Sec.~\ref{s:2-3} and Sec.~\ref{s:5-3-3}). 
Obviously, the event had to be "small enough" (in terms of its stellar scale) to not demolish the entire solar system.

\subsection{Challenges in Understanding of Planetary Structure of Solar System}
\label{s:2-3}

A "local" nucleogenetic event capable of meaningfully enriching the solar system would likely have  left some other noticeable traces.  
Indeed, there are many features in the solar system structure that are not easily explainable at present.  
Within the proposed {\em expanded paradigm} of the solar system evolution, these features become naturally interpreted as the consequences of the event.  

For example, it has been recognized for a long time that the existing bimodal (or even trimodal) structure of the solar system is perplexing:  
\guillemotleft Assuming that planetesimals formed everywhere in the disk with comparable masses ... the subsequent process of planet growth by pebble accretion should favor the bodies closer to the Sun ... In other words, giant planet cores should have formed in the inner disk and Mars mass embryos in the outer disk!%
\guillemotright  {} 
\cite{Morbidelli_2016}  
Moreover, while the rocky objects are thought to have formed by accretion (from dust grains into larger and larger bodies), 
two competing models exist about formation of the giants---
the core accretion model and the disk instability model. 
In either case, reconciliation of formation of two classes of planets has not been successful yet. 
The {\em core accretion} model 
\cite{Safronov_1969},  
\cite{goldreich1973formation}, 
\cite{Wetherill_1989}, \cite{weidenschilling1993protostars}, \cite{Lissauer_1993}  
presumes that rocky, icy cores of giant planets accreted in a process very similar to the one that formed the terrestrial planets and then captured gas from the solar nebula to become gas giants. This model explains why the giants have larger concentration of heavier elements than the Sun has, but numerical simulations yield formation times that are way too long (unless the mass of the primordial nebula is increased).   
The {\em disk instability} model 
\cite{Cameron_1978}, \cite{Boss_1997}, \cite{Mayer_2002} 
posits that spontaneous density perturbations in the primordial disc could have caused clumps of gas to become massive enough to be self-gravitating and form the Sun and the planets.  
Formation scale is then much more rapid, but the model does not readily explain the observed chemical enrichment of the planets. 

Within the proposed in this paper expanded paradigm of the solar system evolution, the nucleogenetic event occurred {\em after} giant gaseous objects had formed (via disk instability), and only later (in a separate evolution step)
the post-event nuclear-fission products---debris (pebbles)---accreted into the "rocky" objects (planetesimals, terrestrial planets, meteorites, asteroids, and so on). 
The bi-modal planetary structure arises naturally, and the two models of planet formation complement each other without any internal or mutual contradictions.  

Another peculiarity of the solar system is the fact that the orbits of the giants are widely spaced and nearly circular,  
which is unusual \cite{Ford_2001}, \cite{Beer_2004}, \cite{Rice_2014}.   
Remarkably, they also do not exhibit any resonance despite the fact that, as $N$-body studies of planetary formation  and orbit positions indicate, due to the convergent planetary migration in times before the gas disk's dispersal, each giant planet should have become trapped in a resonance with its neighbor \cite{Masset_2001}, \cite{Kley_2000}.  
Some studies have also exposed the possibility that one more giant object initially might have been present in the solar system and later was ejected---dynamical simulations starting with a resonant system of four giant planets showed 
low success rate in matching the present orbits of giant planets  \cite{Nesvorn__2011},  \cite{Batygin_2011}. 
A cataclysmic event in the inner part could indeed explain the apparently "violent" history of the gaseous solar system, scattering the existing giants, disrupting their resonances, and perhaps destroying the "missing" giant (discussed in Sec.~\ref{s:5-3-2}).

\subsection{Summary of Proposed Hypothesis (Nuclear-Fission "Event" within Solar System)}
\label{s:2-4}
 
 In view of the above-mentioned findings (and additional ones discussed in Sec.~\ref{s:5-3} and extensively referenced), 
regardless of whether or not the proposed (nuclear-fission-driven) enrichment mechanism is a meaningful contributor 
to the \emph{galactic} nucleogenesis, the mechanism---if indeed impacted the solar system---may be rather meaningful for the evolution of \emph{our home system}, and hence it is worth contemplation. 
  
{\bf Hypothesis:}  
In brief, our hypothesis%
\footnote{%
The first attempts to express this idea were undertaken in References \cite{Pavlova_1992} and \cite{Tito_2013}.}  
proposes that: 
 a  (nuclear-fission-driven) nucleogenetic event  occurred "locally"---in the inner part of the solar system---at the time currently believed to be the "birth" of the solar system  
(about 4.56~Ga ago, based on meteoritic data), 
which enriched the solar system with exotic isotopes and altered its composition and planetary structure.  
The hypothesis includes presumptions that  
(1) the solar system was formed {\em before} the event and initially had only giant (mostly hydrogen-helium) objects,   
and 
(2) the nuclear-fission products (debris) from the event evolved into the "rocky" objects in the system (terrestrial planets, asteroids, and so on) and also enriched the pre-existed hydrogen-helium objects (the Sun and the gaseous giants). 
Thus, the evolution of the solar system occurred in two stages. 

{\bf Scenario:}  
Because there exist no natural and appropriately-scaled "sources" of nuclear-fission in the solar system or its vicinity,  
the logical conclusion is that---for the fission-event to take place within the system---the (non-demolishing) nuclear-fission-capable object had to arrive from afar. 
Thus, we suggest that such object could be a compact super-dense stellar "fragment" 
born in a distant galactic cataclysm---for example, in a tidal disruption of a neutron star by a black hole,
perhaps the supermassive black hole at the center of our galaxy (which could have catapulted the fragment with hyperbolic velocity; see, for example, the Penrose effect). 

Although for such fragment a number of fission-triggering mechanisms may perhaps exist, we envisioned and  
present in this paper  just one.  
Indeed, if even one such mechanism is identified as feasible, it means that the proposed event is {\em plausible}, 
that is, not impossible, not forbidden by the laws of nature. 

Thus, we suggest that after a long journey---during which the nuclear matter sufficiently cooled down and approached its $T(\rho$)-phase-instability---the fragment accidentally encountered  an "obstacle" 
(some random stellar system, which later became  "the solar system"). 
As the result of the experienced deceleration, the already-quasi-stable inner matter of the fragment decompressed (in localized zones) due to propagating (inside the fragment) waves of compression-then-decompression, thus shifting into the unstable phase-state of "nuclear fog", further decompressed, and  chains of nuclear transformations 
(in each element of the nuclear fog) led to nuclear (not thermo-nuclear) explosion. 

The key presumption of the scenario is that the fragment had "sufficiently" cooled down by the time of the encounter. 
(The other steps in the described sequence are the natural physical phenomena and their consequences.)

{\bf Model:}  
The focus of this paper is on {\em how}  specifically the process of nuclear-fission can be triggered within the so-far structurally-cohesive fragment. We present a model of internal instability of a compact stellar fragment composed of super-dense nuclear-matter (Sec.~\ref{s:4-1}). With this model, we examine perturbation of the quasi-stable nuclear-matter due to deceleration.  
The derived results of the study are the  criterion for internal stability/instability for the proposed stellar fragment and the characterization of its "small size" (Sec.~\ref{s:4-2}).  
We start by outlining key physical processes involved in the proposed scenario.

\section{Key Physical Processes} 
\label{s:3}

\subsection{Nuclear-Fission-Capable Object of Galactic Origin} 
\label{s:3-1}
  
Compact super-dense objects whose inner matter is nuclear matter, with density similar to that of a nucleus, are capable of cataclysmic nucleogenesis, but such events are typically large-scale---a neutron star merger is an example.
For the solar system to survive the cataclysm, the object had to be relatively "small".

\subsubsection{Compact Super-Dense Stellar Fragment}
\label{s:3-1-1}

Generally speaking, a number of exotic compact stars have been hypothesized 
\cite{Shapiro_1983}, \cite{Glendenning_2000}, 
such as:
\guillemotleft 
quark stars%
\guillemotright  {} 
---a hypothetical type of stars composed of quark matter, or strange matter;
\guillemotleft 
electro-weak stars%
\guillemotright  {} 
---a hypothetical type of extremely dense stars, in which the quarks are converted to leptons through the electro-weak interaction, but the gravitational collapse of the star is prevented by radiation pressure;
\guillemotleft 
preon stars%
\guillemotright  {} 
---a hypothetical type of stars composed of preon matter.   
Even 
\guillemotleft 
dark energy stars%
\guillemotright  {} 
 and 
 \guillemotleft 
 Planck stars%
 \guillemotright  {} 
  have been proposed.  
Other objects could perhaps exist billions of years ago. 
Neutron stars are the most commonly considered compact super-dense objects.  
Because their matter is highly neutronized, 
neutron star mergers---in close binaries with another neutron star or with a black hole---
have been proposed as possible sites for the creation of neutron-rich  
$r$-process nuclei \cite{1974ApJ...192L.145L}, \cite{1976ApJ...210..549L}, \cite{Lattimer_1977}, \cite{Eichler_1989},   \cite{1982ApL....22..143S}. 

Hydrodynamic simulations of NS–NS (with 
$M \sim 1-2 M_{\odot}$) and NS–BH mergers have showed that a non-negligible amount of matter may be ejected. 
Numerically, in the hydrodynamical models, the matter is represented by a set of "particles".  
As detailed in Reference \cite{Bauswein_2013}:  
The code evolves the conserved rest-mass density $\rho_{*}$, the conserved specific momentum $\tilde{u_i}$, 
and the conserved energy density $\tau$, whose definitions evolve the metric potentials and the “primitive” hydrodynamical quantities, that is, the rest-mass density $\rho$, the coordinate velocity $v_i$, and the specific internal energy $\epsilon$. The system of relativistic hydrodynamical equations is closed by an equation of state (EOS) which relates the pressure $P = P(\rho,T,Y_e)$ and the specific internal energy $\epsilon = \epsilon(\rho, T , Y_e)$ to the rest-mass density $\rho$, the temperature $T$, and the electron fraction $Y_e$. The temperature is obtained by inverting the specific internal energy $\epsilon = \epsilon(\rho, T , Y_e)$ for given $\rho$ and $Y_e$. Changes of the electron fraction are assumed to be slow compared to the dynamics (see, e.g., Reference \cite{Ruffert_1997}), and the initial electron fraction, which is defined by the neutrinoless beta-equilibrium of cold NSs, is advected according to ($d Y_e /d t ) = 0$ ($d /d t$ defines the Lagrangian, that is, comoving, time derivative) \cite{Bauswein_2013}. 
The EOS of NS matter is only incompletely known, and numerical studies rely on theoretical prescriptions of high-density matter. 
Comparative studies of various EOSs revealed that  the properties of the ejecta (amount, expansion velocity, electron fraction, temperature) are crucially sensitive to EOS: among the 40 EOSs considered in Reference \cite{Bauswein_2013}, "softer" EOSs resulted in systematically higher ejecta masses.  
The numbers of such fluid particles is $\sim 10^5 - 10^6$.  For example, 550,000 particles were used in NS merger studies by \cite{Goriely_2011}; 350,000 nonuniform (comparable to about 1,000,000 equal-mass) particles with an effective mass resolution of about $2 \times 10^{-6} M_{\odot}$ were used in Reference \cite{Bauswein_2013}. In the sets of tests in Reference \cite{Bauswein_2013},  $\sim 10^3-10^4$ particles became unbounded. 

\begin{figure}[h!]
\centering
\includegraphics[width=0.45\columnwidth]{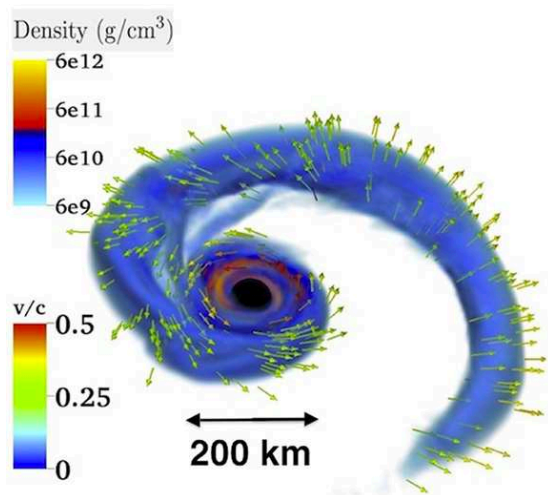}
\caption
[Illustrative simulation of distribution of matter during tidal disruption of a neutron star in neutron star-black hole merger]
{
 Numerical simulation of matter distribution during tidal disruption of a neutron star in neutron star-black hole merger: 
a significant portion 
of the remnant material is unbound \cite{Foucart_2014}.  
}
\label{Fig:2}. 
\end{figure}

Fig.~\ref{Fig:2} (from  \citet{Foucart_2014}) illustrates results of NS-BH merger simulations where 100 particles were tracers.  The position of tracer particle $A$  evolved according to the local 3-velocity: $dx^i_A / dt = v^i (x^i_A)$, 
and fluid quantities at tracer positions were monitored by interpolating from the fluid evolution grid. 
In the depicted simulation, about $0.15M_{\odot}$ was ejected at velocities $\langle v \rangle \sim 0.25c$ \cite{Foucart_2014}. 

The fate of the unbounded particles has usually been considered in the context of galactic enrichment, that is, in 
combination with models of nucleosynthesis.
In such models,  nucleosynthesis calculations are carried out in a post-processing step of the ejecta produced by the hydrodynamical models, using a network of relevant nuclear reactions \cite{Just_2015}.   
Importantly though, because nucleosynthesis studies in mergers are commonly based on simulation data that follow the evolution of the ejecta for timescales shorter, $\sim$ ms, than the $r$-process nucleosynthesis timescale,  $\sim$ s, this makes it necessary to extrapolate the time evolution of thermodynamic properties like temperature and density in order to follow the nucleosynthesis to completion. 
  Thus, it is commonly assumed that the expansion is homologous, $\rho \sim t^{-3}$ \cite{Cowan_2019}. 
  However, this assumption is the assumption for "gas". 
  If some of the particles originated and remained in their "nuclear fluid" state, then such assumption---that 
  \guillemotleft 
the escaping ejecta are assumed to expand freely with constant velocity. The radii of the ejecta clumps thus grow linearly with time $t$ and consequently their densities drop like $1/t^3$%
\guillemotright{} 
 \cite{Goriely_2011}---would not hold for the non-gas particles. 
  The determination of whether the particle originates and remains in the phase-state of nuclear-fluid or gas  is sensitive to the model assumptions about EOS of the neutron-rich matter.%
 \footnote{ 
Additional assumptions are also employed. For example, as noted in Reference \cite{Goriely_2011}, the ejected matter is presumed to be initially cold, but most of it gets shock-heated during the ejection (to temperatures above 1 MeV); its composition is then presumed to be determined by nuclear statistical equilibrium.  
 } 
In Reference \cite{Tito_2018a}, we theoretically demonstrated that fragments---objects composed of dense nuclear matter but smaller (even significantly smaller) than conventional neutron stars (or perhaps other exotic stars)---can indeed exist in a drop-like form, staying as dense as a nucleus, and remaining structurally stable (see Sec.~\ref{s:4}).

\begin{figure}[h!]
\centering
\includegraphics[width=0.45\columnwidth]{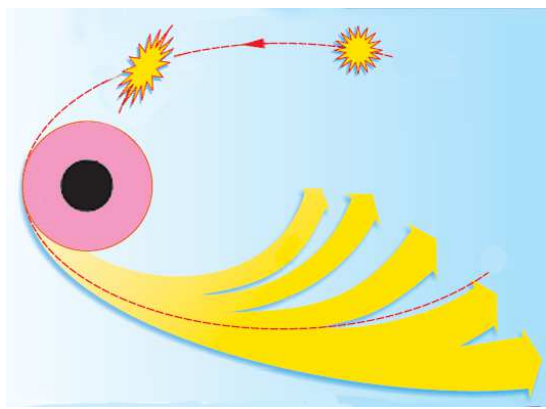}
\caption
[Conceptual illustration of  tidal disruption of a fast-moving star by a massive rotating black hole]
{
Conceptual illustration of  tidal disruption of a fast-moving star by a massive rotating black hole 
(slicing along $\theta = 0$ surface reveals "pink" ergosphere and "black" BH with its outer event-horizon) 
the star may be torn into pieces, some of which become captured, while others catapulted by BH  
(adapted from Reference \cite{Rees_1990}).   
}
\label{Fig:3}
\end{figure}

Furthermore, besides NS mergers, small fragments can be formed and  catapulted  
if a black hole tears a neutron star apart \cite{Rees_1990} without merging with it.  
Indeed,  during the rotating core collapse, one or more self--gravitating lumps of neutronized matter can form in close orbit around the central nascent neutron star \cite{Imshennik_1998}.
The unstable (in the phase-transition and nuclear-reaction sense) member of such transitory multi-fragment system ultimately explodes,
giving the surviving member a substantial kick velocity---as fast as $\sim 1600 \, km/s$ \cite{Colpi_2002}.%
\footnote{
A similar hyper-velocity $\sim 1700 \, km/s$ was recently observed even for a main sequence star kicked 4.8 Myr ago with the implied velocity $\sim 1800 \, km/s$  by the supermassive black hole Sgr~$A^*$ at the galactic center \cite{Koposov_2019}. 
}  
Fig.~\ref{Fig:3} (adapted based on the image from \citet{Rees_1990}) illustrates the scenario.  
\begin{figure}[h!]
\centering
\includegraphics[width=0.45\columnwidth]{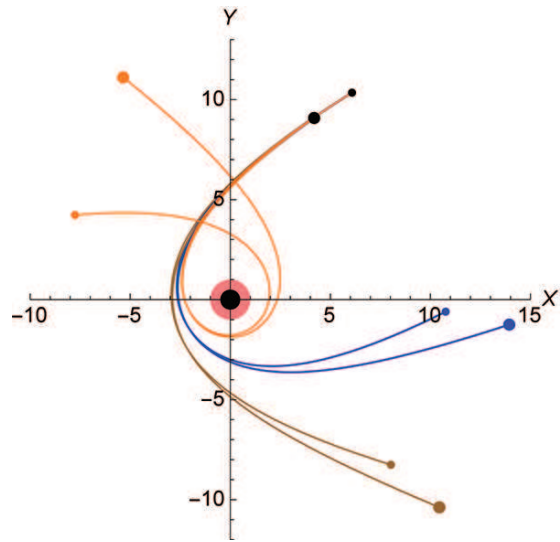}
\caption
[Illustrative simulation of tidal destruction of a stellar body by a fast-rotating massive BH]
{
Illustrative simulation of the process of  tidal destruction of a stellar body by a fast-rotating massive black hole: 
two fragments of the initially cohesive stellar body---depicted as the black dot near dimensionless coordinates (+6; +10)---
follow different, bound or unbound, trajectories depending on the initial parameters of the system \cite{Tito_2018b}. 
}
\label{Fig:4}
\end{figure}

In our prequel study \cite{Tito_2018b}, we conducted an analysis of a set of scenarios where a body approaching the vicinity of a  massive rotating black hole and demonstrated that indeed a tidally-torn fragment can be catapulted by the black hole.  Fig.~\ref{Fig:4} illustrate such possibility (three scenarios depicted).

In the presented in this paper hypothesis, we propose that the solar system encountered one of such fragments---
a clump of nuclear matter  resembling (in essence) a giant "nuclear drop" (a mega-nucleus).

\subsubsection{Instability of Nuclear Matter:  Nuclear Fog} 
\label{s:3-1-2}

For the proposed scenario of enrichment of the solar system due to a "local" (within the system) cataclysm, 
the fragment of nuclear matter (a giant "nuclear drop") had to arrive from afar (i.e., remain structurally-stable during the journey), but then explode (i.e., lose its structural integrity). 
 Such effect can be achieved if upon encounter with the solar system, due to experienced perturbation, the nuclear matter underwent its $T(\rho)$-phase-state evolution---between the phases of quasi-stable nuclear-liquid and unstable (due to onset of nuclear transformations below $\rho_{drip}$) nuclear-gas---{\em through} the spinodal zone of mixed-phase nuclear-fog.

\paragraph{Mixed-Phase (Nuclear Fog) and Spinodal Zones}

\begin{figure}[ht!]
\centering
\includegraphics[width=0.50\columnwidth]{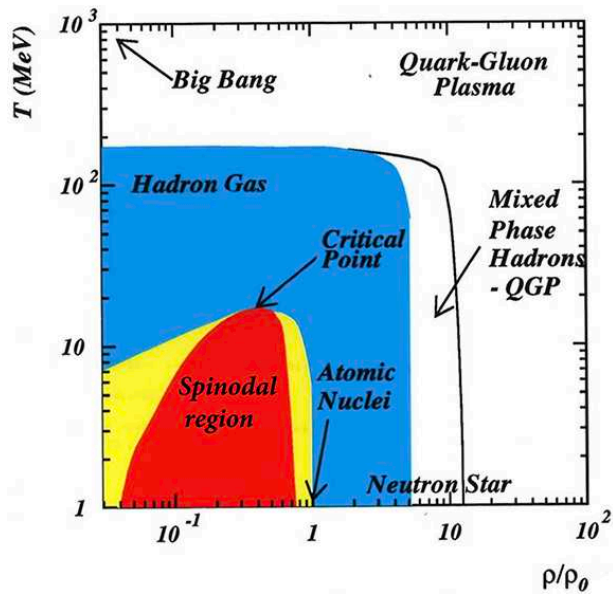} 
\caption
[Phase diagram $T(\rho)$ for nuclear matter]
{
Phase diagram $T(\rho)$ for nuclear matter  
(density $\rho$ is expressed in units of  $\rho_0 \equiv \rho_{nucleus} \simeq 2.85 \times 10^{14} \, g/cm^3 $,  
temperature $T$ is expressed in $MeV$ units, $1 \, MeV \simeq 10^{10} K$): 
the liquid-gas mixed phase region (yellow area, which ends up at the critical point) contains the spinodal region (red area) \cite{Borderie_2019}.  
}
\label{Fig:5} 
\end{figure}

The fact that nuclear matter may exist in the two-phase state has been known for a while \cite{Jaqaman_1983}, \cite{Karnaukhov_2005}. 
Fig.~\ref{Fig:5} qualitatively depicts  $T(\rho)$  phase diagram for nuclear matter. 
The liquid/gas mixed-phase region (yellow area) which ends up at the critical point contains the spinodal region (red area).
Below the critical temperature $T_c$, depending on its density, nuclear matter can exist in nuclear-liquid phase (higher range of densities), or nuclear-gas phase (lower range of densities), or as nuclear fog  which is a mixture of both phases (within the spinodal zone of the density range corresponding to its $T$).

\paragraph{Equation of State (EOS) for Nuclear Matter Permitting Nuclear Fog}

\begin{figure}[ht!]
\centering
\includegraphics[width=0.55\columnwidth]{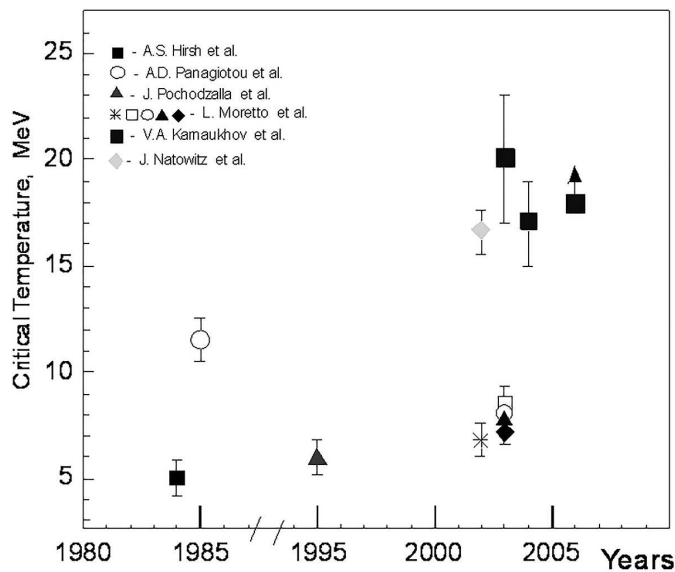}
\caption
[Experimental values of critical temperatures of nuclei ($T_c$)]
{
Experimental values of critical temperatures of nuclei ($T_c$): 
historically consensus $T_c$  has varied, as different measurement techniques produced ranging $T_c$ estimates 
\cite{Karnaukhov_2011}.
}
 \label{Fig:6}
\end{figure}

High-energy nuclear experiments (in terrestrial conditions) have demonstrated that
the matter of a typical heavy-nuclei is characterized by critical parameters, such as  temperature $T_c$ and density $\rho_c$ 
(see, for example, References
  \cite{Jaqaman_1983}, \cite{ Jaqaman_1984}, \cite{Karnaukhov_2006}, \cite{Karnaukhov_2009}, 
\cite{Karnaukhov_2011}, and references therein and within Reference \cite{ Tito_2018a}). 
Critical temperature $T_c$ for the liquid-gas phase transition is a crucial characteristic of the nuclear equation of state. 
Estimating it, however, has been a challenge. 
Over the years experimental studies have provided a range of estimates (see Fig.~\ref{Fig:6}).
Calculations (performed in a number of works, for example, References \cite{Jaqaman_1983}, \cite{Goodman_1984}, \cite{Sil_2004}, \cite{Sauer_1976}, \cite{Zhang_1996}) have determined that depending on the chosen effective interaction  and on the chosen model (see Reference \cite{Jaqaman_1983}, \cite{Jaqaman_1984}, \cite{Csernai_1986}, \cite{M_ller_1995}),
the nuclear equation of state exhibits a critical point at $\rho_c \simeq (0.1 \div 0.4)\rho_0$ and $T_c \sim 5 \div 18 \, MeV$ \cite{Karnaukhov_2006}, \cite{Karnaukhov_2011}.  
Value $T_c = 17.5 \, Mev$ is commonly used and hence we use it in the model calculations presented in Sec.~\ref{s:4}.  

Notably, in laboratory conditions and experiments, parameters of nuclear targets are such that 
$T < T_c$ and $\rho_{nucl} \sim 2 \div 3 \rho_c$. 

\begin{figure}[ht!]
\centering
\includegraphics[width=0.50\columnwidth]{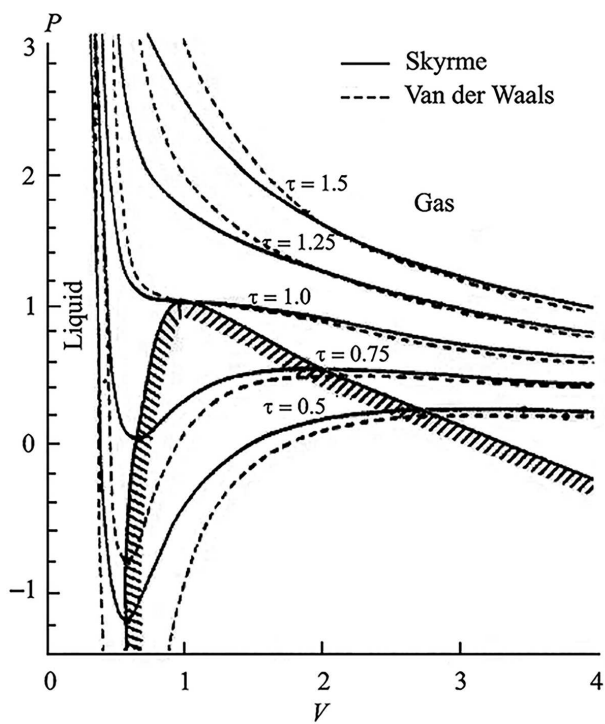} 
\caption
[Equations of state $P(V)$ for a nuclear system interacting through a Skyrme potential and a Van der Waals compressible liquid-gas system]
{
Equations of state $P(V)$ for a nuclear system interacting through a Skyrme potential and a Van der Waals compressible liquid--gas system  (shown in relative units): qualitative similarity is apparent  \cite{Jaqaman_1983}.  
}
\label{Fig:7}
\end{figure}

The equations of state $P(V)$ of a multi--body system of nucleons interacting via Skyrme potential  
is shown in  Fig.~\ref{Fig:7}.
EOS isotherms---$P(V)$ at constant temperature---corresponding to Skyrme effective interaction and finite temperature of Hartree-Fock theory (see Reference \cite{Jaqaman_1983}) exhibit the maximum-minimum structure typical of the Van der Waals-like EOS.  
The very steep part of the isotherms (on the left side) corresponds to the liquid phase. 
The gas phase is presented by the right parts of the isotherms where pressure is changing smoothly with increasing volume. 
Between them lies the mixed zone where two phases can co-exist---for nuclear matter, such mixture is either liquid droplets surrounded by gas of neutrons, or homogeneous neutron-liquid with neutron-gas bubbles. 

In the spinodal zone (marked by the hatched line in Fig.~\ref{Fig:7}), 
where the isotherms correspond to the negative compressibility, that is, $(\partial P / \partial V )_T > 0$,  
random density fluctuations lead to almost instantaneous collapse of the initially uniform system into a mixture of two phases. Obviously, within the spinodal zone, the zone of collective instability (where the square of adiabatical speed is negative) is inside the coexistence zone (where the square of isothermical speed is negative).

To develop our model (discussed in Sec.~\ref{s:4}), which we used to examine stability/instability of the stellar fragment, 
we constructed a nuclear-fog-interpolating EOS that satisfies these (and several other) conditions and properties.

\paragraph{Decompression of Nuclear Matter}

\begin{figure}[h!]
\centering
\includegraphics[width=0.55\columnwidth]{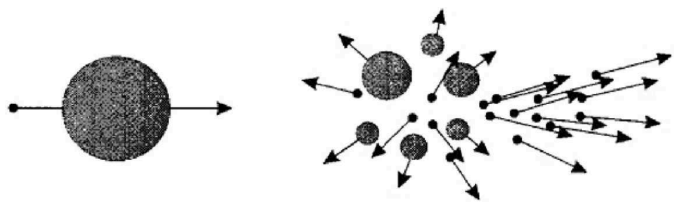}  
\caption
[Conceptual illustration of the process of multi-fragmentation of excited heavy nucleus]
{
Conceptual illustration of the process of multi-fragmentation of excited heavy nucleus \cite{Karnaukhov_2006}. 
Causes of nucleus excitation may vary.   
}
\label{Fig:8}
\end{figure}

If the equilibrium state of the inner nuclear-liquid of the stellar fragment is initially close to the boundary of the liquid/gas phase transition, then the nuclear-liquid phase can decompress into the nuclear-fog phase due to some perturbation   
(see, for example, References \cite{ Tito_2018a},  \cite{Jaqaman_1983}, \cite{ Jaqaman_1984}, \cite{Karnaukhov_2006}, \cite{ Karnaukhov_2009}, \cite{ Karnaukhov_2011}, and references therein).
The matter would then exist as a mixture of two phases of nuclear matter---either liquid droplets surrounded by gas of neutrons, or generally homogeneous neutron-liquid with neutron-gas bubbles.
In such state, the matter can reach substantial further rarification, reducing density by a factor of $10^2$ or more due to hydrodynamic instability.  
At this stage, cascading nuclear {\em fragmentation} of the nuclear-droplets and subsequent {\em fission} of these fragments may start. 
Fig.~\ref{Fig:8} (from \citet{Karnaukhov_2006}) illustrates decomposition of a perturbed heavy nucleus (here due to collision with a swift proton). 

\begin{figure}[h!]
\centering
\includegraphics[width=0.65\columnwidth]{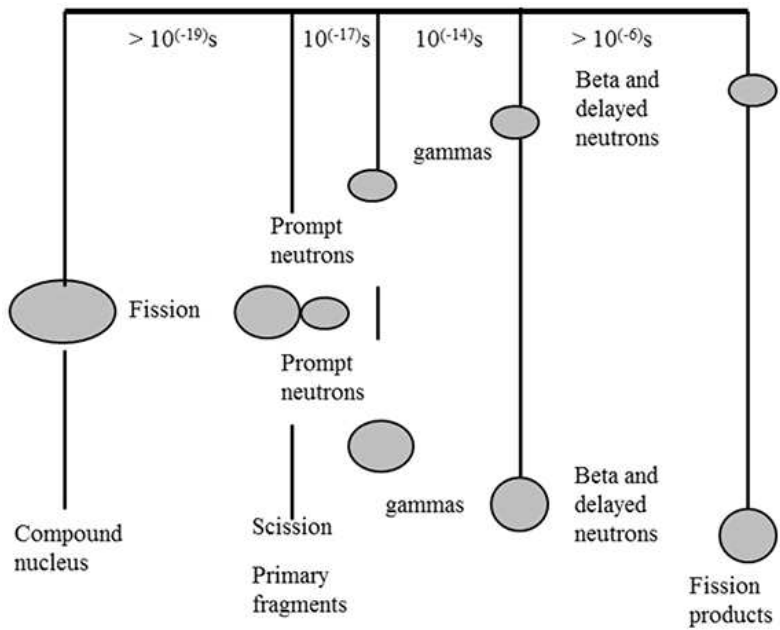}
\caption
[Conceptual illustration of the fission process with characteristic timescales]
{
Conceptual illustration of the fission process with characteristic timescales \cite{Goutte_2015}. 
}
\label{Fig:9}
\end{figure}

Below density $\rho_{drip}$---even if in some small physical domain within the object --
$\beta$-decay becomes no longer Pauli-blocked. 
This process triggers cascading  {\em fragmentation} of these supersaturated mega-nuclei (see, for example, References \cite{Bertsch_1983}, 
 \cite{Heiselberg_1988}, 
\cite{L_pez_1989}).
These reactions, known to release substantial energy ($\sim 1 MeV$ per fission nucleon, as seen in
transuranium nuclei fission events), proceed effectively at the same moments as the $\beta$-decay reactions.
Everything happens very fast, with time scales of the order of nuclear-time scales. 
Fig.~\ref{Fig:9} sketches stages of the fission process. 

\begin{figure}[h!]
\centering
\includegraphics[width=0.55\columnwidth]{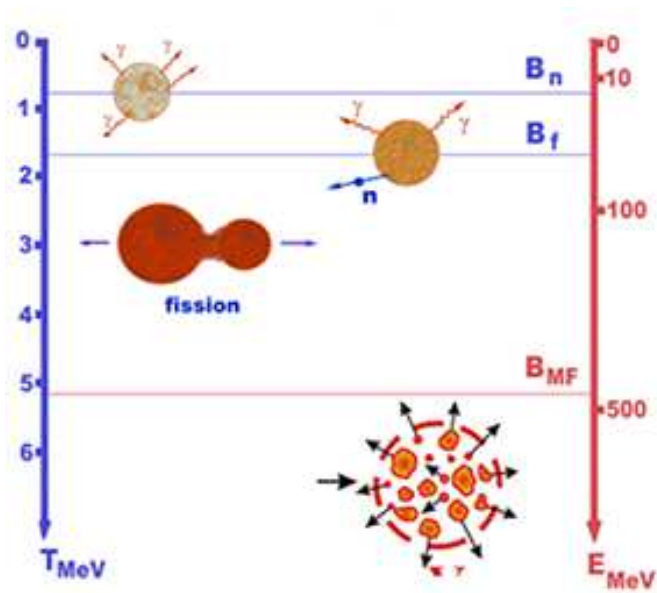}   
\caption
[Decay modes of excited nuclei]
{
Decay modes of excited nuclei, A= 150  
\cite{Karnaukhov_2011}. 
}
\label{Fig:10}
\end{figure}

Generally speaking, at different stages (with respect to applied energy/excitation of heavy nuclei), different types of reactions occur   \cite{Karnaukhov_2011}.
When a heavy nucleus is excited (relatively) weakly, only {\em $\gamma$--emission} occurs.
At a higher level of excitation, {\em neutron-emissions} start taking place.
When even  more energy is applied to the heavy nucleus, it deforms and {\em fission} starts because, as known,
for deformed charged nuclei with parameter $Z^2/A > 50$, electrostatic repulsion starts to exceed 
the surface tension effect. 
And finally, when injected energy is sufficiently high, {\em fragmentation} --
splitting into fragments ("droplets" if the initial nucleus is a mega-nucleus)---occurs,
followed by the cascade of subsequent splitting into fragments and  neutron-, $\beta$-, and $\gamma$- emissions. 
Fig.~\ref{Fig:10} illustrates the processes in relation to excitation energies (barriers) and corresponding temperatures 
resulted from applying the energy needed to trigger the reaction. 

\subsection{Nuclear-Fission-Trigger (Perturbation Due to Encounter with "Obstacle")}
\label{s:3-2}

As already noted in Sec.~\ref{s:2-4}, 
although a number of mechanisms may exist that could perturb the quasi-stable nuclear matter in the stellar fragment, 
we envisioned and  present in this paper just one---localized {\em decompression} inside the fragment due to its {\em deceleration}.   
The details of the model and examination of the process are elaborated in Sec.~\ref{s:4}.
In brief,  we suggest that as the result of the experienced deceleration (Sec.~\ref{s:3-2-1}), 
the quasi-stable (because of cooling during the long journey)
inner matter of the fragment decompressed (in localized zones) due to propagating (inside the fragment) waves of compression-then-decompression (Sec.~\ref{s:3-2-2}), thus shifting into the unstable phase-state of nuclear fog (Sec.~\ref{s:4-2-1}), further decompressed, and chains of nuclear transformations (in each element of the nuclear fog, Fig.~\ref{Fig:10})  led to nuclear (not thermo-nuclear) explosion. 

Conceptually, we can think of three specific "obstacles" within the inner part of the solar system which could  
decelerate the stellar fragment: 
(1) the Sun itself (the edge of it, since the Sun survived); 
(2) a possibly then existing binary companion of the Sun (discussed in  Sec.~5.3.2.A);  
and  
(3) a possibly then existing "inner"-Jupiter located between the Sun and Jupiter (discussed in Sec.~5.3.2.B).

\subsubsection{Effect of Deceleration}
\label{s:3-2-1}

Generally speaking, 
a number of mechanisms contribute to the object's deceleration as it penetrates a medium: 
classical drag \cite{Landau_1987},
dynamical friction \cite{Chandrasekhar_1943},
accretion \cite{Tito_2016},
Cherenkov-like radiation of various waves related to collective motions \cite{Pavlov_2009} generated within the medium \cite{Pavlov_1985}, \cite{Pavlov_1990},
distortion of the magnetic fields, and possibly others.  
Obviously, some deceleration causes would be dominant and some would be negligible.  
Analytical and numerical treatment of the deceleration process can quickly become complex and cumbersome.
However, in the context of the question of whether explosion can be triggered by internal instability, 
here we focus on the {\em effect} rather than the cause of the deceleration.

The effect of (rapid) deceleration---from the moment of "encounter" to the moment of initiation of "structural decomposition" of the stellar fragment---exhibits itself as follows: 
As the stellar fragment (with a nuclear-drop-like shape) encounters an "obstacle", it decelerates and its   
inner matter stratifies – first the compression shockwave propagates from the front point towards the back, then (because the fragment's surface was strain-free due to extreme density contrast between the inner and outer media) the reflected shockwave reverses polarity \cite{Zeldovich_1966} and returns as the wave of decompression.  
Generally, in a nuclear-like medium, the shockwave propagation speed is comparable with the speed of light, so the stratification process develops very quickly. 
During such short time, the shape of the drop does not have time to change because propagation speed of surface perturbations is much slower than the speed of body waves. 
In the proposed scenario, in the zones of decompression, the matter that was previously (thermodynamically) quasi-stable (perhaps due to aging and cooling of the stellar fragment), now became unstable and "preferred" not the homogeneous but the two-phased state (the state of nuclear fog where "nuclear droplets" coexist with "nuclear gas").

\subsubsection{Compression/Decompression}
\label{s:3-2-2}

When a droplet collides with some object (obstacle), inside the droplet---as known---various motions arise, the velocity of which is comparable with the velocity of the droplet. If the droplet's initial velocity is comparable with the "speed of sound" within the droplet's matter, then compressibility becomes apparent \cite{Landau_1987}.

The following phenomena 
arise inside the droplet upon collision:
excitation and propagation of shockwaves of compression and decompression,
interaction of the waves with each other and with free surfaces,
formation and development of radial near-surface cumulative jet,
formation and collapse of cavitation bubbles inside the droplet,
and other complex hydrodynamic phenomena.

\begin{figure}[ht!]
\centering
\includegraphics[width=0.55\columnwidth]{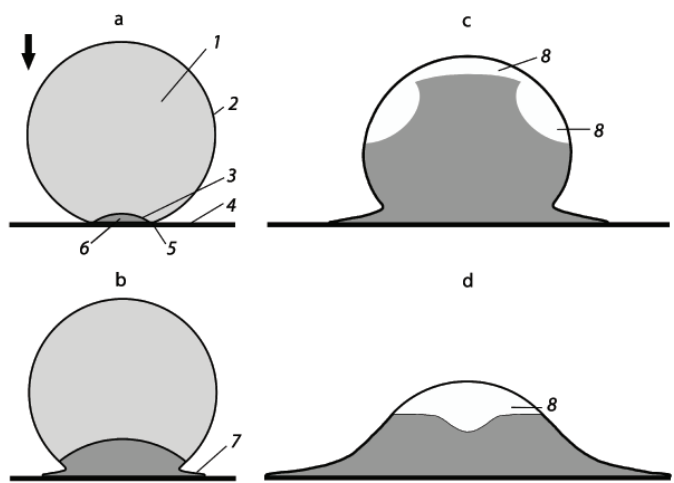}
\caption
[Conceptual illustration of processes occurring when a liquid drop falls onto a surface (encounters an obstacle)]
{
Conceptual illustration of processes occurring when a liquid drop falls onto a surface 
(encounters an obstacle);  adapted from Reference \cite{Chizhov_2000}. 
Panels: (a) before spreading; (b) jet initiation; (c) shockwave approaches the top of the drop, toroidal expansion region is formed; and (d) initiation of vast expansion area with cavitation region. Zones: (1) unperturbed liquid, (2) free drop surface, (3) shockwave, (4) obstacle's surface, (5) contact boundary, (6) compressed liquid area, (7) jet, and (8) cavitation region. 
}
 \label{Fig:11}
\end{figure}

Quantitative numerical simulations of these effects show that results are strongly model-dependent, particularly, on the choice of the model EoS for the droplet's matter. Even the qualitative picture of a high-velocity collision is not yet fully understood. Understanding of many aspects remains incomplete, such as the roles of viscosity and surface tension even in the case of the simplest model EoS of the liquid,
the mechanisms of development and destruction of the cumulative jet,
the estimates of velocity of the radial jet,
the mechanism of formation of cavities,
the strains experienced on the obstacle, and so on.

Qualitatively the process of high-velocity collision can be described as follows (see Fig.~\ref{Fig:11} 
re-drawn based on Reference \cite{Chizhov_2000}):

During the process of interaction of the droplet with the surface of the obstacle, the flow of fluid forms, which develops a strongly-non-linear wave structure and strongly deforms free surfaces.

One of the features of collision of a convexly-shaped droplet  is that at the beginning stage, the free surface of the droplet that does not touch the surface of the obstacle, does not deform. The region of compression is confined to the shockwave that forms at the edge  of the contact spot (Fig.\ref{Fig:11}a).

Furthermore, there develops a near-surface wave. (The front of which is tangential to the front of the shockwave,
and starts from the edge of the contact spot. It is not shown in Fig.\ref{Fig:11}a)

This is explained by the fact that the speed of expansion of the contact spot
$V_0 (t) = V_0 \cot \beta (t)$
 (here $V_0$ is the initial velocity of the drop,
 $\beta (t)$  is the angle between the drop's free surface and the obstacle's surface at moment $t$)
is greater than the speed of propagation of the shockwave within the droplet's medium from time zero to the critical moment $t_c$ when these speeds match --
the speed of the contact spot boundary diminishes from its infinite value at the moment of contact,
but remains  greater than the  speed of the shockwave until the moment $t_c$. Therefore, during this time perturbations expanding from the contact spot do not interact with the free surface of the droplet.
At the edge of the contact spot, compression of the droplet's liquid is maximal.

At the critical moment of time $t_c$, the shockwave detaches from the edge of the contact spot and interacts with the free surface of the droplet, and a reflective decompression wave forms \cite{Zeldovich_1966} which propagates inward (toward the central zone of the drop). The free surface becomes deformed, and a near-surface high-velocity radial jet of cumulative type forms (Fig.~\ref{Fig:11}b). The time of formation of the jet depends on the viscous and surface effects within the liquid near the surface of the obstacle, its velocity substantially exceeds the velocity of collision.

Once the wave is reflected from the droplet's free surface, the change in polarity of impulse occurs \cite{Zeldovich_1966}. The reflective wave of decompression forms a toroidal cavity, the cross-section of which is qualitatively shown in Fig.~\ref{Fig:11}c.

At the final stage of interaction, the wave of decompression collapses onto the axis of symmetry, and forms a vast cavity with most decompression occurring in the region near the axis  (Fig.~\ref{Fig:11}d).

During the propagation of the decompression wave toward the surface of the obstacle, the cavity fills almost the entire volume of the droplet, except for the thin layer near the droplet surface and the zone occupied by the near-surface jet. As the result of development of instability within this thin envelop, the droplet becomes shaped as a "crown", and the matter of the droplet becomes splashed out in small fragments.

\subsection{Nuclear-Fission-Driven Nucleogenesis}
\label{s:3-3}

Once the giant "nuclear drop" lost its structural integrity, and all of the matter significantly decompressed, the process of nuclear-fragmentation and nuclear-fission of the separated mega-nuclei (Sec.~\ref{s:3-3-2}) continued, producing further fragmentation, again and again, which combined with the full set of various captures and decays possible in the environment that was both neutron-rich (due to the matter contained in the stellar object) and $H/He$-rich (due to the matter contained in the gaseous obstacle).  The outcome of such event is the multitude of nucleogenetic cascades (Sec.~\ref{s:3-3-3}).

\subsubsection{Evolution Equations}
\label{s:3-3-1}

For nuclei of each type ($A,Z$) (where $A=Z+N$ is the total number of nucleons, and $Z$ and $N$ are the numbers of protons and neutrons, respectively) the evolution equations for mass fractions $Y(t)$---
or number concentrations $C(t)$, which are related as $Y(t) = C(t) / (\rho N_A)$---
are traditionally written in the form similar to the following (see, for example, References \cite{Panov_2005}, \cite{Panov_2010}, \cite{Korneev_2011}): 

\begin{equation}
\begin{aligned}
 \frac{d Y_{A,Z} }{ dt} = 
%
%
\sum\limits_{A_f,Z_f}^{} 
\lambda_{F_s} (A_f,Z_f) 
W_{F_s}(A_f,Z_f, A, Z) 
Y_{A_f,Z_f}  \\
+ 
\sum\limits_{A_f,Z_f}^{} 
\lambda_{F_n} (A_f,Z_f) 
W_{F_n}(A_f,Z_f, A, Z) 
Y_{A_f,Z_f}  
+ 
\sum\limits_{A_f,Z_f}^{} 
\lambda_{F_{\gamma}} (A_f,Z_f) 
W_{F_{\gamma}}(A_f,Z_f, A, Z) 
Y_{A_f,Z_f}  \\
+ 
\sum\limits_{A_f,Z_f}^{} 
P_{\beta d f} (A_f,Z_f) 
\lambda_{F_{\beta}} (A_f,Z_f) 
W_{F_\beta}(A_f,Z_f, A, Z) 
Y_{A_f,Z_f}  \\
+ 
\sum\limits_{l}^{} 
\sum\limits_{k}^{} 
P_{kl} (A+k,Z-l) 
\lambda_{\beta} (A+k,Z-l) 
Y_{A+k,Z-l}  \\
 +
\lambda_{\gamma \alpha} (A+4,Z+2)
 Y_{A+4,Z+2}  
 +
\lambda_{\alpha \gamma} (A-4,Z-2)
 Y_{A-4,Z-2}  \\
+
\lambda_{\gamma n} (A+1,Z)
 Y_{A+1,Z}  \quad \quad \; \; \,
+
\lambda_{n \gamma} (A-1,Z)
 Y_{A-1,Z}  
 \quad 
 \\
+
\lambda_{\gamma p} (A+1,Z+1)
 Y_{A+1,Z+1}  
+
\lambda_{p \gamma} (A-1,Z-1)
 Y_{A-1,Z-1}  \\
     +
  \lambda_{n p} (A,Z+1)
 Y_{A,Z+1}  \quad \quad \; \; \,
    +
  \lambda_{p n} (A,Z-1)
 Y_{A,Z-1}  
 \quad 
 \\
  +
   \lambda_{p \alpha} (A+3,Z+1)
 Y_{A+3,Z+1}  
 +
   \lambda_{\alpha p} (A-3,Z-1)
 Y_{A-3,Z-1}  \\
 +
    \lambda_{n \alpha} (A+3,Z+2)
 Y_{A+3,Z+2}  
 +
   \lambda_{\alpha n} (A-3,Z-2)
 Y_{A-3,Z-2}  \\
   +
    \lambda_{\nu e} (A,Z-1)
 Y_{A,Z-1}  
 \quad 
 \\
 %
 %
 - 
 [ 
 \lambda_{F_s} (A,Z)
 +
 \lambda_{F_i} (A,Z)
  +
 \lambda_{F_{\beta d f}} (A,Z)
 +
 \lambda_{\beta} (A,Z)
 +
 \lambda_{\alpha \gamma} (A,Z)
 +
 \lambda_{n \gamma} (A,Z)
  \qquad \qquad \quad
 \\
 +
 \lambda_{\gamma n} (A,Z) 
  %
  +
  \lambda_{p \gamma} (A,Z) 
   +
  \lambda_{\gamma p} (A,Z) 
  +
  \lambda_{n p} (A,Z)
    +
  \lambda_{p n} (A,Z) 
  \qquad \quad
  \\
   +
  \lambda_{p \alpha} (A,Z)
    +
  \lambda_{\alpha p} (A,Z)
  +
   \lambda_{\alpha n} (A,Z)
 +
   \lambda_{n \alpha} (A,Z)
   +
   \lambda_{\nu e} (A,Z)
] 
Y_{A,Z} 
 \, . \\
\end{aligned}
\label{Eq:1}
\end{equation}
\begin{equation}
\begin{aligned}
&
\frac{d Y_{n} }{ dt} =  
\lambda_{\nu \nu' n} Y_{\alpha} 
-
\sum\limits_{Z}^{} 
\sum\limits_{A}^{}
\left[
\lambda_{n \gamma} 
- \lambda_{\gamma n} 
+
 \lambda_{n p} 
- \lambda_{p n} 
+
\lambda_{n \alpha} 
- \lambda_{\alpha n} 
- 
\sum\limits_{k}^{} 
k
P_{k} (A,Z) 
\lambda_{\beta} (A,Z) 
\right] 
Y_{A,Z}  \, ,\\
&
\frac{d Y_{p} }{ dt} = 
\lambda_{\nu \nu' p} Y_{\alpha} 
-
\sum\limits_{Z}^{} 
\sum\limits_{A}^{}
\left[
\lambda_{p \gamma} 
- \lambda_{\gamma p} 
+
 \lambda_{n p} 
- \lambda_{p n} 
+
\lambda_{p \alpha} 
- \lambda_{\alpha p} 
\right] 
Y_{A,Z}  \, ,\\
& 
\frac{d Y_{\alpha} }{ dt} = 
- (\lambda_{\nu \nu' n} + \lambda_{\nu \nu' p})Y_{\alpha} 
-
\sum\limits_{Z}^{} 
\sum\limits_{A}^{}
\left[
\lambda_{\alpha \gamma} 
- \lambda_{\gamma \alpha} 
+
 \lambda_{\alpha p} 
- \lambda_{p \alpha} 
+
\lambda_{\alpha n} 
- \lambda_{n \alpha} 
\right] 
Y_{A,Z}  \, .\\
\end{aligned}
\label{Eq:2}
\end{equation}
\noindent
Here, 
$n$ and $p$ denote neutron and proton components; 
$F_s$ denotes spontaneous fission, $F_{n,\gamma}$ denotes $n$-induced or $\gamma$-induced fission, $F_\beta$ denotes $\beta$-delayed fission; 
$A_f$ and $Z_f$ denote the mass and charge number of the mother nucleus, respectively;    
$W$ are the weighting functions of the fission products; 
$P_{\beta d f}$ are the probabilities of delayed fission after $\beta$ decay; 
$P_{k}$ are the probabilities of emission of  $k$ neutrons; 
$P_{kl}$ are the probabilities of (single- or multi-) $\beta$-decays with emission of $l$ electrons and $k$ delayed neutrons; 
$\sum\limits_{}^{} P_{} = 1$ over all channels relevant for nuclide (A,Z);  
$\lambda$ specify the rates of various processes, 
 $\lambda_{ij}$ indicates $i$-induced $j$-release. 
For multi-participant processes (such as various captures), expressions for $\lambda = \lambda (Y, ...)$
include dependence on concentrations of the involved participants (other than the subject nucleus) and on reaction cross-sections. 
For single-participant processes (such as spontaneous fission or decays), expressions for $\lambda$ include dependence on the properties of the specific nucleus. 
(In Eqs.~\ref{Eq:2}, $\lambda_{\nu \nu' n} (t)$ and $\lambda_{\nu \nu' p} (t)$ are the rates of neutron and proton evaporation through inelastic scattering of different types of neutrinos and antineutrinos by $\alpha$ particle.)

This traditional form of kinetic/evolution equations is convenient for designing numerical simulation routines. 
Another form may be constructed to illuminate key features of specific mechanisms, and to juxtapose nucleogenesis models for the proposed event within the solar system  and for traditional stellar cataclysms. 
For example, one can consider  expanding the set of notations to include 
-- in addition to $Y (A=Z+N, Z)$ expressed in terms of ($Z, N$) 
so that $Y (A,Z) \equiv$ $Y^{N}_{Z}$ $\equiv Y_{i}^{k}$ $\equiv Y_{ik}^{}$  
-- 
 the following notations: 
$Y_0^0$ to symbolize  $\gamma$ abundance; 
$Y_{\pm1}^{0}$ to symbolize 
$e^{\pm}$ (electron or positron) abundance; 
$Y_1^0$ and $Y_0^1$  to symbolize 
proton and neutron abundances, respectively.      
Then, a (rather sizable) vector-column can be constructed containing the abundances 
of the mentioned particles and all theoretically-possible nuclides 
(with $N = \overline{0, N}_{limit}$ for every $Z$, where $N_{limit}$ is theoretically not limited, 
although numerically some cutoff is always imposed). 
Every component $Y_{ik}$ of such vector may be visualized as corresponding to a cell on the ZN-plane whose axes extend in  both positive and negative directions (representing particles and anti-particles). 

Then, the set of kinetic equations (Eqs.~\ref{Eq:1} and \ref{Eq:2}) in the leading approximation can be expressed in the form: 
\begin{equation}
\begin{aligned}
\frac{d Y_{\imath k}}{dt} = \Omega_{\imath k}^{\jmath m \, \ell n} Y_{\jmath \ell} Y_{m n } 
+ \Lambda_{\imath k}^{q s} Y_{q s} 
+ \ldots 
\end{aligned}
\label{Eq:3}
\end{equation} 
where summations over indices---$\imath$, $k$, $\jmath$, $m$, $\ell$, $n$, $q$, and $s$---are performed in accordance with the standard rules. 
Coefficients in $\Omega$ characterize the combined production/exodus impact on nuclei ($Z= i, N=k$) 
by all possible channels from multi-participant processes (such as captures), 
coefficients in $\Lambda$ - from single-participant processes (such as spontaneous fission and decays).

This form of kinetic equations reveals that the existence (abundance) of every nuclide may be 
linked, directly or indirectly, to every other nuclide (from the entirety of the rather extensive set)---
but of course, the lifespan of some linkages may be very short,  
justifying the assumption of their zero impact on the outcomes in some  settings
(for example, in the stationary or constrained ones), 
but not necessarily in others (see discussion below).

\begin{figure}[h!]
\centering
\includegraphics[width=0.55\columnwidth]{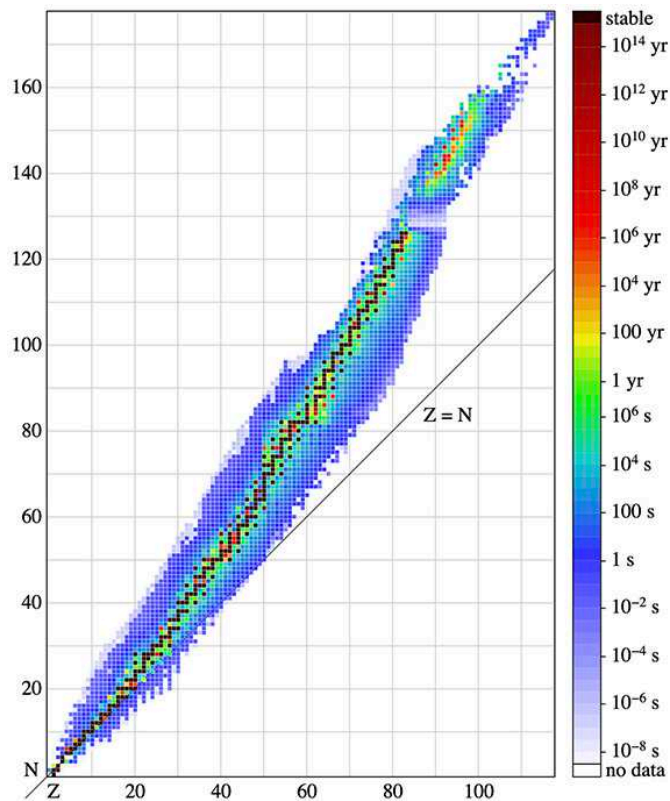} 
\caption
[Half-lives of isotopes]
{
Half-lives of isotopes (Z: number of protons, N: number of neutrons). 
Data from National Nuclear Data Center (NuDat2 database 6/1/2012) \cite{Wiki_2012}. 
}
\label{Fig:12}
\end{figure}

Fig.~\ref{Fig:12} depicts the double-positive quadrant of the ZN-plane, 
the cells of which represent every theoretically-possible nuclide (Z,N), colors denote half-lives of the studied isotopes. 
The vast number of uncolored cells are the unexplored nuclides---at present, they remain beyond the technical feasibilities of observations. 
And, while the question  is still open \cite{Nazarewicz_2018} 
about whether there exists a fundamental limit on the maximum charge $Z$ in a nucleus (to qualify as a chemical element%
\footnote{From \citet{Giuliani_2019}: 
\guillemotleft 
According the report of the Transfermium Working Group \cite{Wapstra_1991}, 
in order to talk about a new element, the corresponding nuclide with
an atomic number $Z$ must exist for at least $10^{-14}$s, which is a reasonable estimate of the
time it takes a nucleus to acquire its outer electrons, bearers of the chemical properties.
Consequently, if for all isotopes of some superheavy element, including isomeric states 
\cite{Heenen_2015}, \cite{Jachimowicz_2017}, 
nuclear lifetimes are shorter than $10^{-14}$s, the corresponding element does not exist. 
On the other hand, in order to define a nuclide, its
lifetime should be longer then the single-particle time scale $T_{s.p.} \approx 1.3 \times  10^{-22}$s 
\cite{Goldanskh_1966}, \cite{Thoennessen_2004} 
that corresponds to the time scale needed to create the nuclear mean field. 
Consequently, there is no chemistry for nuclides with lifetimes between $10^{-14}$s and $10^{-22}$s.%
\guillemotright {}
}), 
no such question (or limit) exists for the number of neutrons $N$. 
(Indeed, in the purest model, a neutron star is essentially a giant-nucleus composed of neutrons.)   
Therefore, conceptually, the rows of cells representing nuclides should  continue upward, 
into the zone of nuclear-fog. 
Indeed, above the depicted domain in Fig.~\ref{Fig:12} the theoretically-predicted super-heavy  "islands of stability" 
-- at ($Z \sim 114, N \sim 184-196$), ($Z \sim 138, N \sim 230$), ($Z \sim 156, N \sim 310$), and ($Z \sim 174, N \sim 410$)---
should be indicated \cite{Greiner_2013}, \cite{Afanasjev_2018}.

Conventional models of {\em nucleosynthesis} (in stellar cataclysmic events) 
consider production of nuclei from lower ($Z,N$) towards higher ones. 
In simulations, numerical models naturally include those nuclides for which data is available, 
thus stopping at about $A \sim 320$ or $Z_{max} \sim 110$  
\cite{Mendoza_Temis_2015}, \cite{Shibagaki_2016}. 
For the models of nucleosynthesis (upward in $A$) such approach is reasonable.  
(Although sensitivity of models to the termination of the $r$-process path and the number of fissioning nuclei that contribute to fission recycling and the freezeout of the $r$-process abundances, has been noted \cite{Shibagaki_2016}.)

In contrast, in the hypothesis discussed in this paper, nucleogenesis starts from the top (Fig.~\ref{Fig:12}). 
The nuclear-fog droplets become fragmented into mega-nuclei, which in turn fission into smaller super-nuclei, and so on. 
Abundant free neutrons, free protons (from the "obstacle"), $\alpha$- and $\beta$- particles, and intense $\gamma$ radiation, allow for a variety of capture/decay-processes to occur concurrently. 
Since fission leads to cascading probabilistic outcomes, then various (downward) nucleogenetic paths can occur, 
even if through the very short-lived nuclei, far away from the islands and valley of stability (see Sec.~\ref{s:3-3-3}).
Eventually the paths finish someplace at the valley of stability.  
However, they may approach it (from the top) by moving not "along the valley", but from the "sides".  
For the {\em fission-driven} nucleogenesis, the conventional reduction of the model nuclei-domain (reasonable in numerical simulations of stellar {\em nucleosynthesis}) is no longer acceptable---
any nuclei-domain-cutoff would fundamentally distort the completeness of the fission-driven model.
The entirety of the ZN-plane cell-population needs to be taken into account, 
as each cell (nuclide) is affected by its parents/neighbors, who are affected by theirs, and so on. 

By explicitly including into consideration the entirety of ZN-plane (via the presence of full $Y$ in the quadratic term), 
Eq.~\ref{Eq:3} also reveals that fission-recycling (upward captures on "seeds" resulted from downward fission) is a natural phenomenon which occurs throughout the entire domain of fissile nuclides (as long as they have not fully decayed, continue to "arrive from the top", or remain being produced by captures). 
For nuclides that are "distant" from the valley-of-stability analogue in the domain of super- and mega-nuclei, 
half-lives are very short (of the order of "nuclear time") but nonetheless the processes do occur, 
even if current observational techniques are not yet capable of detecting them. 

As long as the (post-event) nucleogenesis is still non-stationary, the contribution from these processes cannot be ignored 
(at least not without a good rationale for such simplification, 
which unfortunately cannot be properly analyzed until experimental data become available to allow for such analysis).
Indeed, as Eq.~\ref{Eq:3} reveals,  the stationary solution ($dY / dt = 0$) depends on the values of coefficients $\Omega$ ($\sim \sigma$) and $\Lambda$ ($\sim T^{-1}_{1/2}$), which combine the production and exodus terms for each nuclide 
(segregating the processes into the multi-participant $\Omega$ and single-participant $\Lambda$ groups).
But until the system stabilizes (i.e., production and exodus approximately offset each other), 
individual coefficients in $\Omega$ and $\Lambda$ may significantly differ from zero. 
Paradoxically, the terms corresponding to the shortest living nuclides may create the largest imbalances and, as a result, 
make the greater impact on the overall solution (evolution path).
In fact, referencing the analogy with linear systems of equations, when the determinant of the matrix of coefficients is close to zero, the solution is unstable with respect to small changes in any of the matrix coefficients.
Thus, it is quite possible that the system's evolution---the genesis path---
may pass not through the valley-of-stability but through the regions {\em away} from the valley, 
where nuclides exists very briefly;  however, the actual path is unpredictable. 
This understanding differs significantly from the conventional conception of galactic nucleosynthesis  
{\em along} the valley-of-stability populated with long-living nuclides.  

Another intriguing question is whether---because the system is nonlinear---it may or may not {\em forget} its initial conditions. 
If, as described in the theory of hydrodynamic turbulence, the entanglement of modes can lead to the system's 
"loss of memory" about its initial conditions, then the process may result in the "universal" abundance distribution for nuclei (characterizing the mechanism). 
On the other hand, if the system remembers its initial conditions, then the observed abundance profile can be used 
(in principle, once $\Omega$ and $\Lambda$ coefficients are eventually constructed via experimental and theoretical methods)  to attempt to solve the inverse problem---to find out the initial conditions that led to the solar system chemical composition as we know it.

\subsubsection{Fission of Super-Heavy Nuclei}
\label{s:3-3-2}

In the proposed hypothesis, nucleogenesis starts from the top (Fig.~\ref{Fig:12}), from the very-neutron-rich 
{\em mega-nuclei (nuclear-fog droplets)} that fragment and fission. 
Obviously, no experimental  data for such nuclei exist at present. 
Development of some model---a theoretical description for the process of mega-nuclei fission---faces a number of challenges, seeming unsurmountable at present, as indicated by the studies of fission of the (relatively smaller) {\em super-heavy} nuclei \cite{Giuliani_2019}, \cite{Schmidt_2018},  \cite{Oganessian_2015}. 

Fission of  elements up to actinides is relatively well understood, at least semi-quantitatively \cite{Giuliani_2019}. 
Studies confirm that fission of nuclei with large $A$ can result in multiple fragments. 
Indeed, besides the most familiar binary fission (quasi-symmetric; or asymmetric, also known as cluster-emission), 
ternary (triple) fission and quaternary fission have been observed experimentally \cite{San_Tsiang_1947}.
In most experimentally-known nuclei the probability of triple fission is small because of the high second barrier of ternary fission and long path to the saddle point of ternary fission in comparison with binary fission. 
However,  the barriers decrease as $A$-numbers increase. 
Indeed, experiments with heavy ions \cite{Perelygin_1969} show that the yield of ternary fission fragments of stable or long-lived isotopes of $Th$, $U$ and $Bi$ can be of the same order as in binary fission. 
In the experiments  the masses of ternary fragments were approximately equal.

Super-heavy nuclei ($Z \geq 100$) are much less explored. Not only much experimental data remains missing, but also the theory becomes more complicated as $A$-numbers increase. 

Currently only isotopes forming the so-called "lower superheavy region" (Z = 110 - 113) and "upper superheavy region" (Z = 114 - 118) have been synthesized in nuclear reactions  \cite{Oganessian_2015}, \cite{Oganessian_2017}. (The regions are not yet known to be connected via any nuclear decay chains.) These nuclei cover only a small proton-rich corner of the vast and primarily unexplored territory of super-heavy nuclides (see Fig.~\ref{Fig:12}). There are currently no obvious ways to synthesize neutron-rich super-heavy systems \cite{Giuliani_2019}. 

Experimental exploration of the synthesized nuclei is also challenged by technical difficulties. Fission half-lives of known nuclei generally vary from $10^{-19} - 10^{+24}$~s,  but present experimental techniques for the detection and identification of super-heavy nuclei  are sensitive for fissioning nuclei with half-lives roughly between tens of $\mu s$ up to a few hours at most  ($10^{-5} - 10^{+4}$s) \cite{Khuyagbaatar_2016}. 

\begin{figure}[h!]
\centering
\includegraphics[width=0.55\columnwidth]{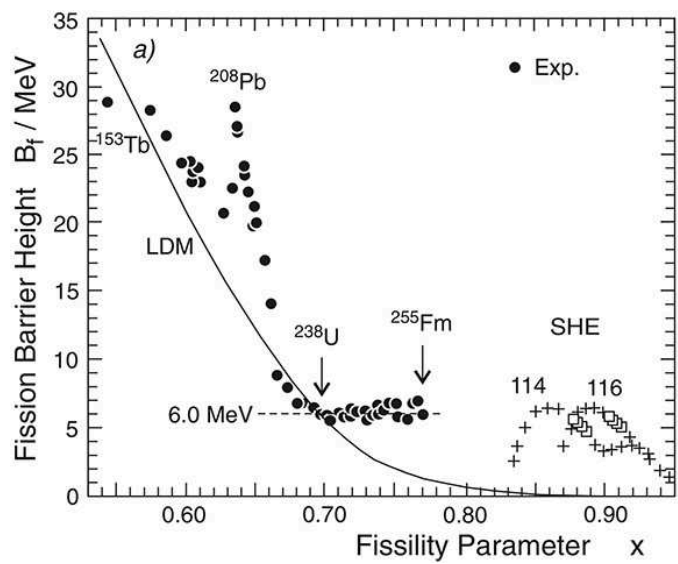}
\caption
[Fission barrier heights as a function of the fissility parameter  $x = (Z^2/A)/(Z^2/A)_{crit}$]
{
Fission barrier heights as a function of the fissility parameter 
$x = (Z^2/A)/(Z^2/A)_{crit}$ at $(Z^2/A)_{crit}$ 
  \cite{Oganessian_2015}.
Black points show experimental data, solid line---calculations in the liquid drop model, 
crosses---calculated fission barrier heights in the macroscopic-microscopic model 
for the isotopes of elements 114 and 116, 
open squares---the same for the nuclei produced in the $^{242, 244}Pu$, $^{245,248}Cm + ^{48}Ca$ reactions.
}
\label{Fig:13}
\end{figure}

Theoretical modeling of nuclear fission---a quantum-mechanical process involving large- amplitude nuclear collective motion---is also enormously challenging  \cite{Schunck_2016}. The familiar picture of penetration through a double-humped fission barrier \cite{Bj_rnholm_1980} undergoes serious revisions in the super-heavy region because the fission barriers decrease or vanish due to increasing Coulomb pressure (see Fig.~\ref{Fig:13}).  (Indeed, the heaviest experimentally-explored nuclei are  known to be barely bound in their ground state---an excitation energy on the order of a few per thousand of their total binding energy is sufficient to induce disintegration into large pieces, releasing significant amounts of energy \cite{Schmidt_2018}.  The revisions in the theory (as $A$-numbers increase) have an impact on fission observables such as the fission lifetimes, the distribution of fission fragments, and the characteristics of the fission spectrum ($n$, $\gamma$, $\beta$, etc.) \cite{Giuliani_2019}.  

The importance of Coulomb pressure increases with increasing system size. 
This favors a reduction of nuclear density around the center and so may give way to exotic nuclear topologies such as bubbles or toroids. 
For superheavy nuclei, such as $^{780}{254}_{526}$ \cite{Nazarewicz_2002} (here $^{A}{Z}_{N}$),   
exotic profiles---bubbles which have a void at the center, and band-like toroids---
may be competitive with normal profiles (similar to densities of stable nuclei). 
Some recent calculations suggest that many such forms are unstable against triaxial distortions and fission 
\cite{Afanasjev_2018}, \cite{Brodzi_ski_2013}.  

Unfortunately, see for example, Reference \cite{Bernard_2011}, a fully microscopic theory of fission has not yet been achieved and is unlikely to be forthcoming in the near future. 


\subsubsection{Nucleogenetic Cascades}
\label{s:3-3-3}

Recall at first that the sequence of evolution of various "drops" in the scenario is as follows: 
\begin{itemize}
\item
The (quasi-stable) compact stellar object (a {\em giant-nuclear-drop}) (outlined in Sec. 3.1 and modeled below in Sec. 4.1) becomes thermodynamically destabilized and thus localized decompression takes place in its interior. 

\item
As the result of decompression, the matter enters the state of nuclear-fog (i.e., charged {\em nuclear-fog-droplets} form). 

\item
As known from experiments on heavy-nuclei collisions, 
the nuclear-fog-droplets fragment further into multiple mini-droplets ({\em mega-nuclei}):  
\begin{quote}
«The classical fog is unstable substance, which transforms finally into liquid "sea" with "atmosphere" of the saturated vapor. 
The {\bf nuclear, {\em charged} fog} is stable in respect to such fortune. But it "explodes" because of the Coulomb repulsion. This event is detected as {\bf multifragmentation}.» \cite{Karnaukhov_2006}
\end{quote}
It is the evolution of these unstable, short-living, mega-nuclei which is discussed in this section.  
This evolution eventually leads to nucleogenesis of various unstable and finally stable nuclides (isotopes of chemical elements).
\end{itemize}

In order to evaluate quantitatively the proposed nucleogenesis, 
we use the nuclear-liquid-drop model \cite{Weizs_cker_1935}, \cite{Bohr_1939}. 
Such approach appears to be appropriate for droplet-nuclei with very high nucleon number $A$. 
In the liquid-drop description of a nucleus with nucleon number $A$ and  proton number (charge) $Z$,
the Weizsäcker formula for  the binding energy $B(A,Z)$ 
combines the volume, surface, Coulomb, asymmetry, and pairing energies: 

\begin{equation}
B (A,Z) \simeq a_v A - a_s A^{2/3} - a_c Z (Z-1) / A^{1/3} - a_{a} (Z - A/2)^2 /A + a_p \delta A^{-3/4} 
\label{Eq:add-1}
\end{equation}

\noindent
where 
$\delta = +1, 0, -1$, respectively, for even-even, even-odd, and odd-odd nuclei.  
The last term is non-essential for very large $A$. 
The coefficient values  
($a_v$, 
$a_s$,
$a_c$, 
$a_{a}$,   
$a_p$) vary somewhat between different models depending on the choice of included experimental data 
(for example, see Reference \cite{Bohr_1998} p. 168, citing References \cite{Mattauch_1965}, \cite{Green_1953}). 
Since at present no experimental means are available to explore mega-nuclei to confirm the values of coefficients $a$, 
the extrapolation of the formula  
into the high-$A$ region may perhaps need adjustment. 

Historically Eq.~\ref{Eq:add-1}
was constructed based on qualitative suppositions. 
It is worth noting the presumptions 
 that the absolute temperature of the nucleus is zero ($T=0$) and that all internal macroscopic collective motions are absent, 
in particular,
the proper rotation of the nucleus. 
Generally, 
Eq.~\ref{Eq:add-1} fits not only  the nuclei of the valley-of-stability, but also  evolving, short-living,    
mega-nuclei. 

Extremum of binding energy $B$ defines the valley-of-stability. 
The 
first derivatives of $B$ 
with respect to parameters $A$, $Z$, and $N$, 
have to be calculated and set to zero,  
in order to find 
the characteristic curves: 
$\beta$-line $Z_{eq,\beta}(A)$  and $n$-line $Z_{eq,n}(A)$. 
The 
equilibrium combinations 
(with respect to corresponding processes) 
of 
cohabiting 
protons $Z_{eq}(A)$ and neutrons $N_{eq}(A)$
are situated along these curves. 
Within such calculations, 
the meaning of expression $(\partial_z B)_A$ is the derivative of $B$ with respect to proton number 
$Z$ at a fixed nucleon number $A$ ($= Z + N$) of a nucleus. 
When this derivative is not zero---
when the nucleus' charge changes, but the nucleon number does not---the corresponding process is   $\beta^{\pm}$-emission. 
Analogously, $(\partial_N B)_Z$ characterizes neutron emission (capture, if possible) at fixed $Z$.

In the calculation of the derivatives, one can use Jacobians  
\cite{Landau_1966},  \cite{Rumer_1977} which  is convenient.
For example, 
 partial derivative $(\partial_Z B)_A$ is expressed via Jacobian as:
\begin{equation}
(\partial_Z B)_A = \frac{\partial (B,A)}{\partial (Z,A)} . 
\end{equation}

\noindent
Standard rules of Jacobian  
manipulations 
lead to expressions:  
\begin{equation}
(\partial_Z B)_N =
\frac{\partial (B,N)}{\partial (Z,N)} =
\frac{\partial (B,A)}{\partial (Z,A)} \, \frac{\partial (Z,A)}{\partial (Z,N)} \, \frac{\partial (B,N)}{\partial (B,A)} =
(\partial_Z B)_A \,  \frac{(\partial_N A)_Z}{(\partial_N A)_B}
\end{equation}

\begin{equation}
(\partial_N B)_A =
\frac{\partial (B,A)}{\partial (N,A)} =
\frac{\partial (B,A)}{\partial (Z,A)} \, \frac{\partial (Z,A)}{\partial (N,Z)} \, \frac{\partial (N,Z)}{\partial (N,A)} =
- (\partial_Z B)_A  \, \frac{(\partial_N A)_Z}{(\partial_Z A)_N}
\end{equation}

\begin{equation}
(\partial_N B)_Z =
\frac{\partial (B,Z)}{\partial (N,Z)} =
\frac{\partial (B,Z)}{\partial (Z,A)} \, \frac{\partial (Z,A)}{\partial (Z,N)} \, \frac{\partial (Z,N)}{\partial (N,Z)} =
 (\partial_A B)_Z \,  (\partial_N A)_Z
\end{equation}

\noindent
Thus, all these four derivatives 
turn to zero if two derivatives  $(\partial_Z B)_A$ and $(\partial_A B)_Z$  turn to zero. 
Equation $(\partial_Z B)_A = 0$ defines  curve $Z_{eq, \beta}(A)$ in the ZA-plane, that is, the $\beta$-line where there is no  $\beta^{\pm}$-emission. 
Analogously, equation $(\partial_A B)_Z = 0$ defines  curve $Z_{eq, n}(A)$ in the ZA-plane, that is, the $n$-line where there is no  neutron-emission. 

For example, 
after differentiation of $B$ 
(with standard parameters $a$ in Eq.~\ref{Eq:add-1}) 
with respect to $Z$, 
the solution of $(\partial_z B)_A = 0$ is 
the equilibrium number of protons in a nucleus, $Z_{eq}$, for a fixed $A$: 
$Z_{eq, \beta} (A) = (A + 0.007 A^{2/3} ) / (2+ 0.015 A^{2/3})$. 
As $A$ increases, the role of Coulomb energy decreases, and the number of neutrons within the formed nuclei starts to exceed the number of protons---the ratio 
tends to $Z/A \sim A^{-2/3}$. 
For "small" $A$, $Z_{eq, \beta} \simeq A/2$  
(in our context, "small" $A$ nuclei are $Fe$ and below.) 
For 
a fixed $A$, when 
$Z < Z_{eq, \beta}$, the nucleus is unstable with respect to $\beta^{-}$-decay.
When  $Z > Z_{eq, \beta}$ (for a fixed $A$), 
the nucleus is unstable with respect to $\beta^{+}$-decay, as well as electron-capture.

\begin{figure}[h!]
\centering
\includegraphics[width=0.55\columnwidth]{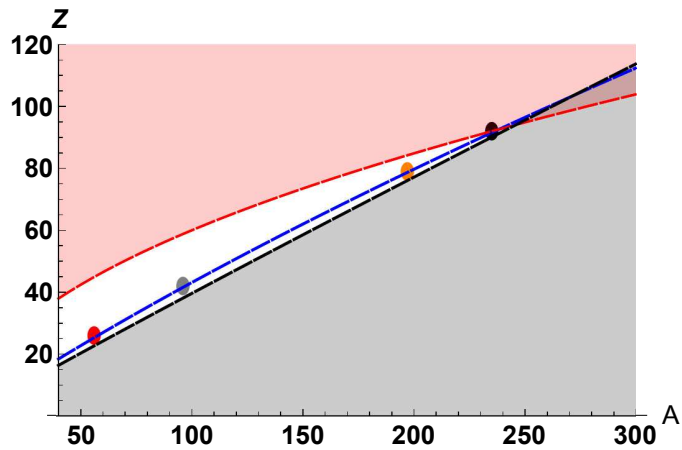}
\caption
[Placement of critical  $\beta$-line, $n$-line, and fission-limit $f$-line]
{Placement of critical lines: $\beta$-line (blue), $n$-line (black), and fission-limit $f$-line (red).
In the shaded gray zone, neutron-emission takes place.
In the shaded red zone, fission takes place.
In the zone below the blue line, $\beta^-$-emissions take place. 
Four dots denote locations of $Fe$ (red dot), $Mo$ (gray dot), $Au$ (yellow dot), and $U$ (black dot).
}
\label{Fig:14}
\end{figure}

Fig.~\ref{Fig:14} plots the  $\beta$-line  and $n$-line, as well as the fission condition  
$Z^2/A > (Z^2/A)_{crit}$ 
for values of $A$ and $Z$ where the known chemical elements are located. 
(Here, the fission curve is drawn through $U$. 
At large $A$, the value for $(Z^2/A)_{crit}$ may change, see Sec.~\ref{s:3-3-2}.) 
In the depicted range of $A$, 
the  $\beta$-line (blue) and $n$-line (black)  are close to each other---apparently defining the valley-of-stability. 
At high $A$, however, the curves diverge. 
In the shaded gray zone, neutron-emissions take place. 
In the shaded red zone, fission takes place. 
In the zone below the blue line, $\beta^-$-emissions take place. 
Placements of four nuclei---$Fe$, $Mo$, $Au$, and $U$---are also depicted for reference. 
Their placements indicate where the 
experimentally measured 
valley-of-stability lies in the AZ-plane.  
The plot also makes is clear why the valley ends with fissile $U$.

\begin{figure}[h!]
\centering
\includegraphics[width=0.55\columnwidth]{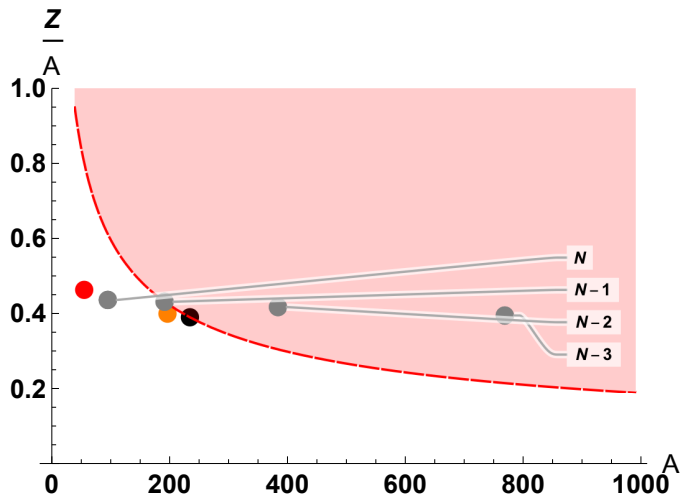}
\caption
[Final stage of fission-driven evolution cascade leading to $^{92}Mo$]
{The final stage of a fission-cascade leading to $^{92}Mo$. 
Gray dots show  $^{92}Mo$ (label N) and its three generations of "ancestors" (N-1, N-2, N-3), each of which simply splits in two, and parametrized adjustment for the "loss" of neutrons due to $n$- and $\beta$- emissions is included at each fission event.
In the shaded red zone, fission takes place.
Placements of $Fe$ (red dot), $Au$ (yellow dot), and $U$ (black dot) are also depicted.
}
\label{Fig:15}
\end{figure}

\begin{figure}[h!]
\centering
\includegraphics[width=0.55\columnwidth]{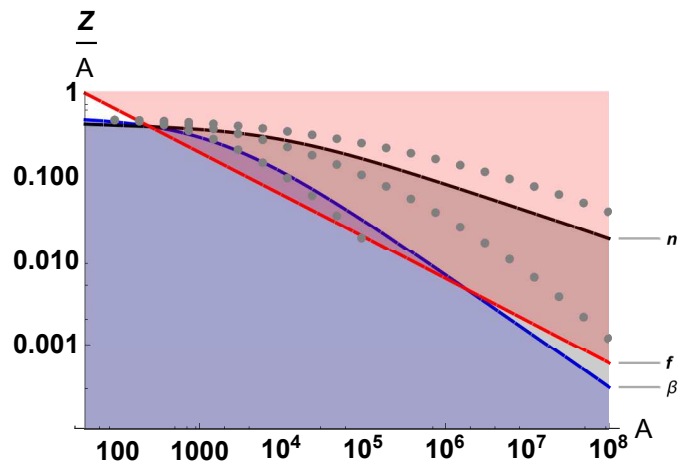}
\caption
[Multi-step fission-driven evolution cascade leading to $^{92}Mo$]
{
Evolution "jumps" in fission cascades (from right to left) leading to $^{92}Mo$ (final left gray dot). 
Each "ancestor" simply splits in two,  and
 parametrized adjustment for the "loss" of neutrons due to $n$- and $\beta$- emissions is included at each fission event. 
Placement of critical lines: $\beta$-line (blue), $n$-line (black), and fission-limit $f$-line (red). 
In the shaded red zone, fission takes place.
In the shaded gray zone, $n$-emission takes place.
In the shaded purple zone, $\beta$-emission takes place. 
Jump-paths are discussed in the main text. 
}
\label{Fig:16}
\end{figure}

Calculations based on Eq.~\ref{Eq:add-1} illustrate (Fig.~\ref{Fig:15}) the final stage of a  
fission-cascade leading to one sample isotope (here for illustration we chose  $^{92}Mo$, one of the problematic $p$-nuclides). 
(The longer cascade with its earlier stages is depicted in Fig.~\ref{Fig:16} with scales capturing very high $A$.) 
Note that the vertical axis is now $Z/A$, not $Z$. 
In this illustration, gray dots show  $^{92}Mo$ (on the left) and its three generations of "ancestors", each of which simply splits in two, 
and parametrized adjustment for the "loss" of neutrons due to $n$- and $\beta$- emissions is included at each fission event. 
Within such parametrization, obviously, 
the transmission-operator 
 (from state $k$ to state $k+1$)   
must be smooth, monotone, tend to zero when $A \rightarrow \infty$.
We constructed the parametrization to arrive at $(Z/A)_{final}$ when $A \rightarrow A_{final}$ 
from $Z/A \sim 0$ when $A \sim \infty$.
Obviously, only experiments can  pinpoint the explicit form of this function.

The necessity of such "adjustment" is intuitively apparent based on several experimental facts (for  experimentally-observed nuclei); 
the   tendencies may be extrapolated. 

Indeed, at every fission event  of $^{235}U$, between 1 and 8 neutrons are emitted 
(the average yield is 2.2 per event).  
The nucleus tends to split in such ways that daughters form  (most) stable structures. 
For $U$, the most likely combination of nucleon numbers is $235 \rightarrow 94 + 138 + 1 + 1 +1$. 
 On the other hand, in $^{235}U$, parameter $N/Z \simeq 1.55$, while for the stable nuclei with $A$ close to the daughters' $A$, such parameter is $1.25 - 1.45$.  
 Therefore, the daughters are oversaturated with neutrons and are unstable with respect to $\beta^-$-decay.  
 The  following sequences of $\beta^-$-decay  have been detected for $^{235}U$ daughters:
\begin{equation}
^{97}_{36}Kr \xrightarrow[]{\beta^-}  \,  ^{97}_{37}Rb \xrightarrow[]{\beta^-}  \, ^{97}_{38}Sr \xrightarrow[]{\beta^-}  \, ^{97}_{39}Y \xrightarrow[]{\beta^-}  \, ^{97}_{40}Zr \xrightarrow[]{\beta^-}  \, ^{97}_{41}Nb \xrightarrow[]{\beta^-}  \, ^{97}_{42}Mo  . 
\end{equation}

\begin{equation}
^{141}_{54}Xe \xrightarrow[]{n, \beta^-}  \,  ^{140}_{55}Cs \xrightarrow[]{\beta^-}  \, ^{140}_{56}Ba \xrightarrow[]{\beta^-}  \, ^{140}_{57}La \xrightarrow[]{\beta^-}  \, ^{140}_{58}Ce . 
\end{equation}

\begin{equation}
^{135}_{52}Te \xrightarrow[]{\beta^-}  \,  ^{135}_{53}I \xrightarrow[]{\beta^-}  \, ^{135}_{54}Xe \xrightarrow[]{\beta^-}  \, ^{135}_{55}Cs \xrightarrow[]{\beta^-}  \, ^{135}_{56}Ba . 
\end{equation}

Besides the sequential beta-emission, a mega-nucleus may perhaps also experience
a multi-beta decay.  
Indeed, 
in available for studies (not mega) nuclei, 
double-beta decays have already been observed experimentally \cite{Barabash_2015} 
and multi-beta (quadruple) decays have been predicted theoretically \cite{Heeck_2013}. 
As discussed in Sec.~\ref{s:3-3-2},  
exotic nuclear topologies such as bubbles or toroids are expected in super-heavy nuclei. 
Therefore, "remote" parts of mega-nuclei may perhaps experience  beta-decays without much coordination with other "remote" parts---nothing is yet known about mega-nuclei.

Because beta-emission is the process determined by weak interactions, while  
neutron-emission---by strong interactions, their timescales differ.  Therefore, it may be expected (in view of the experiments in accessible conditions) that neutron-emission occurs with greater intensity than emission of beta-particles.
This means that the $Z/A$ ratio is not constant but gradually changing: 
a system starting with mega-$A$ and relatively small $(Z/A)_0$  
after fragmenting, fissioning, and undergoing various decays, 
may be expected to evolve towards 
$Z/A$ characteristic for nuclei in the valley of stability ($Z/A \sim 0.4$).

Fig.~\ref{Fig:16} expands the scale of depiction of the same process as in Fig.~\ref{Fig:15}:
both axes are now in logarithmic scales.  
For illustration clarity, in Fig.~\ref{Fig:16},  the evolution 
is depicted 
as "cut-off" for levels higher than $A \sim 10^8$ and for $Z/A$ ratios lower than $ \sim 10^{-3}$. 
Multiple evolution paths may happen when initial $A$ of a mega-nucleus is high: 
\begin{itemize}
\item 
As reminded earlier, in the framework of the proposed hypothesis, the {\em giant-nuclear-drop} (stellar object) became thermodynamically destabilized, experienced localized decompression, where  (charged) {\em nuclear-fog-droplets} ("nfd") formed.  These droplets underwent multi-fragmentation 
(each $A_{nfd} = \sum A_i$)  into (charged) {\em initial mega-nuclei} $A_i$ whose size-distribution is broad and unpredictable.  
For these mega-nuclei 
$(Z/A)_{i,0} \sim 0$ but $(Z/A)_{i,0} \neq 0$.  
No initial $A_{i,0}$ are pictured in  Fig.~\ref{Fig:16}.

\item 
For these initial mega-nuclei $A_{i,0}$    
(which could be depicted below $n$-line and $\beta$-line, close to $A$-axis)  
  $n$- and $\beta$-decays are permitted.  Therefore, below red $f$-line each mega-nucleus sheds neutrons   and electrons---
its $Z/A$ rises---such evolution follows some seemingly-smooth (in log-scale) non-jumping line (not plotted). 
If red $f$-line is never reached, the nucleus evolves within the purple zone until it reaches the valley-of-stability (its neutron-rich "lower Z/A" side).

\item 
When $Z/A$ increase is sufficient to reach 
red $f$-line, then fission-process starts  
(see the start of the lower dotted path, for example;  
 only one daughter-nucleus of each generation is depicted as one gray dot).
During the fission process, the system transitions from  state $k$ to state $k+1$ according to its evolution equation, the general form of which may be written as: 
\begin{equation}
\left( \frac{Z}{A} \right)_{k+1} = 
\left( \frac{Z}{A} \right)_{k} \Big[ 1 + \xi_k (A) \Big] \, .
\label{Eq:12}
\end{equation}
Generally speaking, the structure of multiplication factor $\xi$ for mega $A$ is unknown,  
but obviously 
the function is stochastic and it conforms to certain internal symmetries.
In our scenario, this function can be written as 
\begin{equation}
\xi_k (A) \simeq \frac{\delta N_k}{A_k}  + \frac{\delta Z_k}{Z_k} \, .
\end{equation}
It is determined by the amounts of emitted neutrons $\delta N$ and emitted electrons $\delta Z$, 
which follows from the obvious definition: 
$(Z/A)_{k+1} =  (Z_k + \delta Z_k)/(A_k - \delta N_k)$.
The physical meaning lies not in the absolute values of  quantities $\delta N$ and  $\delta Z$, which are obviously positive, but in the ratios and the combination.
Functions $\delta Z$ and $\delta N$ are random.
For example, 
as mentioned, 
experimentally-measured $\delta N$ in one fission event of $^{235}U$ may vary from 1 to 8, yielding the multi-event average of 2.2. 
Therefore, due to this randomness, the system (from a state located on the red $f$-line) 
may jump into the red zone, slide along the red $f$-line, or return into the no-fission zone (below red $f$-line).  The final decisive judgment about these paths and their choice, belongs to experimental studies.
Fig.~\ref{Fig:16} shows 3 possibilities where the system continues within the red zone once fission starts.
The upper jump-path is for the case when the system evolves in the fission-zone.
The middle jump-path is when the system can also have intensive neutron-losses.
The lower jump-path is when an additional channel opens---beta-emission.
The idea of the evolution equation in the form Eq.~\ref{Eq:12} originated historically in exploration of processes of neutron-multiplication in nuclear-reactors and related applications.

\end{itemize}

This discussion illustrates the fundamental expansion of possibilities that fission-driven nucleogenesis offers: in capture-driven {\em nucleosynthesis}---only {\em some} nuclides can be produced; in fission-driven {\em nucleogenesis} with a sufficiently high initial $A$---{\em any} stable nuclide on the valley-of-stability may be formed via the endless set of combinations of various decays and fission-paths.  Such paths may go through even short-living nuclides' ($Z,A$)-addresses  
and eventually converge towards the valley-of-stability.

 It is important to note that in the process, the "stays" at each (Z,A)-address (except the final one) can be very short-lasting, even extremely short-lasting.  
 As long as the nuclide is briefly formed, and then fissions away, the process can continue. 
 Indeed, as Fig.~\ref{Fig:12} indicates, the half-lives  dramatically shorten 
for explored isotopes located far 
 away from the valley-of-stability. 
(Sec.~\ref{s:3-3-1} 
already elaborated various considerations about the importance of the short-living nuclides in the fission-driven nucleogenesis.)

"Jumping" fission is the fastest evolution process. 
In Fig.~\ref{Fig:16}, it only takes a dozen-like number of steps to "create" 
the final stable nucleus 
(the most left dot) even starting from $A \sim 10^8$. 
Consequently, the entire "event" takes $N \times \tau_f$, where $\tau_f \sim 10^{-19}$s (note that
 $\tau_{n} \sim 10^{-17}$s for prompt neutrons and  
 $\tau_{\beta, n} \sim 10^{-6}$s for beta-particles and delayed neutrons) per \cite{Goutte_2015}, see Fig.~\ref{Fig:9}.

 Fig.~\ref{Fig:16} was also constructed to illustrate  one other important point. 
The $n$-, $\beta$-, and $f$-lines are not "solid walls"---they are merely demarkation indicators of domains where the processes can occur. 
 Because fission creates "jumps" in $Z$ and $A$, the evolution path for the fission process {\em can jump over} the $n$-line (and $\beta$-line, of course) 
 and approach the valley-of-stability from the "higher Z/A" side. 
 This is where $p$-process nuclei are located. 
 
 Indeed, to refresh the memory: 
 some of 
 proton-rich nuclides cannot in principle be synthesized through sequences of only neutron-captures ($s$- or $r$-processes) and $\beta$-decays. 
The term \emph{$p$-process} is used to generally describe any process synthesizing such {\em $p$-nuclei} (even when no proton-captures are involved). 
Some (although not all) 
elements have  $p$-process isotopes (one or several)---
$Se$, $Kr$, $Sr$, $Mo$ (2), $Ru$ (2), $Cd$ (2), $Sn$, $Te$, $Xe$ (2), $Ba$ (2), $La$, $Ce$ (2), $Sm$, $Dy$ (2), $Er$, $Yb$, $Hf$, $W$, $Os$, $Pt$, and $Hg$ \cite{Anders_1989} 
-- they cannot be attributed to $s$- or $r$- nucleosynthesis.
Relative to their non-$p$-process counterparts (other isotopes of the element), 
$p$-isotopes are sometimes $10^{1} - 10^{3}$ times less abundant \cite{Anders_1989}.   
$Mo$ is unusual in the set because its two $p$-isotopes ($^{92}Mo$ and $^{94}Mo$) is $\sim 26\%$ of total $Mo$.   
Current models struggle to explain the "excessive" (relative to models) meteoritic abundances of $p$-nuclei with $A < 100$ (as well as with $150 \leq A \leq 165$) \cite{Rauscher_2013}.   
This is why the simulation in Fig.~\ref{Fig:15}  and Fig.~\ref{Fig:16} used $^{92}Mo$ as an example. 

It is apparent 
from Fig.~\ref{Fig:16}
that to produce a $p$-nucleus via fission-driven nucleogenesis, the evolution path from a mega-nucleus must "jump" over the $n$-line to the "higher Z/A" side. 
The fact that intuitively this is  not the most expected pattern is perfectly consistent with the meteoritic data showing only "tiny" amounts of $p$-isotopes relative to their non-$p$-counterparts.
 Intuitively, the (probabilistically) more likely patterns  seem to be the ones approaching  the $n$-line from the "lower Z/A" side and---
 once the process settles at the valley-of-stability  
---
 producing
 isotopes with relatively more neutrons, that is, those which are currently presumed to be produced by $s$- and $r$-capture nucleosynthesis. 
Using the example of $Mo$ ($Z=42$), it has 7 natural/observationally stable isotopes with $A$ = 92, 94, 95, 96, 97, 98, and 100 \cite{Anders_1989}. 
Overall, including short-lived isotopes, 33 of $Mo$ isotopes are known, ranging in $A$ from 83 to 115.  
 
Fig.~\ref{Fig:16} demonstrates that conceptually {\em every} exotic isotope detected in the solar system
may be "created" by the proposed nucleogenesis. 
Indeed, tracing evolution-paths backwards---from any such ("end") isotope 
towards mega-nuclei---
shows that there can always be drawn some "path" to some "initial" mega-(Z,A)-address.  
Therefore, the  masses of all such "initial" mega-nuclei combine into  
(multi-fragmented droplets of nuclear-fog within) 
the macro stellar object---a "giant-nuclear-drop"---which arrived to the solar system from afar as proposed in the hypothesis. 
(Recall that the total mass of the terrestrial planets, presumed to be produced in the proposed local for the solar system nucleogenetic event, is $\sim 10^{-5} M_{\odot}$. 
Additionally, the Sun and gaseous giants appear to be enriched; Sec.~\ref{s:5-3-3-5} and Sec.~\ref{s:5-3-3-6} discuss their enrichment.)

It is also important  to remember that 
fragmentation/fission is a probabilistic process---
for every type of fissile nucleus, the fission outcomes are unpredictable and random.  The spectrum of daughter-nuclei is dispersed. Each of the daughters then undergoes its own fissioning, also unpredictable and random,  with a different own spectrum of unpredictable daughters.  
Indeed, 
as experiments have shown, 
during (induced) fission of simply-heavy $^{235}U$, its fragments are asymmetric, 
and are distributed in the range $A = 72 - 161$ and $Z = 30 - 65$. 

\begin{figure}[h!]
\centering
\includegraphics[width=0.70\columnwidth]{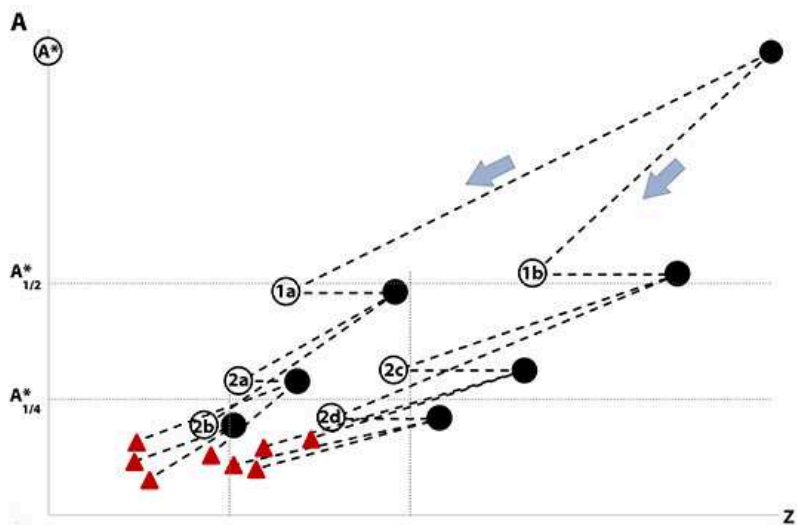}
\caption[Dispersion in nucleogenetic cascades]{
Dispersion in nucleogenetic cascades. 
For visual simplicity, only fission and $\beta$-decays are depicted.
}
\label{Fig:17}
\end{figure}

The dispersion in ZA-plane becomes significant very quickly. 
Fig.~\ref{Fig:17} illustrates the process.  
Note that Fig.~\ref{Fig:17} uses ZA-plane (not ZN-plane as in Fig.~\ref{Fig:12} or (A, Z/A)-plane as in Fig.~\ref{Fig:16}). 
For visual simplicity, we omit the processes of neutron and $\alpha$-particle emission, focusing only on $\beta$-decays and fission. 

Fig.~\ref{Fig:17} depicts the slightly asymmetric fission of the nucleus  ($Z,A^*$) 
into two fragments, 1a and 1b. 
Then, each of the fragments also experiences beta-decays (thus increasing its charge, while maintaining the total nucleon number)  
and undergoes asymmetric fission into two fragments (1a into 2a and 2b, and 1b into 2c and 2d) .
Each of the second-tier fragments again experiences beta-decays and asymmetric two-fragment fission, 
producing together eight third-tier fragments (depicted as red triangles), which continue to fission (no-longer depicted).  
Naturally, fission asymmetry (different at each step) may be great or small, and mega-nuclei may split into more than two fragments. 
Also, alpha-decays (not pictured in Fig.~\ref{Fig:17} for visual simplicity) occur at some stages in the cascade
-- adjusting ($Z,A$) by ($-2,-4$) for each decayed $\alpha$-particle; 
and neutron-emissions adjust ($Z,A$) by ($0,-1$) for each emitted free neutron.  
Nonetheless, despite such simplifications, the provided illustration helps conceptually visualize the cascading process.
Overall, the process continues 
(in the permitted zones, see Fig.~\ref{Fig:14} and Fig.~\ref{Fig:15})
until the evolution paths enter the valley-of-stability. 

One important insight from Fig.~\ref{Fig:17} is the inevitability of the broad dispersion of the isotopes' characteristics, distribution, and abundances, at the end of such cascading nucleogenesis. 
Even after just three steps, it is apparent that the produced fragments do not cluster in one specific and predictable place on the ZA-plane---such as a $1/8$ fraction of ($Z_1,A^*$), for example,---but instead disperse rather broadly and randomly  along both $Z$ and $A$ axes. 

Obviously, 
without {\em any} experimental data or theoretical indications for estimates on  key reaction parameters (as discussed in Sec.~\ref{s:3-3-2}), any numerical simulations of abundance distributions driven by (inherently probabilistic) fission reactions would be completely meaningless.
(Not to forget that the choice of initial conditions would likely strongly influence the outcome.)

\section{Model and Results} 
\label{s:4}

\subsection{Model of Fission-Capable Stellar Fragment}
\label{s:4-1}

The proposed in this paper hypothesis suggests that a nucleogenetic "event" occurred in the inner part of the solar system at the early stage of its evolution, perhaps as a result of an encounter with a stellar object capable of nuclear fission. 
A number of complex physical processes would be involved in the realization of such event.  {\em The focus of this paper is on the examination of one of these processes: {\bf perturbation of thermodynamically quasi-stable nuclear matter}.} 

Several aspects of the proposed model are fundamentally different from the existing models of nuclear matter in cataclysms involving neutron stars (and leading to nucleosynthesis):

(1) As discussed in Sec.~\ref{s:3-1-1}, traditional NS-merger models presume that the escaping nuclear-matter ejecta clumps expand freely with constant velocity, thus their radii grow linearly with time $t$ and consequently their densities drop like $1/t^3$ \cite{Cowan_2019}, \cite{Goriely_2011}---in other words, the clumps decompress as "gas". 
However, generally-speaking, the determination of whether the ejecta clump originates and remains in the phase-state
of nuclear-gas or nuclear-liquid is sensitive to the model assumptions about EOS of the neutron-rich matter. 
(Indeed, the NS-merger studies have generally noted sensitivities of model results to their chosen expressions for EOS.) 
In our model we presume that some fragments are ejected from the scene of the cataclysm (two types of cataclysms are distinguished in Sec.~\ref{s:3-1-1}) in the structurally-stable nuclear-liquid form---as giant "nuclear droplets". 
To make this presumption, we rely on the analysis \cite{Tito_2018a} which examined properties of such fragments (compact super-dense objects) with the equation of state permitting the state of nuclear fog.  The nuclear-{\em liquid} droplets do not decompress as $1/t^3$.  

(2) Decompression of nuclear-liquid may (and does in experiments) occur via nuclear-fog (Sec.~\ref{s:3-1-2}).
The nuclear-fog phase of nuclear matter is traditionally not considered in NS-merger models.  
In our model, we focus on  the fragment that was ejected from the scene of the cataclysm (in the structurally-stable nuclear-fluid form, as a giant "nuclear droplet"), traveled for a long time, and cooled down along the way so its thermodynamical state became close to the bounds of the instability (spinodal) zone for nuclear matter. 
The properties of nuclear matter near and within the spinodal zone must be described by the equation of state that is capable of capturing the effects of phase transitions within the two-phase zone, but most traditionally-used forms of EOS  do not possess such ability.  
Therefore, to examine destabilization of the fragment due to perturbation of thermodynamically quasi-stable nuclear matter, we again rely on the EOS from our work \cite{Tito_2018a}.

\subsubsection{Equation of State (EOS) with Nuclear Fog Interpolation}
\label{s:4-1-1}

The expression for the EOS---that is, pressure $P$ as function of basal thermodynamical parameters---was constructed from the expression for free energy $f = F_1 / T_c$ (per one particle) based on the following qualitative arguments \cite{Tito_2018a}: 
\begin{enumerate} 
\item
The expression for $f$ must obviously contain a term that determines the rest mass of nucleon (term with $m_1$).
This expression should contain the term associated with "repulsion" due to the hard core inside the nucleons (term $\sim a_3 z $, where $z$ is dimensionless density).
This is a "universal" term in the sense that it provides "non--violation" of the principle of causality: the "adiabatic speed" of propagation of elastic perturbations in a medium should not exceed the speed of light.
\item
The free energy $f$ must contain a "thermal term" that depends on the temperature  of the medium. This positive correcting term can be "quadratic" in the temperature  $\sim \theta^2 $ when the temperature of the Fermi system is smaller than the degeneracy temperature $\theta_F$, or it has the dependence $\sim \theta \ln [\theta ^ {3/2} / z] $ when the temperature is greater than $\theta_F$. 
\item 
In the intermediate range of temperatures, the temperature dependence is obviously more complex.
But this is not important at this stage of consideration since its role is to "make a small correction" to the density value for which the pressure becomes equal to zero. 
\item
Finally, a term assuring transition from the domain where the medium has net traits of "fluid" to the domain where the medium is more similar to "gas" must be present in the free energy $f$.
This term must have a less steep dependence on density $z$ than a linear one, to not violate the causality principle in the  domain of high densities. 
\end{enumerate}
Combining all the terms, the dimensionless free energy expression (per  one particle) is constructed as  
\begin{eqnarray}
f =  m_1  +  \frac{a_3}{2} z + a_1 \, Li_2  (- a_2 z)
- \theta \ln \bigg( \frac{\theta^{3/2}}{z} \bigg)  .  \quad \label{Eq:4}
\end{eqnarray}
where in order to satisfy condition~(3), 
function $Li_2 (-x)$ is the polylogarithm function of argument $x$. In limit cases, $Li_2 (-x) \simeq - x + x^2 / 4 - x^3 / 9 + \, ...$ for $x \rightarrow 0$, and $Li_2 (-x) \simeq \pi^2 / 6  - (1/2) \ln^2 (1 / x) + (1 / x) \, ...$ for $x \rightarrow \infty$.
For the critical temperature $T_c = 17.5 \, Mev$, the numerical values of parameters were found to be 
$a_1 = 1.225, a_2 = 1.841, a_3 = 1.074, \, z_1 = 5.5$, yielding the factor of incompressibility $\kappa= 16.533$ (i.e., $K = 289.3 \, Mev$, which is close to the experimentally measured value).

The dimensionless pressure and volume density of internal energy were then obtained 
  (after standard manipulations\footnote{
  Dimensionless pressure $p$, entropy $s$, and the volume density of internal energy $\varepsilon$, are calculated as
$p = z^2 \partial_z f$ , 
$s = - \partial_{\theta} f$, 
$\varepsilon =  - z \theta^2 \partial_{\theta} (f/\theta)$  
from the basal thermodynamical expression $d f = - s d \theta - p d ({1}/{z})$. 
Here, the free energy $f$ is given as a function of density $z$ and temperature $\theta$, 
therefore, the other thermodynamical quantities depend on $z$ and $\theta$ as well.
  })  
from Eq.~\ref{Eq:4} as 
\begin{eqnarray}
p = \frac{a_3}{2} z^2 - a_1 z \ln (1 + a_2 z) + z \theta,  \label{Eq:5}\\
\varepsilon = m_1 z  + \frac{a_3}{2} z^2 + a_1 z \, Li_2 (- a_2 z) + \frac{3}{2} z \theta.  \label{Eq:6}
\end{eqnarray}

\noindent
This EoS satisfies the following conditions: 
(a)   its form permits the existence of the critical point where $\partial_z p = \partial_{z z} p = 0$;
(b)  pressure  $p (z)$ may turn to $0$ at some density $z_1 \neq 0$,
thus permitting the existence of a compact structure with boundary condition $p(z_1)=0$ 
at the free surface%
\footnote{
Superposition of the encountered (Solar System’s) gravitational field onto the fragment’s proper gravitational field would modify the equilibrium conditions for the traveling stellar object.
}, 
that is, at the boundary of the medium with vacuum%
\footnote{
In Eq.~\ref{Eq:4}, we neglected surface effects for a drop-like configuration, that is, the surface free energy $F_S$ (proportional to the surface $S$ of the drop)  
is omitted. Since $F_s \sim S \sim A^{2/3}$ and the volume free energy $F_V$ is proportional to drop volume (i.e., $F_V \sim V \sim A$), then 
for large systems, when $ln A \gg 1$, the volume term in the total free energy expression dominates. 
Recall that Eq.~\ref{Eq:4} is the free energy per one particle. 
};  
(c) the critical density $\rho_c$ (in usual units) must be of order of $(0.1 \div 0.4) \, \rho_0$, that is, $z_1$ is of order of $\simeq (3 \div 7)$;
(d) compressibility factor $K$ must be in the range of $\sim (240 \div 300) \, Mev$ (see discussion in Reference \cite{Tito_2018a});
(e) the principle of causality must be respected, which means that the adiabatical sound speed (i.e., the speed of 
transmission of information and energy of small perturbations) that follows from the model is 
always smaller than the speed of light ($V_s^2 < 1$).
Obviously, the expressions for pressure (Eq.~\ref{Eq:5}) and internal energy (Eq.~\ref{Eq:6}) per volume unit of the hot system (defined by Eq.~\ref{Eq:4}) contain terms which are proportional to effective temperature---
just as exhibited in the VdW-Skyrme model for the empirical nuclear EoS (in the form of polynomial approximation) 
\begin{equation}
\label{Eq:7}
P = \frac{T}{m} \rho - A_1 \rho^2 + A_2 \rho^3 ,
\end{equation}
proposed (see \citet{Jaqaman_1983})  
to explain observed experimental data (Fig.~\ref{Fig:7}).

\subsubsection{Structural Stability of "Small" Super-Dense Compact Objects}
\label{s:4-1-2}

The model produces, as elaborated in Reference \cite{Tito_2018a}, the mass--radius relationship  (with relativistic corrections) 
expressed in terms of  mass $M$ (measured in $10^{-3} M_{\odot}$ units) and  radius $a$ (measured in $10^{-6} R_{\odot}$ units):
\begin{eqnarray}
\frac{M}{10^{-3} M_{\odot}} = 0.211 a^3 \int_0^1 d \xi \, \xi^2 \, z(\xi; a, \theta) 
= 0.168 a^3 m(1; a, \theta) 
\equiv 0.168 \frac{4 \pi}{3} z[1] a^3 F(a, \theta, z[1]).
\label{Eq:8}
\end{eqnarray}
  Here dimensionless density  
$z  = z (1) + G z_1 (\xi) + G^2 z_2 (\xi) +  G^3 z_3 (\xi) + \; ...$, and 
  dimensionless mass 
$m = ({4 \pi}/{3})z(1) \xi^3 (1 + G \mu_1 (\xi) + G^2 \mu_2 (\xi)  + G^3 \mu_3 (\xi)+ ... )$. 
Also, 
$z(1)$ is dimensionless density normalized on critical density $\rho_c$ at  dimensionless radius $\xi =1$ 
(i.e., at the boundary where pressure is assumed zero; $p(z(1))= 0$), 
dimensionless parameter $G = \gamma_N R_b^2 \varepsilon_c / c^4$, 
$R_b$ is radius of the "body", 
$\varepsilon_c$ is critical energy density, 
$c$ is the speed of light, 
$\gamma_N$ is the newtonian gravitational constant.
Quantities $z_i$ and $\mu_i$ may be considered as "add-ons" - small perturbations of the basic state.
Parameter $F (a, \theta, z[1])$ defining the mass--radius relationship depends on dimensionless temperature $\theta = T/T_c$ normalized by critical temperature $T_c$, thus capturing the dependence of stellar fragment characteristics on its cooling history. 

As discussed in Reference \cite{Tito_2018a}, stationary spherical configurations---that is, structurally stable "small" stellar fragments---
  may exist   but only if the boundary condition for zero pressure is respected (for some radius).

\subsection{Results}
\label{s:4-2}

\subsubsection{Thermodynamical Criteria of Instability}
\label{s:4-2-1}

If a system is thermodynamically unstable, the rapidity of development of small spontaneous perturbations of density is determined by the parameter called "adiabatical sound speed". This parameter (dimensionless here) for relativistic fluid is calculated using expression $V^2_s =( \partial p / \partial \varepsilon )_s$ where $p$ is pressure and $\epsilon$ is internal energy per particle.
Quantity $V^2_s$ is calculated in condition that entropy per  particle, $s$, is constant.
However, pressure and internal energy are frequently given as functions of density $z = \rho / \rho_c$ and temperature $\theta = T / T_c$.
In this case, it is natural to calculate $V_s^2$ using Jacobians  and their properties (see References \cite{Landau_1966}, \cite{Rumer_1977} for details):
\begin{eqnarray}
V^2_s \equiv \bigg( \frac{\partial p}{\partial \varepsilon} \bigg)_s \equiv \frac{\partial (p, s)}{\partial (\varepsilon, s)}
=
\frac{ p_z  - s_z  ( s_{\theta} )^{-1} p_{\theta} }{ \varepsilon_z  - s_z  ( s_{\theta} )^{-1} \varepsilon_{\theta}}
.
\label{Eq:9}
\end{eqnarray}
Once the expression for free energy $f$---the equation of state (EoS)---of the model is known, then pressure $p$, entropy $s$, and internal energy $\epsilon$, as well as all derivatives in Eq.~\ref{Eq:9}, can be found.
Then $V_s^2$ can be calculated using standard procedures.

\begin{figure}[ht!]
\centering
\includegraphics[width=0.55\columnwidth]{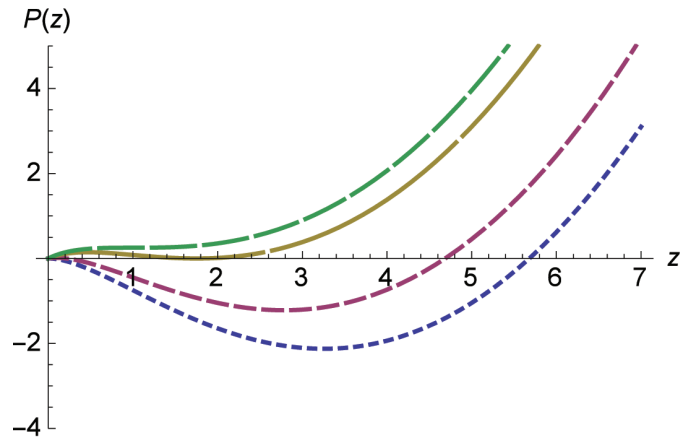}
\caption
[Model simulation: pressure vs density for the model of nuclear-drop-like object with equation of state 
with nuclear-fog interpolation]
{
Pressure $p (z, \theta)$  as a function of normalized density $z = \rho / \rho_c$, 
for the model of nuclear-drop-like object with equation of state 
with nuclear-fog interpolation
\cite{Tito_2018a}.  
Several values of normalized temperature $\theta = T / T_c$ are shown:
 $\theta =0$ (lowest line), $\theta= 0.3$ (second line from bottom), $\theta = 0.8255$ (second line from top) which contains the point where $p = \partial_z p = 0$, and the critical isotherm $\theta = 1$ (upper line) which contains the point where $\partial_z p = \partial_{z z} p = 0$. The lowest curve represents the hypothetical case where the thermal term in the expression for free energy is omitted.
All curves below the critical isotherm, that is, when $\theta < 1$, possess two turning points ($z_1 < z_2$) where  $(\partial_z p )_{z=z_i} = 0$, that is, $s^2 (z_i) = 0$.
In the domain $0 < z < z_1$, the matter is in its gas state. In the domain $z > z_2$,  the matter is in its liquid state.
Between $z_1$ and $z_2$, lies the zone where the gas and liquid phases co-exist.
}
\label{Fig:18}
\end{figure}

\begin{figure}[ht!]
\centering
\includegraphics[width=0.55\columnwidth]{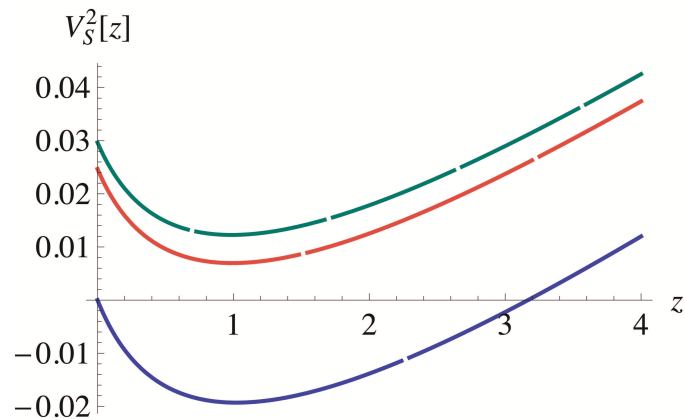}
\caption
[Model simulation:  $V^2_s (z)$ for the model of nuclear-drop-like object with equation of state with nuclear-fog interpolation]
{
Square of adiabatical sound speed $V^2_s (z)$, normalized by the speed of light, as function of normalized density $z$, for the model of nuclear-drop-like object with equation of state with nuclear-fog interpolation
\cite{Tito_2018a}.
Several values of normalized temperature $\theta = T / T_c$ are shown: critical isotherm $\theta = 1$ (upper line), $\theta \simeq 0.84$ (above but close to touching horizontal axis), and $\theta =0$ (lower line).
Domain with $V^2_s (z) < 0$ (where sound speed $V_s(z)$ is imaginary, i.e. the system is unstable) is the  
instability zone (within the spinodal zone)  
where small spontaneous initial perturbations of density grow exponentially fast once triggered.
Development of instability in homogeneous medium leads to formation of two--phase pockets where liquid (drops) and gas (vapor) states co-exist.
Only the states with temperatures below some temperature $\theta_*$ (unique for the medium), for which the curve $V^2_s (z)$ touches the horizontal axis in plane ($z, V^2_s$), may experience such instability.
For the states with $\theta > \theta_*$, the speed of sound is always real ($V^2_s (z) > 0$) and the matter remains in its mono-phase state.
}
\label{Fig:19}
\end{figure}

\begin{figure}[ht!]
\centering
\includegraphics[width=0.55\columnwidth]{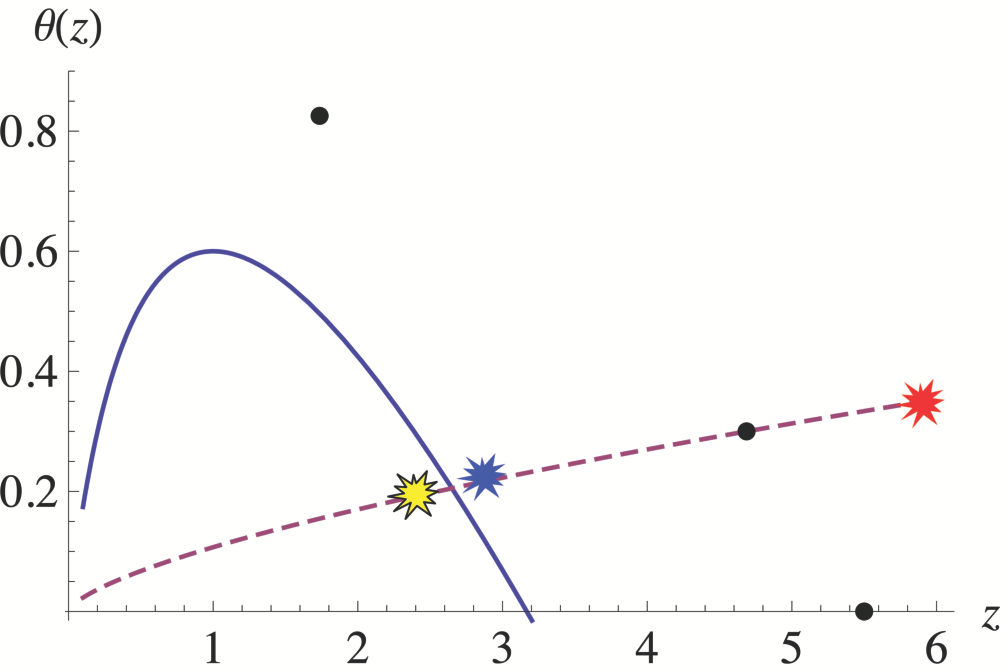}
\caption
[Model simulation: Instability region for the model of nuclear-drop-like object with equation of state with nuclear-fog interpolation]
{
Instability region (inside the solid line boundary) for the model of nuclear-drop-like object with equation of state 
with nuclear-fog interpolation \cite{Tito_2018a}.
Inside the region, $V_s^2 < 0$; outside the region, $V_s^2 > 0$.
On the ($\theta, z$)-graph, pressure points $p = 0$ 
(for the same cases as illustrated in Fig.~\ref{Fig:18} and Fig.~\ref{Fig:19}) 
are shown as black dots---their coordinates are $(5.5, 0)$, $(4.7, 0.3)$, and $(1.74, 0.83)$. 
Any process that cools and/or decompresses the stellar fragment 
from its initial mono-phase state  ($z_0, \theta_0$) 
into the state of nuclear-fog (within the unstable spinodal zone)   
can trigger development of collective instability and fragmentation within the nuclear matter. 
Adiabatic line $\theta = \theta_0 (z / z_0)^{2/3}$ is drawn to guide the eye.  
Red, blue, and yellow stars represent stages of stellar fragment evolution: 
red---"hot" state after its birth, 
blue---"cool" state at the time of encounter with an obstacle, 
yellow---"nuclear-fog" (explosive) stage localized by internal wave of decompression during deceleration caused by the encounter. 
}
 \label{Fig:20}
\end{figure}

 Plots of functions  $P(z)$ and $V^2_s$ for several illustrative cases are shown in Fig.~\ref{Fig:18} and Fig.~\ref{Fig:19}. 
The domain of inner matter where $P(z) < 0$ and $V_s^2 < 0$ is the {\em instability} zone ($V_s^2 < 0$) within the 
{\em spinodal} region ($(\partial P / \partial \rho )_T = 0$) illustrated on the ($z, \theta$)-plane (Fig.~\ref{Fig:20}).
When $V^2_s < 0$, the system becomes unstable with respect to small spontaneous perturbations (fluctuations).

Domain with $V^2_s (z) < 0$ (where sound speed $V_s(z)$ is imaginary, that is, the system is unstable) is the instability region (within the spinodal zone) where small spontaneous initial perturbations of density grow exponentially fast once triggered.
Development of instability in homogeneous medium leads to formation of two--phase pockets where liquid (drops) and gas (vapor) states co-exist.
Only the states with temperatures $\theta \, (\equiv T / T_c)$ below some temperature $\theta_*$ (unique for the medium), for which the curve $V^2_s (z)$ touches the horizontal axis in plane ($z, V^2_s$), may experience such instability.
For the states with $\theta > \theta_*$, the speed of sound is always real ($V^2_s (z) > 0$) and the matter remains in its mono-phase state.

Fig.~\ref{Fig:20} plots the instability region defined by its boundary condition $V_s^2 = 0$ (solid line). 
Inside the domain, $V_s^2 < 0$; outside the domain, $V_s^2 > 0$.
Any process that can "push" the nuclear matter from its initial "liquid" state  ($z_0, \theta_0$) into the instability region, 
ends up triggering instability development.
For a mega-nucleus, such instability leads to fragmentation. 
Sharp (straight-line) deceleration and resulting (localized) decompression (for example, $\rho_0 \rightarrow \rho_0 / 2$)
can serve as the trigger.

\begin{figure}[ht!]
\centering
\includegraphics[width=0.55\columnwidth]{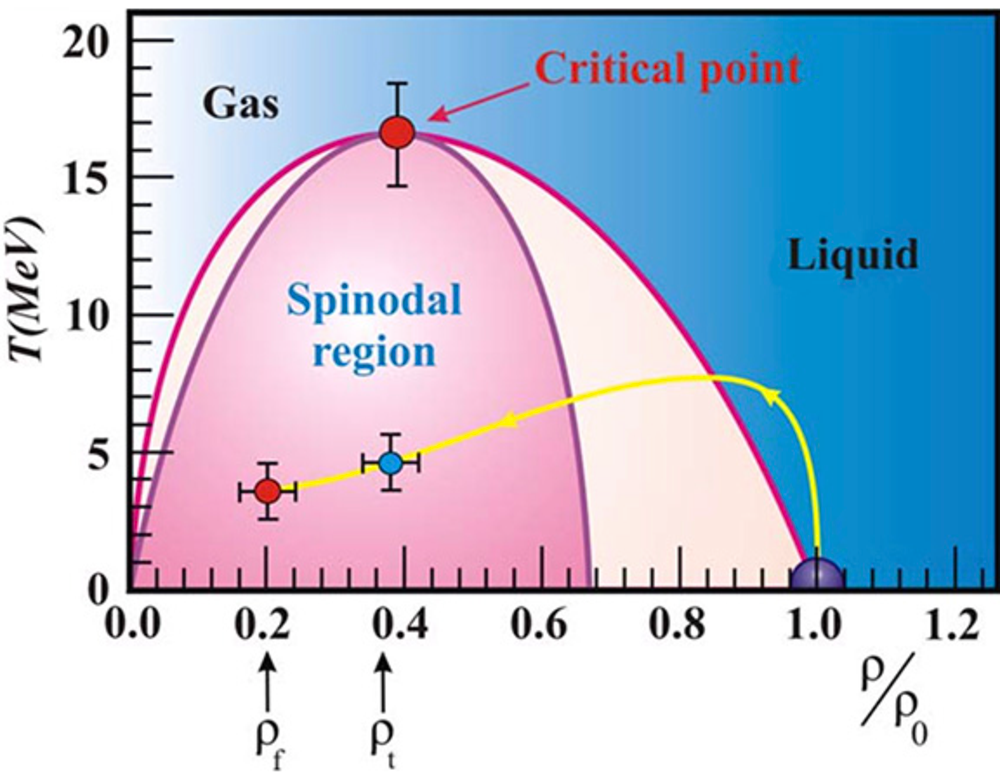}
\caption
[Illustrative nuclear-fog (spinodal) region in $p(8.1GeV) + Au$ collisions / nuclei fragmentation experiments]
{
Experimental data and analysis of fragmentation in $p(8.1GeV) + Au$ collisions, and proposed spinodal region for the nuclear system  \cite{Karnaukhov_2005}. The arrow line shows the path of the system from the starting point at $T=0$ and $\rho_0$ to the multi-scission point at $\rho_f$. The points for the partition and freeze-out configurations are located at $\rho_t$ and $\rho_f$. 
}
 \label{Fig:21}
\end{figure}

Indeed, experiments on fragmentation of heavy nuclei have described  the shift into the spinodal region from (bombardment-induced) states with greater density and temperature. 
Thus, as illustrated in Fig.~\ref{Fig:21} and described in the underlying analysis \cite{Karnaukhov_2005}: "One can imagine that a hot nucleus (at T = 7-10 MeV) expands due to thermal pressure and enters the unstable region. Due to density fluctuations, a homogeneous system is converted into a mixed phase consisting of droplets (IMF) and nuclear gas interspersed between the fragments. Thus the final state of this transition is a nuclear fog [5], which explodes due to Coulomb repulsion and is detected as multifragmentation." (IMF stands for intermediate mass fragments.) 

When comparing Fig.~\ref{Fig:20} and Fig.~\ref{Fig:21}, the following differences are important to note:  

Besides the obvious difference in units---Fig.~\ref{Fig:21} uses  $T$ and $\rho/\rho_0$, while 
Fig.~\ref{Fig:20} uses $\theta = T/T_c$ and $z = \rho/\rho_c$ where $\rho_c \sim (0.1 \div 0.4) \, \rho_0$---
 in the scenario involving the stellar fragment, the initial temperature is not $T=0$ as in the $p(8.1GeV) + Au$ experiments, but some "initial" $T_0$. 
At the first stage of the stellar fragment's evolution,  $T_0$ represents the fragment's temperature after its production-cataclysm (red star in Fig.~\ref{Fig:20}).  
During the journey through space, rather than receiving additional energy and expanding, the fragment instead is presumed to have been cooling down. 
Thus, its trajectory on the ($\theta, z$)-plane, quasi-adiabatic or not, is sketched (as transition from red star towards the unstable phase zone) in Fig.~\ref{Fig:20}.  
The adiabatic line  $\theta = \theta_0 (z / z_0)^{2/3}$ is drawn to guide the eye. 
At the focal stage of the proposed scenario, $T_0$ represents the temperature that the stellar fragment possessed immediately before the encounter with the solar system (blue star in Fig.~\ref{Fig:20}).  
As discussed, as the result of deceleration, compression-then-decompression wave propagation within the  fragment 
would locally rarify the inner matter and shift the localized pocket's  phase-state into the instability zone (illustrated as transition from blue star to yellow star in Fig.~\ref{Fig:20}).   

Another distinction between Fig.~\ref{Fig:20} and Fig.~\ref{Fig:21} is that  the boundary confining 
the {\em instability} zone ($V_s^2 < 0$) 
used in Fig.~\ref{Fig:20} is not the same as the boundary confining the {\em spinodal} zone sketched in Fig.~\ref{Fig:21}. 
Indeed, the boundary of the two-phased spinodal zone is traditionally defined by condition 
$(\partial P / \partial \rho )_T = 0$.  
In our study---because the key mechanism in consideration involves 
propagation of waves of compression/decompression and development of instability---
the boundary of the instability zone (within the spinodal zone) is defined by condition  
$(\partial P / \partial \rho )_S = 0$.  
The adiabatic derivative (which is by definition the square of adiabatic speed of sound $V_s^2$)  
may be close but is not identical to the isothermal derivative.

\subsubsection{Mass and Radius of Stellar Fragment}
\label{s:4-2-2}

\begin{figure}[ht!]
\centering
\includegraphics[width=0.55\columnwidth]{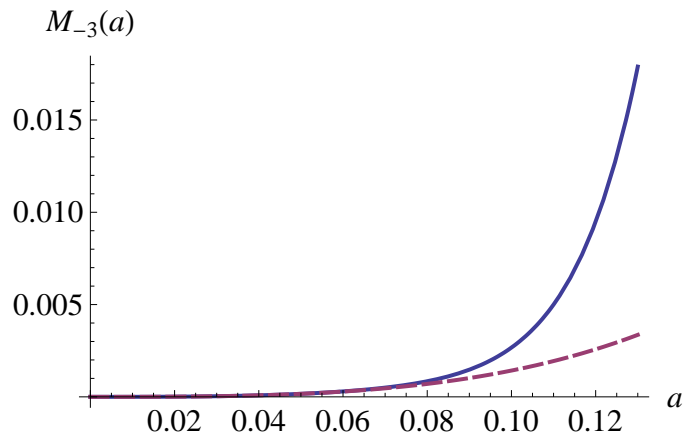}
\caption
[Model simulation: mass--radius relationship  for stellar fragment]
{
Mass--radius relationship  (in $10^{-3} M_\odot$ units for mass and $10^{-6} R_\odot$ for radius $a$) for a 
stellar fragment (with temperature $\theta =0.824$). Dashed line plots the relationship in the newtonian approximation.
}
\label{Fig:22}
\end{figure}

Mass--radius relationship $M_{-3} (a)$ (mass in $10^{-3} M_\odot$ units) for the stellar fragment with temperature near critical temperature $T_c$ is shown in Fig.~\ref{Fig:22}.

Recall that the mass of all terrestrial planets is about $2M_{\bigoplus} \leq 10^{-5} M_{\odot}$. 
Furthermore, the gaseous giants are thought to contain a minimum of $10 M_{\bigoplus}$ of "heavy" elements
(see Sec.~\ref{s:5-3-3-6}), 
and perhaps even  $\sim 20 M_{\bigoplus}$ for Saturn and probably Jupiter \cite{Guillot_2005}. 
Additionally, the Sun itself is enriched 
with a meaningful amount of both volatile and refractory elements 
\cite{Vagnozzi_2019} (see Sec.~\ref{s:5-3-3-5}). 
However, not  all of the "enrichment mass" had to arrive from afar (as the stellar fragment) 
since it is not unreasonable to suppose that 
the matter from the "obstacle"---the binary companion or "inner"-Jupiter---
also participated in the post-nucleogenetic chemical reactions 
and became a part of the debris 
which eventually formed the terrestrial planets and enriched the gaseous giants and the Sun.%
\footnote{This point is moot if the "obstacle" was the edge of the Sun.} 
Indeed, prior to the proposed event, the composition of the Sun and gaseous planets was presumably similar 
to the solar system neighborhood, already enriched by various galactic sources 
(see, for example, References \cite{Luck_2015} \cite{Luck_2016}, \cite{Luck_2018}).  
Thus, if the effective dispersion of the destroyed-object's matter and its subsequent (perhaps partial) 
absorption by the Sun and the existing giants was meaningfully non-homogeneous, then relative elemental  
abundances in the recipients could be impacted to the degree that we currently perceive as enriched or depleted (relative to each other or to the solar neighborhood stars%
\footnote{
Spectral analyses cannot resolve abundances of all elements, nor isotopes, detected in the meteorites and terrestrial samples of the solar system. Thus, only a limited number of elements are considered in stellar spectra analyses,  
for example, 28 elements in References \cite{Luck_2015} \cite{Luck_2016}, \cite{Luck_2018}.
}). 

In other words,  the {\em primary significance} of the fragment's contribution to the solar system enrichment is not so much in the addition of some mass to the solar system, but in 
(1) the nucleogenesis of the detected but difficult-to-explain exotic elements, 
such as $p$-elements, or actinides, or various short-lived isotopes (listed in Sec.~\ref{s:2-2}) 
down to $^{7}Be$ (with $T_{1/2} \sim 53$~days; which formed the detected $^{7}Li$)---all of which are essentially negligible portions of the solar system's mass; 
(2) the creation of peculiar mixing and characteristics of components in meteorites (discussed in upcoming 
Sec.~\ref{s:5-3-3-2}) 
which at present are used as baselines for the origin of the solar system;  
(3) the (direct and indirect) impact on the composition (and dynamics) 
of the gaseous giants (see Sec.~\ref{s:5-3-3-6}) 
and of the Sun  (see Sec.~\ref{s:5-3-3-5}); 
and most apparently,   
(4) the creation of the terrestrial planets (thus explaining the puzzling bi-modal planetary structure) from the combination of the matter from the fragment and the "obstacle".

\begin{figure}[ht!]
\centering
\includegraphics[width=0.55\columnwidth]{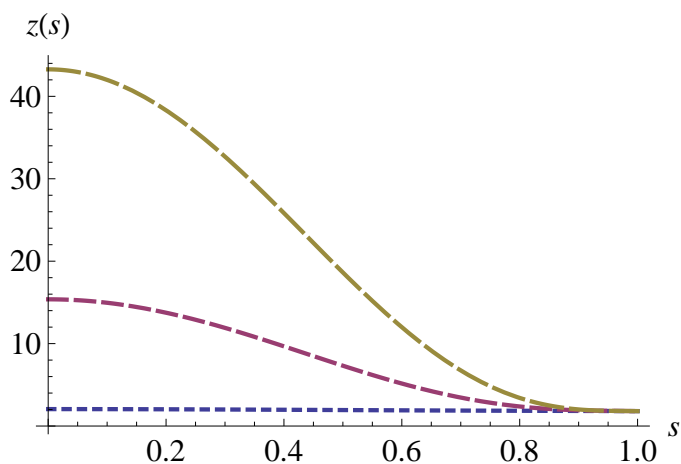}
\caption
[Model simulation: density distribution for stellar fragment (relativistic effect)]
{
Density distribution for a stellar fragment (with temperature $\theta =0.824$) as a function of the distance from the center, $s$, for several values of the fragment's radius $a$: $a = 0.05$ (the lower line), $a = 0.1$, and $a = 0.12$ (the upper line).
}
\label{Fig:23}
\end{figure}

If $M \sim 10^{-5} M_{\odot}$ is taken as the first-order proxy for the stellar fragment mass, 
Fig.~\ref{Fig:22} reveals that according to our model its radius could be $\sim 0.12 \times 10^{-6} R_{\odot}$, 
which is only about a hundred meters or so. 
Notably,  more massive fragments do not appear to be significantly larger in terms of radius.  
(The density increase is governed by relativistic effects \cite{Tito_2018a} and is illustrated in Fig.~\ref{Fig:23}.) 
Even if $M \sim 10^{-3} M_{\odot}$ (i.e., $10^2$ times greater), then $R$ increases by $\sim 4.6$, that is, $R \sim 400-500$~m (even if no density increase is considered).

\noindent
Unfortunately,  stellar fragments with diameters of the order of a few hundred meters are virtually impossible to non-intentionally detect astronomically (perhaps until they explode someplace, and even then, such explosion is a mere nuclear explosion, not thermo-nuclear).

\section{Discussion}
\label{s:5}

\subsection{Structural Disintegration of Compact Super-Dense Stellar Fragment}
\label{s:5-1}

In this paper we proposed a hypothesis that a nucleogenetic "event" occurred in the inner part of the solar system at the early stage of its evolution, perhaps as a result of an encounter with a compact stellar object capable of nuclear fission. 
A number of complex physical processes are involved in the realization of such event.  
In this paper, we focus on the question of 
 {\em whether internal instability within super-dense nuclear-matter (stellar fragment) can trigger nuclear explosion} and 
limited our study to one specific mechanism: 
{\em deceleration-triggered decompression} of the nuclear matter within the stellar fragment upon its encounter with an "obstacle" (upcoming Sec~\ref{s:5-3-2} discusses the obstacle).  

While a number of mechanisms may contribute to the stellar fragment's deceleration upon its encounter with the obstacle,  
in the context of the question of whether explosion can be triggered by internal instability,
the {\em "strength" of deceleration should be defined not in the kinetic sense, but in the thermodynamic sense.}  
(In other words, the exact nature of the deceleration cause is not important, 
and the absolute kinetic magnitude of deceleration is not important.) 
If the pre-encounter $T(\rho)$-phase-state of the nuclear liquid is rather close to the boundary of the 
two-phase (spinodal) zone of instability, then deceleration with even "weak" magnitude in the kinetic sense 
can in fact be a "significant" perturbation capable of triggering  "sufficient"  density stratification 
pushing the system into nuclear-fog.  
Fig.~\ref{Fig:20} illustrates the concept. 

The lower are the nuclear-drop-like object's initial inner density and temperature "at birth" (red star in Fig.~\ref{Fig:20}),  
the less time will it take for its ($T, \rho$)-phase-state to shift (via cooling) towards the nuclear-liquid/nuclear-gas phase-transition boundary. 
The closer is the ($T, \rho$)-phase-state to the boundary (blue star in Fig.~\ref{Fig:20}) 
immediately before the encounter, 
the smaller is the minimal perturbation (deceleration) magnitude necessary to produce decompression "sufficient" for triggering the cascade of nuclear transformations (transition from blue to yellow star in Fig.~\ref{Fig:20}).

The smaller (less massive) is the stellar fragment from the start, and the older it becomes, then the lower is its 
temperature (and inner density). 
Unfortunately, even though theoretical plausibility of existence of small stable objects (spherical configurations) with the above-mentioned nuclear-drop-like properties has been demonstrated \cite{Tito_2018a}, 
such small  ($R \sim 100$~m for $M \sim 10^{-5} M_{\odot}$) objects
are difficult if not impossible to detect at great distances.

\subsection{Likelihood of Stellar Encounter} 
\label{s:5-2}

\paragraph*{The Concepts of Likelihood}
\label{s:5-2-1}

The very thought of an event occurrence 
often brings up a question of its likelihood. But in any context, it is very important to be clear what the term 'likelihood' is meant to describe.

The first kind of likelihood is 'plausibility', which inquires, in essence, whether the laws of physics permit the occurrence of the event in the first place. Understanding how a combination of various mechanisms can produce the event in question yields conclusion that the event is {\em plausible}---in other words, {\em not impossible}, not forbidden by the laws of physics.

The second kind of likelihood is 'statistical probability', which is about statistical odds of mental repetition of a {\em similar} event, not about whether the {\em first} (prior) event can happen. Questions about statistical probability always {\em imply} that the first event can or did happen. The concept of {\em statistical} probability of an event is connected with the concepts of the most {\em expected} outcome, the {\em frequency} of repeated events, and other related concepts. 
Generally speaking, the "statistical odds" have nothing to do with the question of whether the  event proposed in our hypothesis 
could indeed have happened 4.6~Gyrs ago. Such event would have been (was) the {\em first} event.  (And hence the only relevant inquiry is its plausibility.) 
And we humans should be very happy that the odds of the {\em second} event happening in our solar system are low.

Also, when talking about probabilities, it is important to remember the difference between "expectation" and "realization". 
For an "encounter", and especially for a "collision", the often-used word "target" can mean two different things: 
the {\em intended-goal / specific aim for the path} (like the rope for hanging that Clint Eastwood's hero was shooting from afar to release his co-conspirator in the movie 
\guillemotleft 
The Good, the Bad, and the Ugly%
\guillemotright {}) 
and the {\em accidental-result / random obstacle on the path} (like the hole that is left in a wall by a blind-man's accidental gunshot).  
Using these metaphors, we can say that our scenario is not about "whether a bullet can hit the distant rope", 
but instead we note that "the hole in the wall looks like it came from a bullet", so what kind of bullet must that have been and what might have happened.  
For accidental-results (obstacles), post-event, statistical odds are irrelevant.  
Upon realization, P=1.

\paragraph*{The Meaning of Numbers}
\label{s:5-2-2}

Nonetheless, let's look at the probability numbers for additional insight. 
The "frequency of collisions", $\nu \equiv \tau^{-1} = n \langle \sigma V \rangle$, gives indication about the \emph{chance}
of the occurrence of the event (collision) during some increment of time. Here, $n$ is concentration of the obstacle population, 
$\sigma$ is interaction cross-section, and $V \times 1$ is  the distance covered by the moving object over the unit of time. 
Properly speaking, 
expression $P = \nu \Delta t = \langle n \sigma V \rangle \Delta t$ is defined over the {\em large} number of possible realizations (where symbol $\langle ... \rangle$ denotes  statistical averaging, which is equivalent to ergodicity). Similar estimation is made, for example, for collisions between (microscopical) molecules of gas in a (macroscopical) container. 
Time increment $\tau$ should be compared with the full time of the experience 
$\Delta t$ (the traveling time of the stellar fragment). 
If $\Delta t \ll \tau$, that is, $P = \nu \Delta t = \langle n \, \sigma V  \rangle \Delta t \ll 1$, 
it can be said then that a collision of the stellar fragment with one of the (potential) obstacles during its journey most likely would {\em not} occur.

In our scenario, $V  \Delta t \sim  3 \times 10^4$ light-years (distance from the center of our galaxy to the solar system).
This is the distance that a traveling  object  with velocity $V \sim 3 \times 10^{-3}$ of light-speed, that is, $10^3$ km/sec, would cover in $10^7$ years---not too long of a time in comparison with the age of the universe ($\sim 10^{10}$ years). 
Assuming $n \sim 1^{-3}$ light-years$^{-3}$ (based on the average distance between stars in the central part of our galaxy $\sim 1$ light-year) 
and  
$\sigma \sim (10^{-4})^2$ light-years$^2$ (which corresponds 
roughly to the area within Jupiter's orbit, implying that a collision may in fact "perturb" the object and the system, and thus end the journey), 
the frequency of collisions is then $P \sim 10^{-4} \ll 1$.  
Even this (higher-end) estimate shows  that the object {\em could have reached} 
the current location of the solar system  
(which back then was just one of many random stellar systems) 
in about ten million years, \emph{without an encounter} 
with another system along the way---indeed our galaxy is very "scarcely populated". 
Such long journey would have allowed the object to sufficiently cool down, 
so its nuclear inner state could have approached its thermodynamical instability threshold 
-- this condition allows for the "successful" explosion. 
In order words, {\em lower "encounter" odds actually imply higher "success" odds in our scenario.} 

\paragraph*{The Fate of Other Stellar Fragments}
\label{s:5-2-3}

If our hypothesis is correct and long-lasting structurally-stable nuclear-drop-like fragments are ejected from cataclysms 
(such as destruction of neutron stars by black holes, Sec.~\ref{s:3-1-1}), 
it implies that the fragments' lives end in two possible ways: 
(1) they run into some obstacles (and explode), or 
(2) they do not run into obstacles. 
In the latter case, the fragments continue to cool down until their $T(\rho)$ phase state approaches the 
boundary of instability (Fig.~\ref{Fig:20}) so closely that even a minor perturbation---due to encountered variations in interstellar medium density, or due to an encountered propagating ejecta or shock wave from some other stellar cataclysm, perhaps---would trigger their explosions, in the vast "emptiness" of space.
In view of the discussion above, the second scenario appears to be the most likely fate for the majority of the fragments, thus perhaps contributing to the spacial heterogeneity of galactic enrichment. 
Hence, a suggestion may be made that these fragments contribute to the   observed deviations from the scaled solar $r$-process pattern in stars  which remain puzzling both observationally and theoretically, as well as for the "actinide boost" which is so far unexplained  (see, for example, Reference \cite{Jacobson_2014}).

\subsection{Implications for Formation of Sun and Planets}
\label{s:5-3}

The reason we advanced the  presented in this paper hypothesis 
(that a fission-driven nucleogenetic event occurred in the inner part of the solar system) 
is to offer a potential solution to the numerous yet-unresolved puzzles in isotopic abundances, meteoritic data, 
as well as peculiarities in the solar system's composition and planetary structure. 
Therefore, no discussion is complete without mentioning at least some of the details of some of these puzzles. 
The intent of this section is to connect the implications of the hypothesis with observational and experimental data  
and insights of the existing models. 
(The list of data and models is obviously incomplete because the topic is too grand to be addressed in one paper.)
Some of the mentioned data (such as the meteoritic data utilizing field-specific terminology) may at first seem 
not directly related to the proposed in this paper hypothesis, but 
the puzzles that the data reveal may be directly resolved, or the data interpretations may be directly impacted, 
by the processes that are proposed to have happened within the framework of the hypothesis. 
We also cite the evidence indicating that our hypothesis is {\em consistent} (or not inconsistent) with various available data 
and insights from existing models.

Importantly, one of the key propositions of the hypothesis is the {\em separate} rather than concurrent 
formation of the gaseous and terrestrial planets.
Historically,  two-stage scenarios of the solar system evolution have never been contemplated. 
Thus, first, we discuss the differences in the presumptions and implications between the traditional (concurrent) 
and proposed (two-stage) evolution scenarios, 
and cite the evidence supportive or consistent with the two-stage scenario.

\subsubsection{Protonebula} 
\label{s:5-3-1}

Protonebula is the starting point of the solar system formation paradigm. 
The primordial molecular cloud is generally presumed to be cold (with $T \sim 5-30$K) and consist mostly of $H$ and $He$  ($\sim 10:1$ by the number particles) with "grains of dust" that came from distant stellar sources. 

The nebula steadily flattened down by intrinsic rotation (due to conservation of angular momentum and under influence of magnetic fields) resulting in differentially rotating gas-dust disk of hundreds of astronomical units across forming around the collapsed central core where the pressure and temperature progressively grew up until thermonuclear reactions were ignited \cite{Marov_2018}.

\subparagraph{Dust.} 

One important difference between the traditional and proposed scenarios is in the presumed amount of "dust" contained in nebula/disk.
Traditionally, all of the material that eventually went into the rocky objects of the solar system is presumed to have been mixed in the nebula as dust at the beginning of the collapse. 
In the  scenario proposed in this paper, the matter that led to formation of the "rocky" material arrived from afar (as part of the stellar fragment composed of super-dense nuclear matter which upon encounter exploded in nucleogenetic cascades and dispersed within the solar system) but {\em after} the Sun and gaseous giants had already been formed.  
Therefore, the nebula from which the Sun and the giants formed did not contain the same amount of "dust" as traditionally presumed. 
The content and size distribution of solid particles strongly influence the disk thermal regime, viscous properties, turbulence flow patterns, disk medium opacity, chemical transformations in a gaseous medium and, ultimately, 
its evolution including the processes’ dependence on the radial distance from the protosun and the early subdisk formation \cite{Marov_2018}.

\subparagraph{Thermal Regime.} 

The second difference between the traditional and proposed scenarios is in the presumed thermal regime during the disk-stage formation. 
Traditionally, it is presumed that the solar disk started "hot" (> 1400 K) such that all material was vaporized--- 
this assumption is based on the meteoritic evidence. 
In the proposed scenario, absent the great amount of dust, it would be reasonable to presume that the disk started 
colder.

Indeed, in contradiction to the traditional "hot" thermal regime presumption, there exists evidence pointing at the "cold" temperatures in the solar disk: the widespread presence of presolar grains (albeit in extremely dilute quantities  \cite{Zinner_2014}, \cite{Zinner_1998}, the high levels of deuterium enrichments seen in Oort cloud comets and meteorites \cite{Mumma_2011}, \cite{Alexander_2012}, and the presence of highly volatile CO in comets  \cite{Pontoppidan_2014}.  (The current explanations for the evidence of cold temperatures are that either 
the nebula remained cold partially or that this material was provided to the disk at later cooler stages \cite{Pontoppidan_2014}.) 

If the nebula started and remained cold for much longer than traditionally modeled, then (after the nebula heated up)  
condensation lines / snow line would have formed much closer to the Sun than conventionally presumed.  
A different disk evolution, including the processes’ dependence on the radial distance from the protosun and the early subdisk formation, would have resulted.

\subparagraph{Age of Solar System.}

The third difference between the traditional and proposed scenarios is the presumed age of the solar system. 
The current estimate,  4.56~Ga, is derived based on dating of isotopes in meteoritic samples  \cite{Amelin_2002}. 
If meteoritic data is set aside---because in the framework of the proposed two-stage formation scenario 
the meteorites formed as debris from the explosive nucleogenetic event---
the implication is that the age of the Sun and the gaseous giants is greater than 4.56~Ga. 
This also implies that characteristics of the nebula from which the solar system formed 
would be representative of the galactic epoch that is earlier than traditionally presumed.

\subsubsection{Pre-Event Formation of Gaseous Solar System}
\label{s:5-3-2}

If---according to our hypothesis---the initial protonebula of the solar system did not contain the same "dust" 
(in terms of the amount, chemical composition, and size distribution of grains) as traditionally presumed, 
and consequently its viscous properties, turbulence flow patterns, opacity, thermal regime, chemical transformations in a gaseous medium, and so on, would have been different, then it is not implausible that another gaseous giant (now missing) could have also formed.

\paragraph*{A. Binary Companion} 
\label{Companion}

Studies of the collapse have found that single, binary, or multiple protostar systems can form.  
When fragmentation did occur, binary star systems were the typical outcome, along with a few higher order systems 
\cite{Price_2007}, \cite{B_rzle_2010}, \cite{Boss_2013}. 
Indeed, observations show that in the solar neighborhood only about half of solar-type star systems are single-star systems \cite{Raghavan_2010}. 
Fragmentation into both types of binaries, wide or close, can occur.  Studies have shown that 
a wide binary could occur provided that the initial magnetic cloud core rotated fast enough, while close binaries resulted when the initial magnetic energy was larger than the rotational energy \cite{Machida_2008}. 

All these studies indicate that  the solar system could have formed initially as a binary system---
a close binary would be required for the nucleogenetic event to take place in the inner part of the system. 
In fact, observations have detected a number of the so-called "spectroscopic binaries" (a generic term for all binaries that have separations so close that their orbital motion is measurable with radial velocity techniques), and the large majority of them has a period less than $\sim 4,000$ days
(i.e., within the orbital period of Jupiter) \cite{Carney_2003}. 

In the close binary system, the Sun could have been the larger or the smaller companion. 
After the destruction of the companion by the proposed cataclysm, 
the Sun could have absorbed the companion's mass 
(possibly reduced if some matter escaped the gravitational pull of the system). 
Such outcome could also possibly explain 
the $7^o$  misalignment between the Sun's rotation axis and the north ecliptic pole (see, e.g., Reference \cite{Beck_2005}). 

Unfortunately, the exact answer to the question of whether or not the Sun had a binary companion is not yet obtainable, because as well known, the formation of the protosun and solar nebula is largely an initial value problem.
In principle, one can predict the basic outcome by calculating the flow of matter subject to the deterministic laws, 
{\em if given detailed knowledge of the particular dense molecular cloud core that was the presolar cloud}.
Specific details of the outcome cannot be predicted, as there appears to be an inevitable amount of stochastic, chaotic evolution involved, for example, in the orbital motions of any ensemble of gravitationally interacting particles \cite{Boss2006PresolarCC}. 

\begin{figure}[h!]
\centering
\includegraphics[width=0.45\columnwidth]{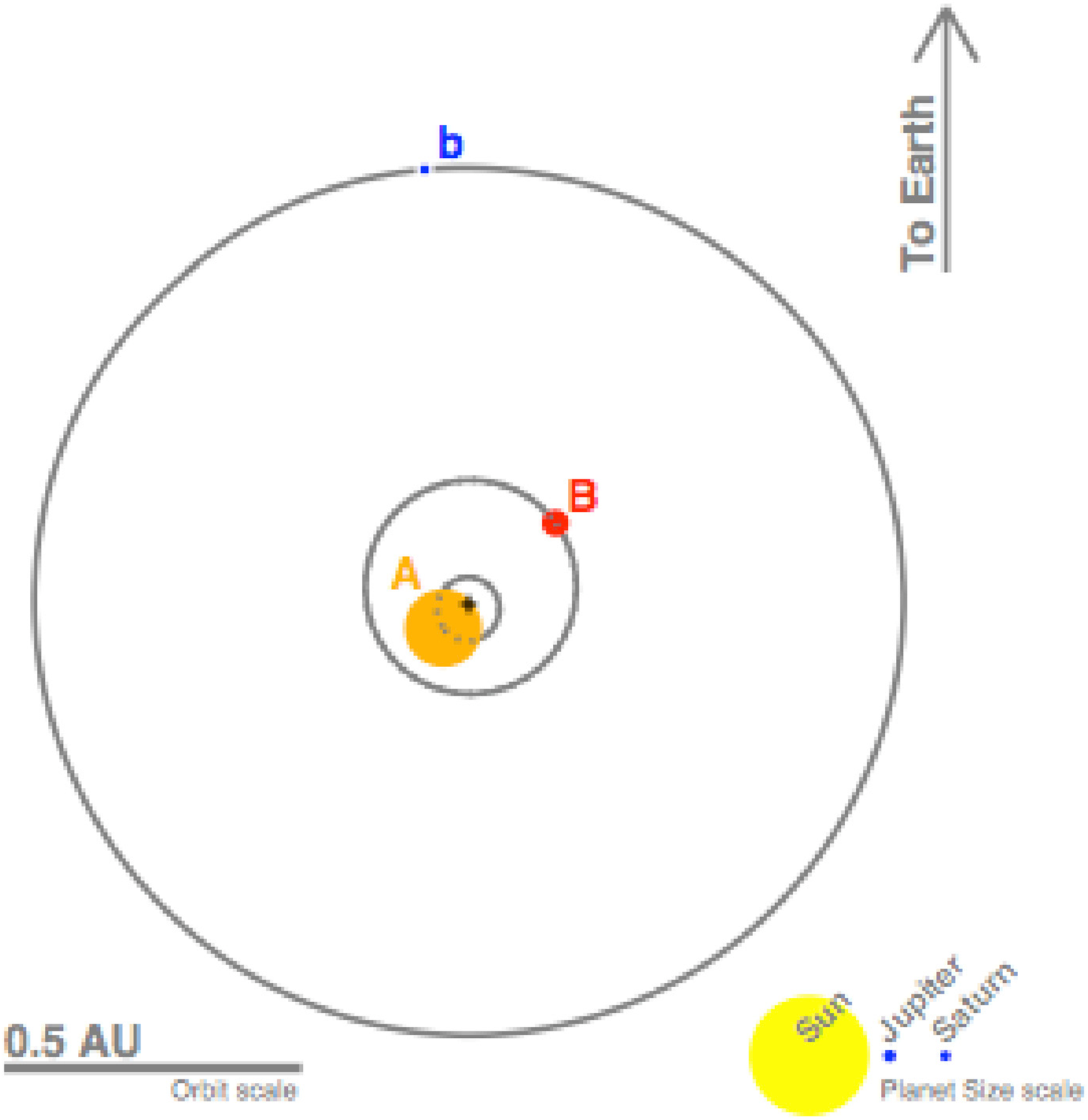}
\caption
[Binary system: scale diagram of the Kepler-16 system]
{
Scale diagram of the Kepler-16 system \cite{Doyle_2011}. 
Gray curves represent the current orbits. 
The sizes of the bodies (including the Sun, Jupiter and Saturn) are in the correct proportions to one another, but they are on a scale 20 times larger than the orbital distance scale.
The binary and circumbinary planet orbital planes lie within $0.4^{\circ}$ degree of each other, 
so the orbits are essentially flat. 
The planet’s orbital eccentricity is nearly zero, while the orbital eccentricity of the binary star system is about 0.16. Symbol "+" marks the center of mass of all three bodies. 
}
\label{Fig:24}
\end{figure}

However, recently, data from the {\em Kepler} spacecraft revealed the detection of a planet whose orbit surrounds a pair of low-mass stars. 
The planet is comparable to Saturn in mass and size, and is on a nearly circular 229-day orbit around its two parent stars. The eclipsing stars are 20\% and 69\% as massive as the Sun, and have an eccentric 41-day orbit. The motions of all three bodies are confined to within $0.5^{\circ}$ of a single plane, suggesting that the planet formed within a circumbinary disk \cite{Doyle_2011}. 
Fig.~\ref{Fig:24} (from Reference \cite{Doyle_2011}) depicts the diagram of the system.  
This evidence is supportive of the idea that the object that we currently call the Sun could have been a part of a binary system.  
In the framework of the proposed in this paper hypothesis, the binary companion could have been the "obstacle"---the  trigger for the nucleogenetic  explosion---destroyed as the result of the encounter. 

\paragraph*{B. "Inner"-Jupiter}
\label{"Inner"-Jupiter}

Alternatively, it is plausible that another giant planet (an "Inner"-Jupiter perhaps) originally existed at the "first" orbit.
Indeed, observations seem to indicate that the protocloud mass distributes between primary stars, companions, and giant planets according to some pattern: giant planets appear to be less frequent around lower-mass stars and more frequent around higher-mass stars (e.g., Reference \cite{2010PASP..122..905J}),  and meanwhile, brown dwarf companions are more frequent around low-mass stars and brown dwarfs in the same separation range (e.g., Reference \cite{Joergens_2008}). 
Just as a reminder, the mass constraint for a body to become a star (to ignite regular nuclear fusion reaction in the interior) is $M \geq 0.08 M_\odot$. Bodies with $ M< 0.01 M_\odot$ are regarded as planets (this threshold is 10 times larger than mass of Jupiter), while bodies in the intermediate range of mass ($0.01 M_\odot \leq   M \leq  0.08 M_\odot$) are called brown dwarfs \cite{Marov_2018}.

As the result of the perceived need to create a {\em concurrent} formation theory for both gaseous and rocky objects in the solar system, the "disk instability" model \cite{Cameron_1978},  \cite{Boss_1997}, \cite{Mayer_2002}  
---where gas giant protoplanets form rapidly through a gravitational instability of the gaseous portion of the disk---has been neglected and the "core accretion" paradigm has been given dominance. 
In part this is also due to the complexities of understanding the conditions under which disks can fragment  \cite{Helled_2014}.
Among the complexities is  also the fact that fragmentation is markedly affected by the presence of significant magnetic fields \cite{Li_2014}  which are challenging to model. 

Disk instability posits  that protoplanetary disks undergo self-compression in a dynamically unstable situation and lead to a transition from a smooth regular disk to an ensemble of clumps in orbit around the Sun. Such clumps may be regarded as candidate precursors of protoplanets. (Nucleated instability relies on the additional gravity field of a planetary embryo, a "core" to trigger gas compression and "envelope" growth in otherwise stable nebulae.) 
Three-dimensional instability is needed to produce a clump in a sheared Keplerian disk that is under the influence of considerable solar tides. Results depend on assumptions regarding the radial temperature profile in the disk 
and the particulars of the numerical technique and resolution used \cite{Boss_2000}, \cite{Pickett_2000}. 
Disk instability model possesses many advantages \cite{Boss_2004}: 
    \begin{itemize}
\item Disk instability can produce self-gravitating protoplanets with cores in about 1000 years, so there is no problem with forming gas giant planets in even the shortest-lived protoplanetary disks. 

\item Disk instability is enhanced in increasingly massive disks, and so it should be able to form planets at least as massive as Jupiter, given that Jupiter-mass clumps form even in disks with masses of about 0.1 the Sun's mass. 

\item
Disk instability sidesteps any problem with Type I orbital migration, and with gap-limited mass accretion, because the clumps form directly from the gas without requiring the prior existence of a solid core subject to Type I drift that can disappear before opening a gap. Once they are formed, the clumps quickly open a disk gap, preventing Type I motion with respect to the gas, but only after most of the protoplanet's mass has already been captured. Thereafter the protoplanet migrates with the disk; in the case of the solar system, little orbital migration appears to be necessary, implying a short lifetime for the solar nebula. 
\end{itemize}
   
 So far (and unfairly) disk instability has been ruled out as the leading model for formation of solar system giants {\em because} of their enrichment with various elements with respect to the solar abundances (such as, for example, Jupiter's enrichment with  $N$, $Ar$, $Kr$, $Xe$; see \cite{Lunine_2004}).
But in the framework of the two-stage formation paradigm, the enrichment of the gaseous giants is presumed to have occurred as the result of the nucleogenetic event, after they formed from the primordial (low-dust) nebula.
Therefore, the constraints imposed by the enrichment data should not apply to their formation scenario. 
As the result, disk instability undoubtedly should be restored as the leading model for their formation. 
And in the framework of disk instability  another giant---an "inner"-Jupiter---could (conceptually) form between the Sun and Jupiter. 

Indeed, as already mentioned in Sec.~\ref{s:2-3}, studies have pointed out that  
dynamical simulations starting with a resonant system of four giant planets showed 
low success rate in matching the present orbits of giant planets \cite{Nesvorn__2011},    
 \cite{Batygin_2011}. 
Furthermore, the orbits of the solar system's giant planets are widely spaced and nearly circular,  
which is unusual \cite{Rice_2014}  
and they also do not exhibit any resonance despite the fact that, as 
$N$-body studies of planetary formation  and orbit positions indicate that  
due to the convergent planetary migration in times before the gas disk's dispersal, 
each giant planet should have become trapped in a resonance with its neighbor \cite{Masset_2001}, \cite{Kley_2000}.  
It has already been suggested that another giant might have existed in the early solar system, 
but subsequently disappeared (was ejected, as proposed), 
thus explaining the implied "violence" that unraveled the planetary resonances and reconfigured the orbits. 

Also, specific angular momentum is one of the main parameters defining the collapsing nebula's fate---its formation into a single- or multi-star system and architecture as a planetary system \cite{Marov_2018}.    
In a single-star system, during accretion of matter from the nebula, an angular momentum is transferred to the star, accelerating its rotation, while an inverse process occurred in the course of a nebula stretching at later stages of accretion gas envelope decay, to satisfy the angular momentum conservation law. 
But in the solar system the distribution of mass and angular momentum is peculiar: 
while the Sun comprises 99.8\% of the whole solar system mass, 
it only comprises about 2\% of the angular momentum. 
It is not yet clear how the angular momentum redistribution in early solar system history has occurred 
\cite{Marov_2018}.%
\footnote{
According to Reference \cite{Marov_2018}:
\guillemotleft 
[...] nowadays it is mainly associated with the presence of a partially ionized disc medium and with the action of electromagnetic forces or the emergence of local shear turbulence in a poloidal magnetic field driven by the magnetorotational instability—MRI (Balbus  Hawley, 1991, 1998; Bisnovaty-Kogan  Lovelace, 2001; Suzuki, Ogihara, Morbidelli, Crida,  Guillot, 2016). Additionally, turbulence in the ionized gas medium can be triggered by hydrodynamical instability, such as baroclinic instability and act throughout the whole $\alpha$ disc (Bitsch, Johansen, Lambrechts,  Morbidelli, 2015). Also, potential contribution to the viscous accretion and hence the loss of angular momentum owing to stellar wind outflow was suggested. Note that this mechanism, which was called magnetothermal disk winds from a protoplanetary disk, would decrease the role of MRI 
(Bai, Ye, Goodman,Yuan, 2016).%
\guillemotright{}  
}
Perhaps, this is an indication that initially the solar system   was formed with a different number of giant objects 
and configuration of orbits, 
and the proposed event has altered the system's angular momentum.  
(However, the  exact resolution of this issue is not apparent.  Additional research is needed to consider the problem within the proposed paradigm.)

\subsubsection{Post-Event Dispersion and Accretion of Debris (into "Rocks", later Planets)}
\label{s:5-3-3}

\paragraph{First Planetesimals:  \guillemotleft Meter-Size Barrier\guillemotright {} Problem in Accretion Model}
\label{s:5-3-3-1}

In the paradigm of concurrent formation of rocky and gaseous planets of the solar system, 
the core accretion model \cite{Safronov_1969},  \cite{goldreich1973formation}, \cite{Wetherill_1989}, \cite{weidenschilling1993protostars}, \cite{Lissauer_1993} has been the leading model of planet formation. 
As noted, one of the reasons for its dominance is the difficulties which the disk instability model runs into while trying to satisfy direct and indirect constraints imposed by the "rocky" data.
However, there is a number of challenges that the accretion model faces as well (see, e.g., References \cite{Morbidelli_2016}, \cite{Helled_2014}).    Perhaps the most vexing problem is the growth of the first planetesimals  \cite{Morbidelli_2016}.  

Traditionally, accretion of all solids in the solar system---from meteoritic components to terrestrial planets---
is presumed to have started with interstellar dust grains mixed in the protodisk. 
Because interstellar dust grains are micrometer-sized  they are presumed to  remain well-coupled to the gas 
 and move along with the gas  \cite{Cassen_1996}.  
Once collisional coagulation and grain growth begins, solid particles are believed to move with respect to the gas, suffering gas drag and additional radial migration as a result \cite{Weidenschilling_1988},  \cite{Weidenschilling_2004}, \cite{Weidenschilling_2006}, \cite{Cuzzi_2006}.  

In the classic model of planet accretion, dust particles settle to the midplane of the disk, collide with each other, and form aggregates held together by electrostatic forces \cite{Dominik_1997}. 

\begin{figure}[h!]
\centering
\includegraphics[width=0.99\columnwidth]{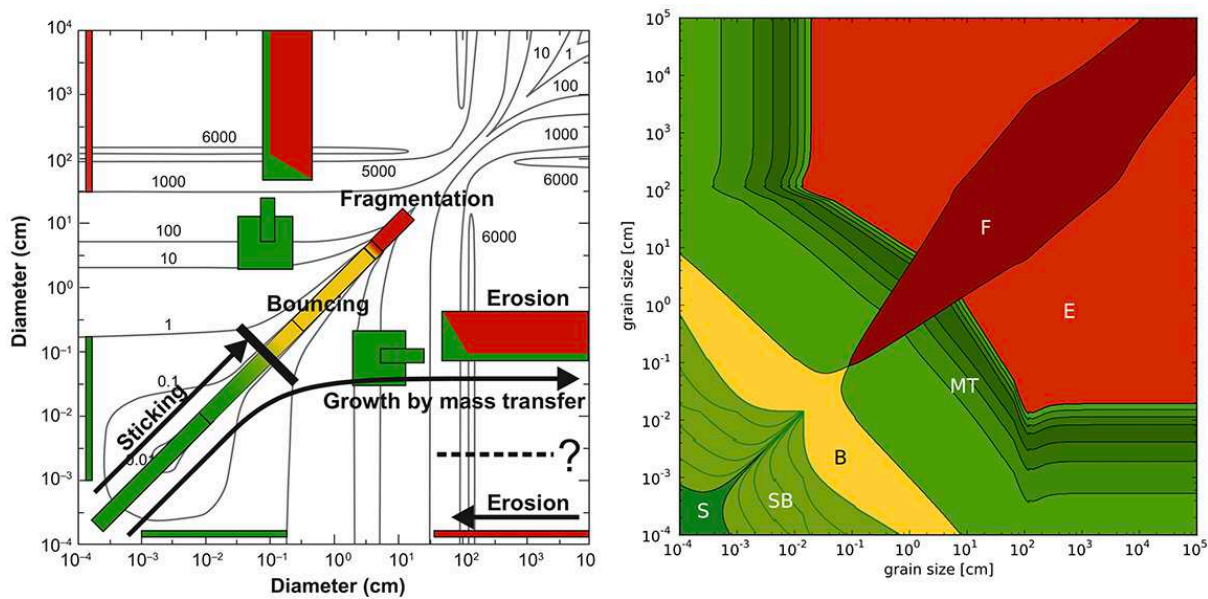}
\caption
[ \guillemotleft Sticking Barrier\guillemotright{} in accretion models (dust collisions in protoplanetary disks)]
{
 Schematic representation of the outcomes of dust collisions in protoplanetary disks   
   \cite{Testi_2014}.
 Left panel:  the collision velocities ( cm/s)  between two dust agglomerates with sizes indicated on the axes 
 in a minimum-mass solar nebula model at 1 AU \cite{Weidenschilling_Cuzzi_1993}. 
 The green, yellow, and red boxes denote the explored parameter space and results of laboratory experiments. 
     Green  represents sticking or mass transfer, yellow bouncing, and red fragmentation or erosion. 
”Sticking” growth \cite{Zsom_2010}  is prevented by bouncing. 
 A possible direct path to the formation of planetesimals is indicated by the arrow ”Growth by mass transfer”. 
 Right panel: the  parameter space for collisions outcomes between bare silicates grains 
 used by numerical models of dust evolution  \cite{Windmark_2012}. 
}
\label{Fig:25}
\end{figure}

\begin{figure}[h!]
\centering
\includegraphics[width=0.65\columnwidth]{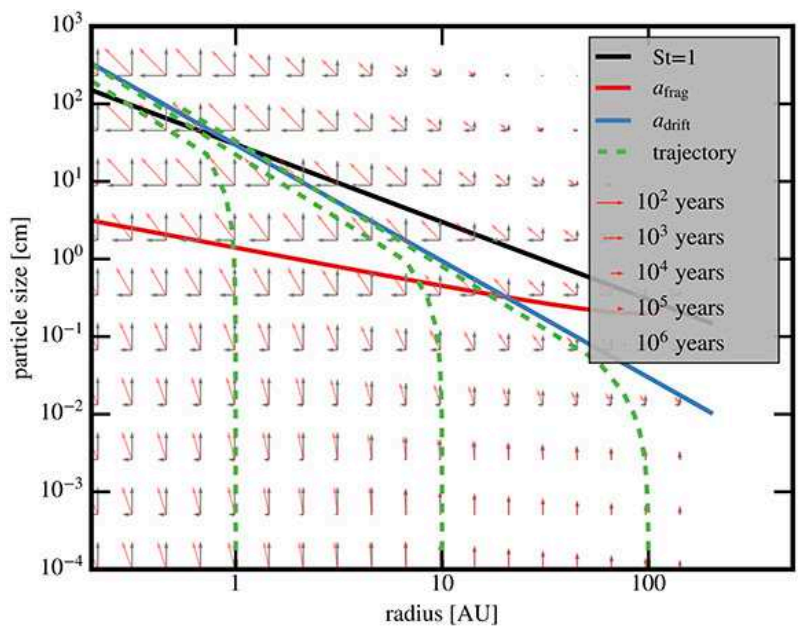}
\caption
[ \guillemotleft Meter-size Barrier\guillemotright{} in accretion models (growth of dust particles in protoplanetary disks)]
{
Comparison of growth and drift time scales  in a protoplanetary disk \cite{Birnstiel_2016}. 
Trajectories for the growth (vertical arrows) and inward aerodynamic drift (horizontal arrows) were calculated at each point in a specific disk model  \cite{Birnstiel_2013}. 
The arrow length is inversely proportional to the relevant logarithmic timescales. 
The  $a_{frag}$ curve shows the fragmentation barrier above which growth is inhibited. 
The dashed curves show the trajectories of particles starting at different orbital radii (calculated neglecting fragmentation). 
The $a_{drift}$ curve is the drift barrier. 
The solid black curve denotes where the Stokes number is equal to 1.
}
\label{Fig:26}
\end{figure}

However, when silicate grains grow to a size of about a millimeter, or icy particles (in the icy part of the disk) up to a few decimeters in size, they start to bounce off each other instead of accreting \cite{Zsom_2008}.   
The so-called bouncing barrier is clearly seen in experiments (see  Fig.~\ref{Fig:25}).
At these sizes, particles migrate rapidly in the disk toward the star due to gas drag  (see Fig.~\ref{Fig:26}).
This radial drift produces large relative velocities among particles of different sizes and hence disruptive collisions. 
Even if collisions were not disruptive and particles could continue to grow, eventually meter-size boulders would migrate so rapidly to be lost into the star before they can grow significantly further \cite{Weidenschilling_1977}. 
This is the well known meter-size barrier. 
     
     A number of solution ideas have been proposed 
     (see, for example, References \cite{Morbidelli_2016}, \cite{Testi_2014}, \cite{Johansen_2014}). 
     Beyond the snow line, the regime of collisional coagulation might work 
     because of the tendency of icy particles to form very fluffy aggregates 
     \cite{Okuzumi_2012}, \cite{Kataoka_2013}.
     Clumping due to turbulence in the disk, with subsequent self-gravity, might work 
     \cite{Johansen_2007}, \cite{Cuzzi_2010}, 
      but it requires turbulence within the disk 
      (yet the very existence of turbulence and its properties are far from being firmly established) 
      and particles must be much larger than that allowed by the bouncing barrier, at least in the inner disk 
      \cite{Morbidelli_2016}. 
      As summed up by Reference \cite{Morbidelli_2016}: 
      \guillemotleft How particles can grow to this size in the inner disk is an open problem.\guillemotright{} 

But in the proposed in this paper hypothesis the solids (of terrestrial planets and asteroids) 
are presumed to form not from the nebula dust, but from the  "debris" produced by the 
proposed fission-event in the inner part of the solar system---the explosive nucleogenetic event. 
The resulting mass-distribution of the (local) debris would certainly be different from the one resulting from sticking and clumping nebula dust,  and it would not be unreasonable to presume that  large-size aggregates (at minimum exceeding the "meter-size barrier") might form post-explosion in sufficiently abundant numbers. 
Although more analysis is needed to fully examine the process, it seems intuitively apparent that bottom-up growth of particles would not be the dominant mechanism in the proposed event. 

In this regard, a curious fact is that there appears to exist a relative lack of asteroids smaller than 100 km in diameter (compared to a direct extrapolation from the largest sizes)  \cite{Pfalzner_2015}, 
and    a similar lack of small Kuiper belt objects   \cite{Sheppard_2010}. 

It has been already proposed that asteroids may have had birth sizes larger than 100~km, in agreement with gravitational collapse models    (rather than accretion model), while the smaller asteroids that are found in the asteroid belt today    may be considered as mainly collisional fragments \cite{2005Icar..175..111B}. 

In other studies, models have shown that the high solid/gas ratio required for chondrule-sized particles to clump in the asteroid belt can only be reached in the late stage of the disk, after a substantial fraction of the gas has been removed \cite{Throop_2005} and if particles are regenerated by some process, for instance, in planetesimal-planetesimal collisions \cite{Morbidelli_2016}. 

 Another  implication stemming from the proposed hypothesis is that some of the known meteorites (especially those classified as irons) may perhaps be reconsidered as representing the direct debris from the proposed nucleogenic event.
For example, analyses of more than 400 collected pieces of  the Sikhote-Alin meteorite revealed that they  
 were  internally-uniform  chunks composed of $88\% \, Fe$, $5\% \, Ni$, and $2\% \, Co$   \cite{Scott_2013}, 
 \cite{Caporali_2016}, \cite{Plavcan_2012},  
collectively weighing more than 23~tonnes, and the largest individual piece being 1745~kg.  
At present, iron meteorites are understood to represent pieces of asteroidal cores \cite{Greenberg_1984} 
and hence are seen as analogues for the Earth's core,  but with the caveat that the analogy 
  \guillemotleft 
must be approached with caution because iron meteorites come from relatively small bodies whose cores formed under relatively low-pressure conditions%
  \guillemotright  {} 
\cite{White_2013}. 

It is important to note that the proposed scenario {\em integrates well} with the traditional theory of accretion 
for rocky objects of the inner part of the solar system (meteorites, asteroids, terrestrial planets, and so on) 
because it helps the existing numerical models overcome the meter-size barrier problem: 
instead of starting with the (sub-micron) dust-distribution as the initial condition, simulations can start with the debris-distribution containing population of "particles" with much larger sizes. 
Studies have shown that once planetesimals are formed, as long as they are embedded in a disk of gas and drifting particles and they are big enough (typically larger than 100 km), they can keep growing by accreting the particles via a combination of gravitational deflection and gas drag 
 \cite{Ormel_2010},  \cite{Johansen_2010},  \cite{Lambrechts_2014},  \cite{Levison_2015}.   
This process is called pebble accretion and it is now fairly well understood and recognized to play a fundamental role in planet formation \cite{Morbidelli_2016}. 
By definition, pebble accretion operates only after that the first planetesimals are formed.  
 Our hypothesis offers  a plausible theoretical rationale for the formation of the first planetesimals.  

An important distinction of the accretion process in the proposed scenario is the absence of "protodisk medium" during  the formation of "rocky" objects.  
As the hypothesis suggests, the process of formation of all gaseous giants in the solar system had been already completed by the time the nucleogenetic event occurred,   
and by then, the space between the Sun and the gaseous giants should have become similar to its present state ("empty"). 
Therefore, during the accretion of the debris into planetesimals and subsequently into the planets, 
no viscous medium effects would play role.

\paragraph{Meteoritic Data}
\label{s:5-3-3-2}

{\bf (1) Isotopic Anomalies.}
\label{anomalies} 
Planetary materials at all scales, from nanometer-sized presolar grains to terrestrial planets, 
contain the so-called nucleosynthetic anomalies 
 \cite{Dauphas_2016}. 
The term "anomaly" describes those isotopic variations which  are found to depart from the geochemically "expected" isotopic variations  that can arise from processes such as 
phase equilibrium, unidirectional reactions, radioactive decay, and nuclear transmutations induced by particle irradiation \cite{Dauphas_2016}.  

With respect to all anomalies involving post-$Fe$ nuclei, the proposed in this paper event offers a resolution because it creates a different definition of the "expected" isotopic composition.  
The traditional "expectations" are based on the conception that the post-$Fe$ nuclei where produced by the capture-driven nucleosynthesis in distant stellar production-sites---these sites have specific production signatures. 
Hence, the "anomalies" are the mismatches between the "actual" meteoritic data and the "expected" outcome---the superposition of the outputs of nuclei-production models (for the many production-sites proposed as responsible for specific isotopes), the proposed patterns of nebula mixing, and the proposed geochemical processes.
The fission-driven nucleogenesis, however, has a {\em distinct}  signature. 
The fission-driven nucleogenesis has 
never before been considered---unfortunately its signature is 
not yet constrained, thus no numerical simulations can be constructed at this point for the reasons elaborated in detail in Sec.~\ref{s:3-3}, but theoretically  the {\em entire} inventory of the enrichment isotopes 
(those whose abundances are above the "original" pre-event composition of the solar system) 
can be produced by the fission-driven nucleogenesis. 
(To fully understand how and why a thoughtful review of all subsections in Sec.~\ref{s:3-3} is necessary.) 
Since the event is capable of producing all the isotopes that form the "anomalies", 
there is no need to presume the multitude of stellar cataclysms (supernovae, AGB stars, and so on) that is currently presumed in the traditional framework, nor to presume their complex mixing patterns within the nebula (which would have by-then condensed into the fully-formed gaseous objects in the framework of the proposed scenario).   
Therefore, 
once the signature of the fission-driven nucleogenesis is accepted as the ultimate "expectation", 
the isotopic measurements become no longer characterized as "anomalies", by definition, because then the data simply match the expectation. 

\begin{figure}[h!]
\centering
\includegraphics[width=0.99\columnwidth]{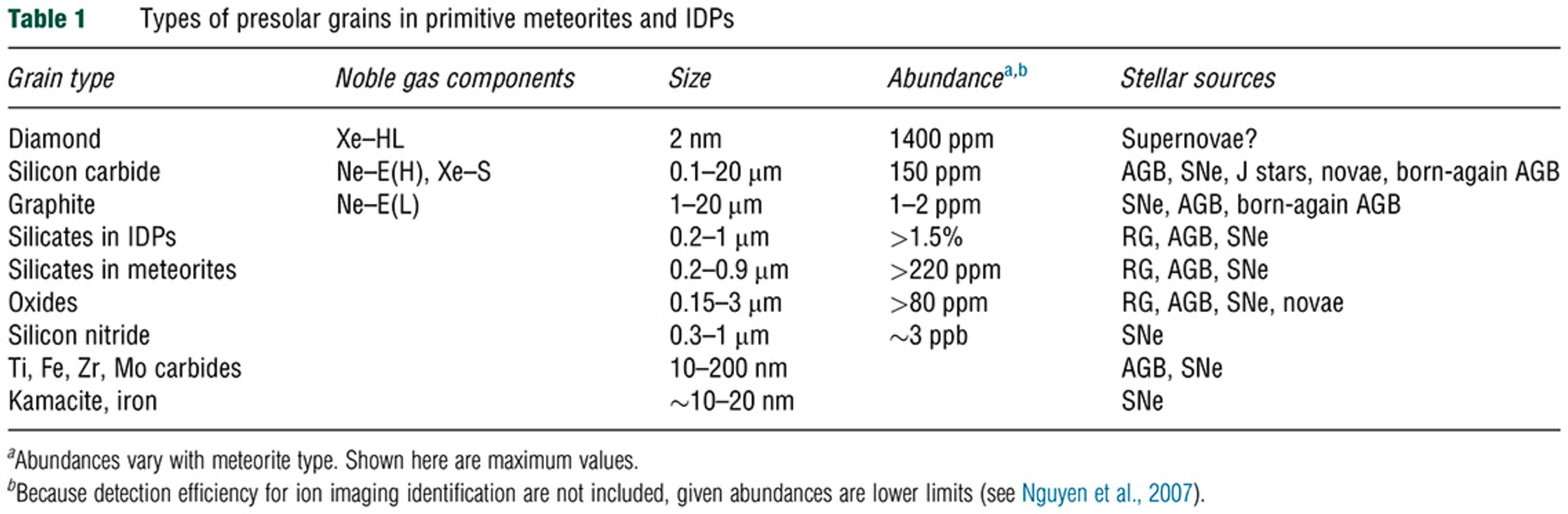}
\caption
[Presolar grains in primitive meteorites]
{
Types of presolar grains in primitive meteorites and interplanetary dust particles (IDPs) and their sizes, approximate abundances, and proposed stellar sources; from Reference \cite{Zinner_2014}. 
}
\label{Fig:27}
\end{figure}

For example, at present, microscopic grains found in primitive meteorites are considered as high magnitude 
anomalies in all measured isotopic ratios. 
The grains are presumed to have been formed in stellar outflows of late-type stars and in the ejecta of stellar explosions and had survived the formation of the solar system. And although models of the assigned stellar production-sites (see Fig.~\ref{Fig:27}, Table 1 from Reference \cite{Zinner_2014}) can qualitatively reproduce some isotopic signatures of individually-studied presolar grains, certain characteristics cannot be explained quantitatively (for extensive review, see Reference \cite{Zinner_2014}).\footnote{
For example, as noted in Reference \cite{Zinner_2014}:
\guillemotleft 
Although multizone mixing models can qualitatively reproduce the isotopic signatures of X grains (Yoshida, 2007; Yoshida and Hashimoto, 2004), several ratios, in particular the large $^{15}N$ excesses and excesses of $^{29}Si$ over $^{30}Si$ found in most grains, cannot be explained quantitatively and indicate deficiencies in the existing models (Lin et al., 2010). The latter is a long-standing problem: SN models cannot account for the solar $^{29}Si/^{30}Si$ ratio (Timmes and Clayton, 1996).%
\guillemotright{}
}

Furthermore, analyses of meteoritic components  imply that contributions from multiple stellar cataclysms must be presumed (Fig.~\ref{Fig:27} is just one example illustrating the multitude of needed cataclysms; 
in some individual micro-sized samples to explain the data the presence of {\em four or more} nucleosynthetic components is required, see, for example, Reference \cite{Dauphas_2016} p. 32).   
Thus, the theory was advanced that the solar system formed as part of a star cluster \cite{Pfalzner_2015} 
 and therefore was enriched by multiple stars.  
The challenge is however that stellar clusters are potentially dangerous environments for planetary systems  \cite{Marov_2018}.
The birth cluster provides strong background radiation fields (UV can evaporate the nebula, X-rays and EUV photons 
ionize the matter affecting formation process), 
and 
the birth cluster disrupts star and planet formation through dynamical processes (including disk truncation and perturbations of planetary orbits due to passing stars). The observed lack of severe disruption provides an upper limit on the density of the birth cluster and the time that the solar system lived in such an environment \cite{Adams_2010}. 

The multi-source enrichment theory also faces a timing challenge.  
To be able to provide the observed abundances of radioactive isotopes, 
multiple supernova must have been located \emph{not too far} from the solar nebula, 
but the distance had to be {\em great enough} so that the shockwave of matter from the supernova did not destroy the nebula. 
For the stars with $M \sim 25 M_{\odot}$ shown to provide the best ensemble of short-lived radioactive nuclei, 
this {\em optimal range is quite narrow}, $\simeq 0.1 - 0.3$~pc \cite{Adams_2010}.  
But stars within the cluster typically form within 1-2~Myr \cite{Hillenbrand_1997} and  the clusters disperse in about 10~Myr or less \cite{Allen_2007}. 
Since stars with mass $M \sim 25 M_{\odot}$ burn for $\sim 7.5$~Myr before core collapse \cite{Woosley_2002}, 
to fit the supernova enrichment scenario the Sun must have formed several Myr after the progenitor \cite{Adams_2010}. 
If located $\sim 0.2$~pc  from the progenitor, the early solar nebula could have been evaporated by the progenitor radiation \cite{Gounelle_2008}.  
One way to reconcile this is to assume that the trajectories of the early solar nebula and 
the progenitor approached the $0.2$~pc separation just before the supernova explosion  \cite{Adams_2010}. 
Such timing requirement lowers the odds for the supernova enrichment theory \cite{Williams_2007}. 
The scenario in which multiple supernovae satisfied such trajectory and timing requirements  has even lower odds.  
Furthermore, as studies show, when constraints imposed by the spread of CAIs condensation ages are taken into consideration, if the detected short-lived radionuclides were produced by  multiple stellar sources, all of these injection events, as well as the subsequent mixing of isotopes in the protodisk, 
had to occur within the {\em time-span of only about 20,000 years}  \cite{Gritschneder_2011}.

In the framework of the proposed local nucleogenetic event (with different production-signature than any other stellar production-sites such as supernovae, AGB stars, etc.), 
as just discussed at the beginning of the section, 
all of the anomalies have one natural explanation---the entirety of the detected isotopes was simply produced by the event, these isotopes represent its "signature", and therefore, they are not "anomalies". 
Indeed, other than the fact that the "presolar grains" have the "non-solar isotopic compositions"---and hence it is believed that they "condensed in the outflows of stars that lived before the solar system was formed" (see, for example, Reference \cite{Rauscher_2013})---there exists no "direct" evidence that they in fact arrived from afar;  they could have just as well formed as the result of the proposed cataclysm in the inner part of the solar system, once such cataclysm is considered. 
The proposed fission-driven event is not only capable of producing all types of nuclides (see Fig.~\ref{Fig:17} and Sec.~\ref{s:3-3-3}),  
including those commonly presumed to be formed in the $s$- and $r$-nucleosynthesis and $p$-processes 
(importantly, with different from all other production-sites signature), but also, in the fission-driven event, the nuclear transformation chains follow stochastic (probabilistic) paths thus producing somewhat varying combinations of abundances in individual final "pieces" (nano-grains, for example).  
Thus, "locally" such mechanism may indeed produce "anomalous" outcomes in comparison with some expected or "homogenized" compositions. 

Consider, for example, the presolar nano-diamonds which carry the Xe-HL component \cite{Lewis_1987}.  
The Xe-HL signature is made by enhanced light and heavy stable nuclei: $^{124,126}Xe$ (Xe-L) and $^{134,136}Xe$ (Xe-H). $^{124,126}Xe$ are the $p$-nuclei currently believed to be produced by the $\gamma$-process, while $^{134,136}Xe$ are believed to be formed by the $r$-process in core collapse supernovae, condensing in CCSNe ejecta. 
Xe-L cannot be disentangled from Xe-H since the diamonds carrying the two components are well mixed. 
   As stated by Reference \cite{Pignatari_2016}: 
\guillemotleft 
The corresponding process cannot be explained so far. 
Furthermore, diamonds are carbon-rich grains, while the $\gamma$  process is activated in oxygen-rich stellar layers, where carbon-rich dust should not form [...] 
Last but most importantly, the isotopic ratio of the Xe-L isotopes is not consistent with the same ratio of these $p$ nuclei in the solar system [...] 
It is unknown why they are different.%
\guillemotright{} 
The proposed in this paper fission-driven event can produce both $p$ and $r$ (and other) nuclei {\em simultaneously} and in locally {\em varying} amounts (for details, see Sec.~\ref{s:3-3-3} and Fig.~\ref{Fig:17}).

\vspace{6pt} 

{\bf (2) Evidence of Rapid Heating and Cooling.} 
Chondrules exhibit features that indicate that they have been heated quite rapidly and cooled rapidly as well, at rates between at 5-3000 K/hr \cite{Lofgren_1986}, \cite{Desch_2002}. 
In most cases, peak temperatures were maintained for only minutes and they apparently cooled completely in hours to days. 
Experiments have shown that chondrules would have evaporated if they existed in the liquid state any longer than this. 
Though cooling was rapid, it was considerably slower than the rate that would have resulted from radiative cooling in open space.  
These observations indicate they formed very quickly, and may never have reached equilibrium 
 \cite{Desch_2002}, \cite{Hewins_2005}.   
The rapid cooling experienced by chondrules is seen as indication that ambient temperatures were low. 
The lack of volatile loss from chondrules implies high vapor pressures of volatiles in the chondrule forming region.
So that evaporated volatiles remained in the vicinity of chondrules, the volume of gas per chondrule must not exceed 
$\sim 0.1$~m$^3$ or, equivalently, the chondrule density was $> 10$~m$^{-3}$.
So that volatiles not diffuse away from the chondrule-forming region, the chondrule-forming region must have been 
$> 10^2 - 10^3$~km in extent \cite{Cuzzi_Alexander_2006}. 

Within the traditional paradigm, the main problem with explanations of how these melts formed is that at the low pressures that must have prevailed in the solar nebula, liquids are not stable: solids should evaporate rather than melt 
\cite{Hewins_2005}.  
Various mechanisms for melting of chondrules have been proposed: 
interaction with the early active Sun, through jets or magnetic flares; melting by lightning; melting by planetesimal impacts;
passage of solids through nebular shocks.
The leading hypothesis so far has been that chondrules were produced in shock waves in the solar nebula \cite{Desch_2002}. 
However, if so, 
\guillemotleft 
such shock waves must have been common in the inner solar system because 40\% or so of the dust that ultimately formed the asteroids and the terrestrial planets was processed into chondrules
\guillemotright{} 
(\cite{White_2013} p. 451).

Also, the texture of most types of CAIs---sub-millimeter to centimeter-sized clasts consisting primarily of calcium- and aluminum-rich minerals, which are present in essentially all chondrite classes except CI,
and which have been found to have had contained exotic short-lived radionuclides, 
such as $^{10}Be$, $^{26}Al$, 
     $^{36}Cl$, $^{41}Ca$,  
$^{60}Fe$, and many others, and even $^{7}Be$ \cite{Chaussidon_2002} (with half-live of 53 days) 
 --
indicate that they have experienced complex histories that included episodes of melting, evaporation, reaction with 
gas, and finally, aqueous alteration and/or metamorphism on parent bodies. 
In general, the formation of CAIs is a 
\guillemotleft 
major unsolved problem in meteoritics%
\guillemotright{} 
\cite{Desch_2010}. 

In the framework of the proposed hypothesis, 
the mechanism  by which meteoritic components and bulk meteorites formed, 
and the environment in which their formation occurred, are quite different from the traditionally considered models. 
The nuclei that formed CAIs, chondrules, and so on, resulted from the proposed nucleogenetic event (in the inner part of the solar system), not from the gaseous nebula via condensation and subsequent growth of dust particles. 
The "debris" from the event would have formed "clumps" of various sizes, 
in which fission of super-heavy nuclei and decays of radioactive nuclei would have continued (until their eventual end-points). 
As discussed in detail in Sec.~\ref{s:3-3-3}, the fission-driven cascades of nuclear transformation chains 
are capable of producing all exotic nuclides, thus explaining their presence in various individual components of the meteorites.  
The energy release associated with nuclear-fission can explain the melting features, while the 
explosive dispersion of the "debris" into the "emptiness" of space (without any nebula gas, because the gaseous objects are presumed to have been already fully formed) can explain the subsequent cooling. 

   Here it is important to note that within the conventional meteorite chronology 
\guillemotleft 
it appears 
the meteorite parent bodies formed at around $4.567 \pm 0.001$~Ga, and there is some evidence 
that high-temperature inclusions and chondrules in carbonaceous chondrites may have formed a few million years earlier than other material%
\guillemotright{}  \cite{White_2013}. 
 Hence, at the first glance, the consistency of the suggestion---that the proposed event is responsible for the melting/cooling features of the "chronologically separately formed" meteoritic components---
  might be questioned, but such quick judgment would (incorrectly) rely on the chronological estimates which in the framework of the proposed hypothesis would need to be revised. 

Indeed, as well known, in a given chemical system 
    (in the absence of additional "sources" of parent-nuclei or daughter-nuclei) 
the isotopic abundance of a radioactive daughter-element 
is determined by four parameters: 
its isotopic abundance at a given initial time, 
the parent/daughter 
ratio of the system, 
the decay constant  $\lambda$ of the parent-element,  
and 
the time elapsed since the initial time \cite{Gast_1960}. 
In other words, the estimate 
of geologic time is based on the expression $R = R_0 + R_{P/D} (e^{\lambda t} - 1)$, or various derivatives of it, where $R_0$ is the initial ratio and $R_{P/D}$ is the parent/daughter ratio, 
hence $t = ln (\Delta R / \Delta R_{P/D} + 1 ) / \lambda$. 
Furthermore, 
two important assumptions 
   are 
built into the use of this expression \cite{White_2013}:  
(1) the system of interest was in isotopic equilibrium at time $t=0$, 
     where isotopic equilibrium in this case means the system had a {\em homogeneous}, uniform value of $R_0$; and  
(2) the system as a whole and each analyzed part of it was {\em closed} between $t = 0$ and time $t$ (usually the present time). Violation of these conditions is the principal source of error in geochronology \cite{White_2013}.  
    
    In the proposed in this paper scenario, {\em these assumptions  do not hold}. 
First, the system is open instead of closed: (a) there exists a lasting nucleogenetic "source" 
(the fission-driven cascades of nuclear transformation chains coming all the way from the "top" mega- and super-nuclei),  
and (b) the dispersed micro- and macro- "debris" dynamically interact (thus combining and dividing parent/daughter nuclei concentrations at each collision and conglomerate-breakup).  
Second, the system remains far from any equilibrium for a long time, both on micro- and macro-scales. 
Therefore, all chronological calculations would need to carefully reconsider the existing age-estimates of each meteoritic data sample.  
Consequently, it is quite possible that in the framework of the proposed scenario all components of meteorites---from the micro-grains, to CAIs, to chondrules, and so on---indeed originated quasi-contemporaneously,   
as the result of the explosive fission-driven nucleogenesis in the inner part of the solar system.
(Notably, in any radiometric age evaluation, clarity is also required about what exactly the  "age" estimate represents because a variety of ages may characterize the meteoritic sample: the age of formation, metamorphism, melting, differentiation, and so on \cite{White_2013}.) 

In fact, in geochemistry it has been already recognized  that fission  can produce variations in the isotopic abundance of 
elements,   thus affecting chronology, but the amount of the daughter nuclei produced is considered to be 
so small relative to that already present in the Earth that these isotopic variations are seen as
immeasurably small  \cite{White_2013}.   
One exception is $Xe$  whose isotopic composition  is explicitly considered as capable of varying 
slightly due to contribution from fission of $U$ and the extinct radionuclide $^{244}Pu$. 
And at least one sustained natural nuclear-fission chain reaction is known to have occurred about 2 billion years ago in the Oklo uranium deposit in Gabon, Africa.  This deposit was found to have an anomalously low $^{235}U/^{238}U$ ratio, 
  and this fact is attributed to the "burn" of some of the $^{235}U$.  
Traditionally, however, in the context of the solar system evolution, 
no local "supply" of nuclei (above $U$) has ever been  presumed, 
hence all models only consider radioactive decays of nuclei from the actinides and below. 
In our scenario, post-event, the "supply" of fissioning and decaying super-nuclei---{\em above} actinides---remains active (until eventually all mega-, super-nuclei became fully decayed, see Sec.~\ref{s:3-3-1} and Sec.~\ref{s:3-3-2}), thus influencing (over a period of time) the abundances of parent/daughter-elements and the environmental conditions during the formation of meteoritic components and bulky objects. 
Indeed, besides creating various parent/daughter nuclei combinations, including super-heavy nuclei, the 
cascades of chains of nuclear transformations (starting from the nuclear-fog "droplets") 
would have provided significant radioactive energy release which would have lasted for some time,  
thus explaining the melting features observed in CAIs and chondrules which were discussed above.

\vspace{6pt}

{\bf (3) Meteoritic Composition as Proxy for Condensable Solar System Composition.} 
One 
other 
important implication of the proposed hypothesis is that data from meteorites (presumed to originate {\em mostly} in the local nucleogenetic event) should not be automatically used as representative of the "initial" solar system composition. 

Indeed, at present, because spectrographic analysis of the Sun does not provide accurate estimates of trace elements (and most elements are trace elements in the solar system), 
 the composition of CI chondrites (which lack chondrules and CAIs) 
is taken to represent the composition of the solar system as a whole for the condensable elements. 
In combination with the results of analyses of solar photosphere and solar wind, 
the meteoritic data is used  to form the "solar system abundances profile" 
 \cite{Lodders_2009}, \cite{Asplund_2009},   
 \cite{Grevesse_2014},  \cite{Scott_2014a}, \cite{Scott_2014b}. 
 This profile is used as the essential  benchmark in astrophysics and many other fields. 
Therefore, the question of whether such benchmark is valid for model-intended purposes is a significant one and requires careful consideration in view of the proposed hypothesis.

Indeed, studies have shown that  $C$, $N$ and $Fe$ abundances tend to be somewhat higher in the Sun than in the B stars, contrary to expectations  \cite{Asplund_2009}. 
In the traditional paradigm, various resolutions have been contemplated concluding that: 
\guillemotleft 
It is unclear whether the solution can be found in the solar or B star analyses or, if a real difference indeed exists, perhaps due to infall of low-metallicity gas to the solar neighborhood%
\guillemotright {} 
 \cite{Asplund_2009}.
The proposed in this paper hypothesis naturally implies that the "excessive" solar abundances are explained by the enrichment by the proposed nucleogenetic event.

\paragraph{$p$-Elements}
\label{s:5-3-3-3}

As noted in Sec.~\ref{s:2-1}, over thirty heavy proton-rich nuclei---which are bypassed by the $s$-process and $r$-process neutron capture paths---have been identified in the meteorites of the solar system: 
 $^{74}Se$, 
  $^{78}Kr$, 
   $^{84}Sr$, 
    $^{92}Mo$, 
     $^{94}Mo$, 
      $^{96}Ru$, 
       $^{98}Ru$, 
        $^{102}Cd$, 
         $^{106}Cd$, 
          $^{108}Cd$, 
           $^{113}In$, 
            $^{112}Sn$, 
             $^{114}Sn$, 
              $^{115}Sn$, 
               $^{120}Te$, 
                $^{124}Xe$, 
                 $^{126}Xe$, 
                  $^{130}Ba$, 
                   $^{132}Ba$, 
                    $^{138}La$, 
                     $^{136}Ce$, 
                      $^{138}Ce$, 
                       $^{144}Sm$, 
                        $^{152}Gd$, 
                         $^{156}Dy$, 
                          $^{158}Dy$, 
                           $^{162}Er$, 
                           $^{164}Er$, 
                           $^{168}Yb$, 
                           $^{174}Hf$, 
                           $^{180}Ta$, 
                           $^{180}W$, 
                           $^{184}Os$, 
                           $^{190}Pt$, 
                           $^{196}Hg$ 
                           \cite{Rauscher_2013}.     
Except for $^{92}Mo$ and $^{94}Mo$ (14.77\% and 9.23\% of total $Mo$) 
and $^{94}Ru$ (5.54\% of total $Ru$),  
their relative abundance is less than 2\% of the respective element. 
In comparison with the more neutron-rich isotopes, the  $p$ nuclei are typically $10-1000$ times less abundant. 
\guillemotleft  Their origin is not well understood%
\guillemotright {}  \cite{Rauscher_2013}.  
  
  At present, massive stars are thought to  produce $p$-nuclei through photodisintegration of pre-existing intermediate and heavy nuclei. This so-called $\gamma$-process requires high stellar plasma temperatures and occurs mainly in explosive $O/Ne$ burning during a core-collapse supernova. 
 Although models of  the $\gamma$-process in massive stars have been successful in producing a large range of $p$-nuclei, significant deficiencies remain \cite{Rauscher_2013}.  
Supplementing such photodisintegration with neutrino processes required in the production of $^{138}La$ and $^{180}Ta$ 
   can perhaps explain the bulk of the $p$-abundances. 
   But deficiencies are found at higher mass, $150 \leq A \leq 165$, and for light $p$-nuclei with $A < 100$ \cite{Rauscher_2013}.   
In other words, 
generally speaking, 
the detected abundances in the solar system {\em exceed} the best combination of the so-far considered model-produced abundances.   
As noted in Reference \cite{Pignatari_2016}: 
\guillemotleft 
After more than fifty years of research, the production of the $p$ nuclei still carries several mysteries and open questions that need to be answered.%
\guillemotright {}  

The proposed in this paper hypothesis offers a completely new mechanism of $p$-nuclei production.   
As discussed in Sec.~\ref{s:3-3-3}, the cascading nuclei-transformation paths can approach the valley-of-stability (Fig.~\ref{Fig:12}) from 
both sides, the neutron-rich (left) side and the proton-rich (right) side.  
The important insight from the process is the intuitively apparent relative dominance of the neutron-rich nuclei production over the 
proton-rich nuclei production. 
The fact that meteoritic data reveal that proton-rich isotopes are typically $10^1-10^3$ times less abundant than their more-neutron-rich counterparts is indeed consistent with the described process.

\paragraph{Bimodal Planetary Structure}
\label{s:5-3-3-4}

At present, it is well acknowledged that the solar system's 
\guillemotleft 
bimodal structure is puzzling%
\guillemotright  {} 
\cite{Morbidelli_2016}.   
The proposed in this paper hypothesis provides a natural resolution to this puzzle: 
after the occurrence of the nucleogenetic event 
the  debris-pieces (in their abundances and trajectories) would obviously cluster  
near the orbit where the event occurred---
in the inner part, near the Sun 
-- eventually forming the terrestrial planets, thus creating the bimodal planetary structure of the solar system.  

Furthermore, the compositional differences between planets that cannot be explained by radially decreasing temperature 
-- such as Venus' noble-gas enrichment and Mercury's iron enrichment 
-- can be naturally explained by not only the non-uniform debris-distribution, 
 but also by the different cascading paths that fission-driven nucleogenesis took 
within debris-fragments at different distances from the Sun.   

A portion of debris could have been absorbed by the Sun, thus perhaps resolving the 
\guillemotleft solar modeling problem\guillemotright{}   
---the conflict between the standard solar models and the internal structure of the Sun, as measured by the helioseismology 
 \cite{Asplund_2009},  
  \cite{Serenelli_2009}, \cite{Mel_ndez_2009}, \cite{Villante_2010}, 
 \cite{Vagnozzi_2017}. 
(See the next section.)
And some debris could have reached the outer system and enrich the gaseous giants, thus explaining their enrichment.  
(See the following section.)

\paragraph{The \guillemotleft Solar Modeling Problem\guillemotright{} }
\label{s:5-3-3-5}

A major issue in solar physics, known as the \guillemotleft solar modeling problem\guillemotright{}, has emerged over the past decade, following a significant systematic downward revision of solar metallicity 
\cite{Asplund_2009},  
\cite{Grevesse_2014}, 
\cite{Scott_2014a}, 
\cite{Scott_2014b}, 
\cite{Asplund_2006}, 
\cite{Caffau_2010}.  
Standard Solar Models (SSM) constructed with these heavy element mixtures 
are in apparent conflict with helioseismic probes of the solar interior, which include the sound speed profile, the radius of the convective zone boundary, and the surface helium abundance (for reviews, see, for example, Reference \cite{Serenelli_2009}).

   Various solutions have been proposed: 
an anomalously large $Ne$ abundance in the photosphere \cite{Bahcall_2005}, 
physical processes not accounted for in the SSM 
\cite{Charbonnel_2005}, 
\cite{2005ApJ...627.1049G}, 
\cite{Castro_2006}, 
\cite{2010ApJ...713.1108G}, 
\cite{2010ApJ...715.1539T}, 
\cite{2011ApJ...731L..29T}, 
\cite{Serenelli_2011}, 
\cite{2016ApJ...821..108Y}, 
\cite{2004ESASP.559..574M},  
\cite{Drake_2005}, 
axion-like particles \cite{2013MNRAS.432.3332V}, 
missing opacity 
\cite{Serenelli_2009},  
\cite{Villante_2010},  
\cite{Christensen_Dalsgaard_2008}, 
\cite{2010ApJ...714..944V},  
\cite{2014ApJ...787...13V},  
\cite{2015PhPro..61..366V}, 
and 
exotic energy transport by captured dark matter 
\cite{2010PhRvD..82j3503C}, \cite{2010PhRvL.105a1301F}, \cite{2010PhRvD..82h3509T}, \cite{2014ApJ...795..162L}, \cite{2015PhRvL.114h1302V}, \cite{2015JCAP...08..040V}, \cite{2016JCAP...11..007V}, \cite{2016PhRvD..94b5001D}, \cite{2017JCAP...03..029G}. 
However, none of these ideas seem to adequately solve the solar modeling problem \cite{2014ApJ...789...60S}. 
Solar wind measurements improved the agreement with helioseismology only for the sound speed at the bottom of the convective envelope and the convective zone boundary itself, whereas the predictions for the sound speed near the core, the surface helium abundance, and neutrino fluxes remained severely discrepant with helioseismological measurements  \cite{Vagnozzi_2019}.  
   At present, the conclusion is that:  
\guillemotleft 
The reason is to be searched for within the huge increase in the abundance of refractory elements ($Mg$, $Si$, $S$, $Fe$), which leads to a hotter core%
\guillemotright{} 
 \cite{Vagnozzi_2019}. 

The proposed in this paper hypothesis provides a natural resolution to this puzzle: a portion of debris 
from the nucleogenetic event went into the Sun.  
A comprehensive study would be needed to quantify the amount and composition of such enrichment.

\paragraph{Enrichment of Gaseous Giants}
\label{s:5-3-3-6}

Enrichment of the gaseous giants of the solar system relative to the composition of the Sun, is not only peculiar, 
but it is one of the key reasons why for the giants the core accretion model overtook gravitational instability as the dominant model of formation. 

As noted earlier, disk instability---where gas giant protoplanets form rapidly though a gravitational instability of the gaseous portion of the disk \cite{Cameron_1978}, \cite{Boss_1997}, \cite{Mayer_2002}---possesses a number of advantages  \cite{Boss_2004}.   
Furthermore, the proposal---that the giants could have been enriched after formation by small body impacts---has already been made  \cite{1998ApJ...503..923B}.  
However, after \emph{Galileo} probe's mass spectrometer (GPMS) measurements of Jupiter's upper troposphere were analyzed, the results influenced the (unfair) verdict: 
\begin{quote}
\guillemotleft 
The proposal to enrich the atmospheres through impacts by small bodies (Boss 1998) is no longer viable, as {\em there are no small bodies we know of that exhibit solar ratios of noble gases, nitrogen, carbon and sulfur.} [...] the comets we know could not have delivered the nitrogen we now find on Jupiter. Thus the GPMS results effectively rule out the gravitational disk instability models for forming Jupiter.\guillemotright{}   
\cite{Lunine_2004} 
(emphasis added)
\end{quote}

Indeed, comets are deficient in nitrogen \cite{Gloeckler_1998}, \cite{krankowsky1991composition},  \cite{1991ApJ...367..641W} and argon \cite{Weaver_2002}. 
But the {\em Galileo} probe's measurements revealed that all of the heavy elements whose abundances could be measured (except helium and neon) are enriched by a factor of $3 \pm 1$, when expressed as a ratio relative to $H$ \cite{Owen_1999}. 
Thus, the reasoning followed that: 
\begin{quote}
\guillemotleft 
If Jupiter's atmosphere is indeed representative of the bulk composition of the planet, this three-fold enrichment implies the addition of at least 12 Earth masses of these solar composition icy planetesimals (SCIPs) to the complement of heavy elements contributed by the nebular gas itself. If this material has also enriched the other giant planets, it must have been the most abundant solid in the early solar system (Owen and Encrenaz 2003). The resulting total of 18 Earth masses of heavy elements is well within the range of 10-43 Earth masses derived from interior models. 
{\em The origin of these unusual planetesimals is difficult to understand.}\guillemotright{}  
\cite{Lunine_2004} (emphasis added)
\end{quote} 

Less information is available about the composition of Uranus and Neptune, but their atmospheres are also significantly enriched with $C$ and possibly $N$ indicating that a significant mass of planetesimals ($\sim 0.1 M_{\bigoplus}$) impacted the planets after they had captured most of their present hydrogen-helium envelopes \cite{Guillot_2005}. 

There are also unresolved puzzles concerning their interiors. 
Although Uranus and Neptune have similar masses and radii, 
Neptune is smaller in radius but greater in mass, and therefore, is more dense than Uranus 
-- which, by the way, rotates on its side, east to west  
-- by about 30\%.
Indeed, the two appear to be quite different internally. 
Neptune has a strong internal heat source (so an adiabatic interior is a reasonable assumption), but Uranus 
is in equilibrium with solar insulation \cite{Pearl_1991}. 
For Uranus, thermal evolution models 
give too long a cooling time for the planet \cite{2003Icar..164..228F}. 
It seems likely that some process (like layered diffusion) 
is inhibiting the heat flow in the planet, and that temperatures in the interior might be much higher than adiabatic.
   With respect to enrichment of Uranus and Neptune,
   the possibility that giant impacts have significantly affected the internal structure of these planets, 
   and have led to the observed dichotomy between them, 
   has also been discussed \cite{1986LPI....17.1011S}, \cite{Podolak_2012}. 

In view of these facts, it is apparent that the  proposed in this paper hypothesis is not only remarkably consistent with the decades old proposal that gaseous giants were  formed by gravitational instabilities 
\cite{Cameron_1978}, \cite{Boss_1997}, \cite{Mayer_2002}  
in the solar nebula (rather than by accretion),   
but also with the proposals 
that they were enriched through impacts by small bodies \cite{1998ApJ...503..923B}, \cite{1986LPI....17.1011S}. 
Moreover, our hypothesis offers the then-missing insight about the origin and the nature of the enriching "bodies"---the debris from the nucleogenetic event that occurred in the inner part of the solar system.  The nucleogenetic signature of the event obviously would have been different from any of the distant cataclysms considered so far, and therefore, the isotopic enrichment that is actually observed would be fully consistent (by definition) with the measured data.%
\footnote{
$^{38}Ar/^{36}Ar$ and $^{13}C/^{12}C$ are the same in Jupiter and the Earth, while $^{15}N/^{14}N$ is distinctly lower 
\cite{Owen_2001}. 
This can be understood if the nitrogen on Jupiter originally reached the planet in the form of $N_2$ rather than $NH_3$ or other nitrogen compounds \cite{Owen_1995}, \cite{Owen_2001}. 
} 
Quantitative amounts, however, are not yet possible to assess (see Sec.~\ref{s:3-3-3}).

\subsection{Exoplanetary Systems Comparison}
\label{s:5-4}

Over the years of observations  more than four thousand exoplanets have been detected \cite{Exo_Confirmed}. 
A range of techniques---radial velocity, transits, microlensing, imaging, timing variations, orbital brightness modulation, and astrometry---has captured a variety of planetary and orbital sizes.     
Planets are found orbiting various types of stars, and planets are found orbiting multiple-star systems. 
{\em Kepler} telescope has found 11 circumbinary systems \cite{2017PAPhS.161...38B}. 

\begin{figure}[h!]
\centering
\includegraphics[width=0.99\columnwidth]{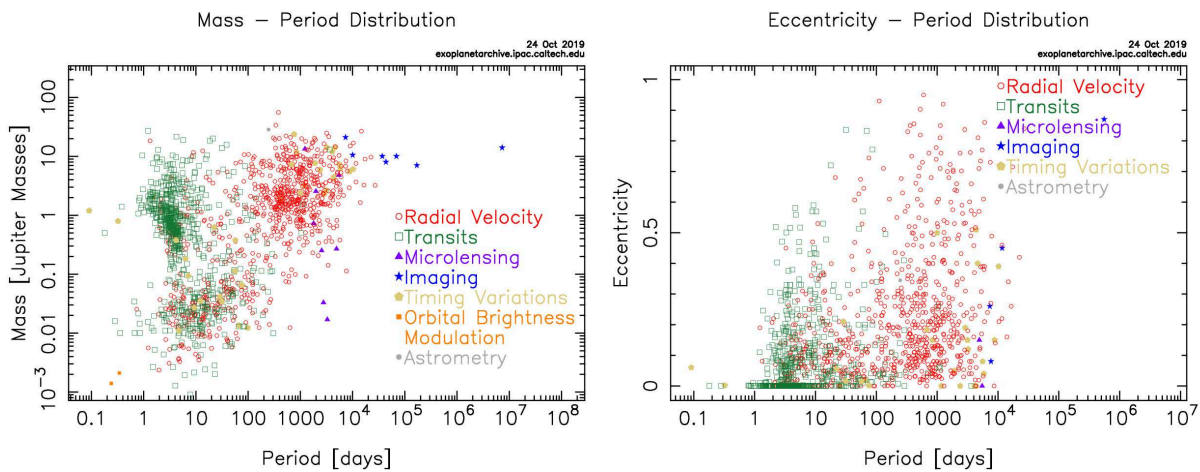} 
\caption
[Exoplanets: mass and eccentricity vs orbital period]
{
Distribution of characteristics of exoplanets (from \citet{Exo_Confirmed}).  
Colors represent different detection methods. 
Left panel: mass plotted against orbital period. 
Right panel: eccentricity plotted against orbital period. 
Most of the planets in the solar system have orbits that are almost circular ($\epsilon \sim 0$).  
Exoplanets, however, show a wide range of eccentricities. 
}
\label{Fig:28}
\end{figure}

Fig.~\ref{Fig:28} plots the detected planets masses (left panel) and eccentricities (right panel) against their orbital periods. 
Even with logarithmic scales, the distributions seem remarkably random---no meaningful patterns are apparent.  
However, unlike solar system planets whose orbits are mainly close to being circular (eccentricity $\epsilon \sim 0$), exoplanets have a wide range of eccentricities.

\begin{figure}[h!]
\centering
\includegraphics[width=0.77\columnwidth]{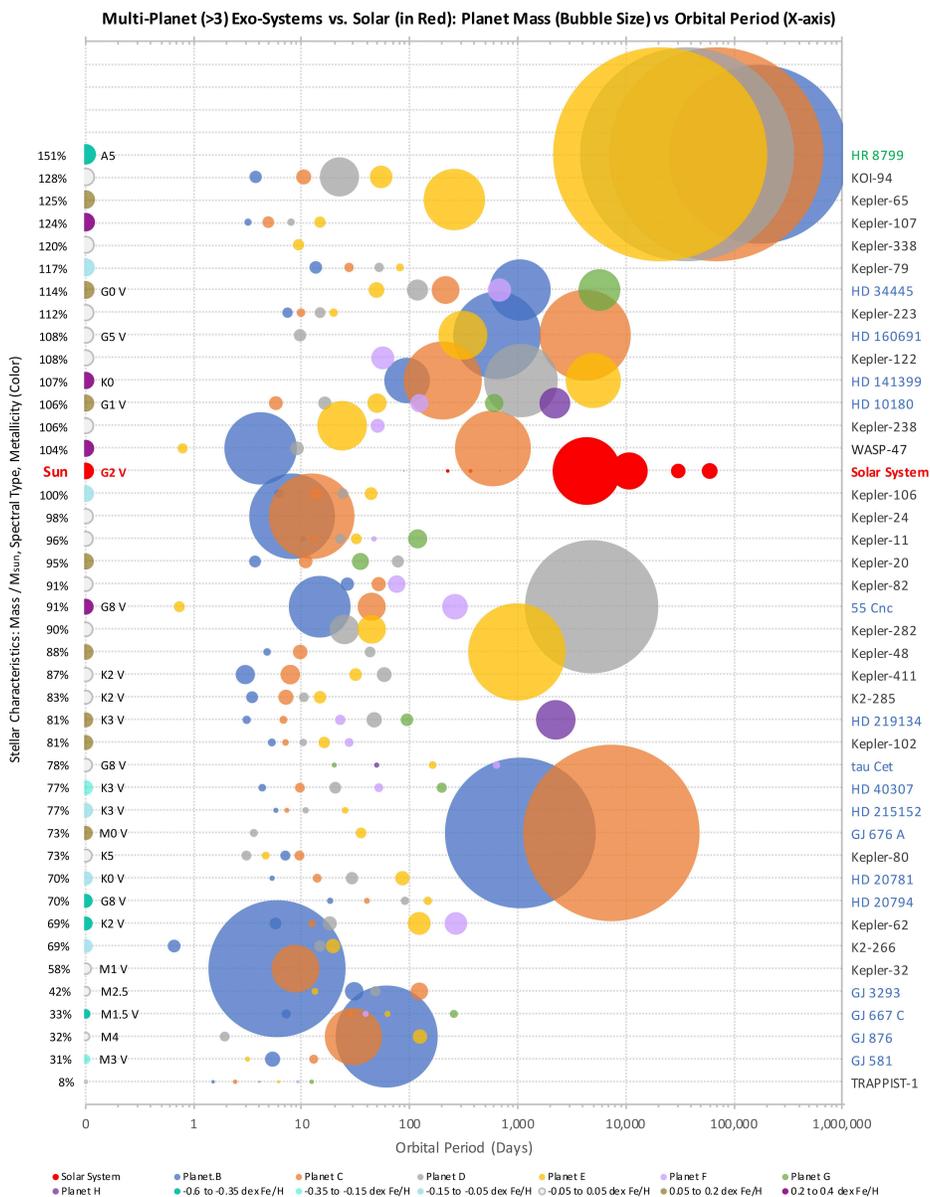}  
\caption
[Exoplanetary systems with more than three planets vs Solar System]
{
Exoplanetary systems with more than three planets (for which planets' masses and orbital periods are available in 
\citet{Exo_Confirmed} as of Nov.~28,~2019). 
Planet masses are to scale relative to one another. 
Planet colors represent their discovery order (in accordance with exoplanet naming convention).  
For comparison, Mercury, Venus, Earth, Mars, Jupiter, Saturn, Uranus, and Neptune, 
are also depicted, in bright red. 
Planetary systems are labeled on the right: green label indicates detection by imaging; blue---by radial velocity; black---by transit   (with exception of planets Kepler-20g and WASP-47c 
detected by radial velocity; 
 and Kepler-82f, Kepler-122f, and Kepler-411e,
detected by transit timing variations). 
Stellar characteristics  noted on the left vertical axis are: $M_{star}$ as percent of the Sun, spectral type (when available), and $Fe/H$ metallicity (colored by dex ranges:  -0.60 to -0.35; -0.35 to -0.15; -0.15 to -0.05; -0.05 to +0.05; +0.05 to +0.20 ; and +0.20 to +0.40). Sizes of stellar bubbles (representing  $M_{star}$) are relative to each other (not to planets).
} 
\label{Fig:29}
\end{figure}

A number of multi-planet systems has been detected. 
Fig.~\ref{Fig:29} compares the solar system and the presently known exoplanetary systems 
--  detected by various methods---here limited to those with more than three planets and for which data about the planets' masses and orbital periods were available.  
The data shows that planetary systems composed of giant planets with orbital periods  $> 10^2 - 10^3$ days  
(at distances greater than 1~AU)  have indeed been observed.  
Therefore, the evidence is {\em not inconsistent} with the proposed in this paper hypothesis that 
the solar system could have been initially composed only of its gaseous giants. 
There are also Jupiter-like planets near 1~AU (with orbital periods between $10^2$ and $10^3$ days), 
so the suggested possible existence of the "first" giant ("inner"-Jupiter) is {\em not inconsistent} with the exoplanetary data.    
It is also apparent from Fig.~\ref{Fig:29} that the characteristics of the terrestrial planets of the solar system are such 
(e.g., smaller than the exoplanets) that they are {\em not inconsistent} with the scenario of their formation from the debris accreting in the inner part of the system (near 1~AU orbit), during the second stage of solar system formation as proposed by the presented hypothesis.

Furthermore, the discovery of a super-Neptune-size planet orbiting {\em close} to a 5- to 10-million-year-old star 
\cite{2016Natur.534..658D} implies that the formation of a giant planet near the host star is possible (or that the planet quickly migrated to its current position) \cite{2017PAPhS.161...38B}. 
Thus, in the proposed in this paper hypothesis, the assumption that another gaseous giant initially could have  been present (during the first-stage) close to the Sun, between the Sun and Jupiter, is {\em consistent} with the evidence.

\newpage

\newpage

\begin{figure}[h!]
\centering
\includegraphics[width=0.90\columnwidth]{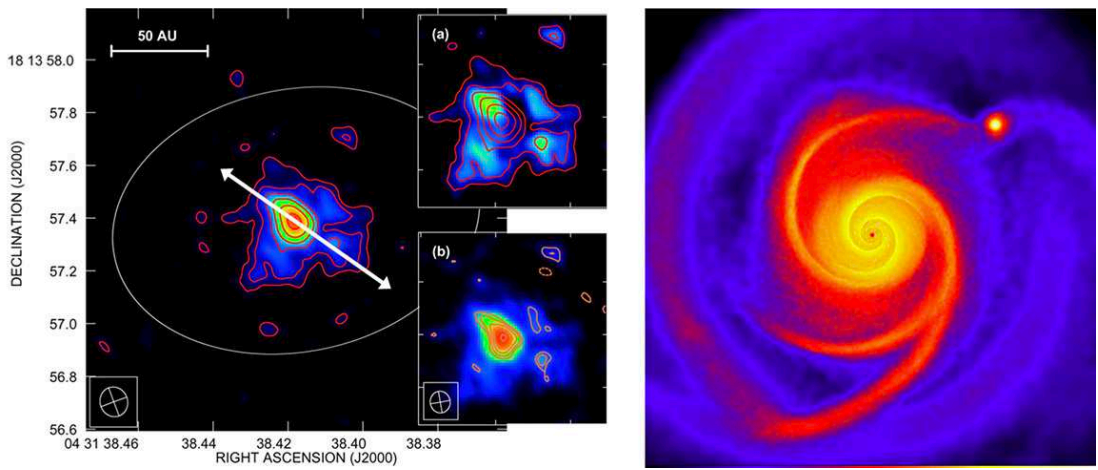} 
\caption
[Observation and simulation of direct gravitational collapse (HL Tau system)]
{
From \citet{70600f34e0e74b3a89899a52a18e7307}. 
Left panel: A radio image of the HL Tau system showing excess emission at $\sim 65$ AU (upper right quadrant of the left-hand images), which could be a protoplanet in formation. 
Right panel: A simulation showing how such an object could indeed form, through direct gravitational collapse, in the outer parts of a disc like that in the HL Tau system. 
} 
\label{Fig:30}
\end{figure}

\begin{figure}[h!]
\centering
\includegraphics[width=0.80\columnwidth]{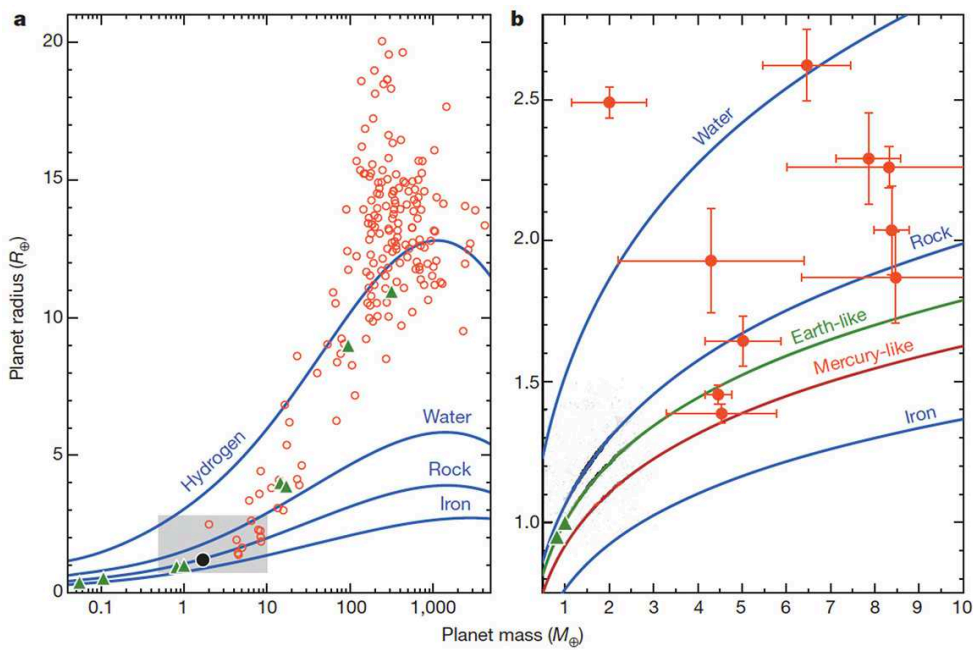}. 
\caption
[Exoplanets: mass-radius relationship vs Solar System planets]
{ 
Mass-radius relationship for the well-characterized exoplanets; 
adapted from Reference \cite{Rice_2014}.  
Solar system planets are marked by triangles. 
Left panel: massive exoplanets are similar in composition to the solar system's gaseous giants. 
Right-panel: zoom to masses and radii similar to that of the Earth. 
Both panels illustrate that the solar system's terrestrial planets are more enriched with heavy elements than exoplanets.      
(Continuous data updates are available at \citet{Exo_Confirmed} and provided images demonstrate the same trends.) 
The graphs also illustrate a degeneracy in that    exoplanets of different masses and compositions can have similar radii. 
}
\label{Fig:31}
\end{figure}

Additionally, the directly imaged observations of a very young system  appear to show evidence for a bound object forming in the {\em outer} parts of the circumstellar disc  \cite{70600f34e0e74b3a89899a52a18e7307}. 
In Fig.~\ref{Fig:30} (from Reference \cite{70600f34e0e74b3a89899a52a18e7307}) the left-hand panel 
shows a radio image of the HL Tau system with excess emission coming from what might be a protoplanet at $\sim 65$ AU (upper right quadrant of the left-hand images). 
The right-hand panel is a numerical simulation of how a disc in such a system may evolve and indicates that it is susceptible to the growth of planetary-mass bodies through {\em direct gravitational collapse} \cite{70600f34e0e74b3a89899a52a18e7307}. 
This evidence offers additional validity to the advocated (in the proposed in this paper hypothesis) position that the initial solar system formed not by accretion but by disk instability.

The observational data and the model calculations, as shown in Fig.~\ref{Fig:31}, 
 suggest that the massive exoplanets are $H/He$-composed and 
that the smaller exoplanets are composed of water and ice \cite{Valencia_2006},  \cite{2007Icar..191..337S} 
with some amount of rocky  ($MgSiO_2$) material \cite{Kuchner_2003}, \cite{L_ger_2004}. 
 (The models cannot fully capture the variety of cases and break the degeneracies in the interpretation of the bulk composition.) 

Fig.~\ref{Fig:31} also shows that the terrestrial planets of the solar system are meaningfully enriched with $Fe$ (and presumably post-$Fe$ material) and appear to be the outliers among the known exoplanets. 
 Therefore,  the exoplanetary data  is  {\em not inconsistent} with the proposed in this paper hypothesis that the solar system's terrestrial planets could have been formed as the result of the proposed fission-driven nucleogenetic event. 
The nuclei-production signature of such event is not yet obtainable, 
but theoretically it is capable and likely of producing a broad range of heavy (post-$Fe$ and post-post $Fe$) elements 
(as well as the lighter elements, of course), thus explaining the distinct compositions of terrestrial planets.


\section{Final Remarks}
\label{s:6}

The hypothesis that we proposed helps explain numerous yet-unresolved puzzles in the solar system 
listed in the front and discussion sections. 
The history of the enrichment of the solar system 
may be better understood if 
an assumption is made that an event of  
nuclear-fission-driven nucleogenesis 
(not capture-driven nucleosynthesis) 
occurred in the inner part of the solar system, 
after the gaseous objects had already formed, but before the terrestrial planets formed 
(by accretion of debris from the event). 
In this paper, we aimed to demonstrate that such event is indeed physically-sound. 

We proposed---based on the detailed review of vast multi-disciplinary literature References \cite{Jacobson_2014}-\cite{L_ger_2004} 
-- a feasible scenario of implementation of such event.  
Figs.1-28 are provided to facilitate comprehension of this extensive multi-disciplinary material for readers with different specialization.  

We proposed a model focused on {\em how} specifically the process of nuclear-fission can be realized within a stellar system (without fully destroying it)---the system which transformed into what one now calls the Solar System. 

We analyzed the key physical processes involved in the scenario: 
the structure of the nuclear-fission-capable object  (Sec.~\ref{s:3-1}),
its origin (Sec.~\ref{s:3-1-1}), 
the condition of instability of nuclear matter that leads to formation of nuclear-fog (Sec.~\ref{s:3-1-2}). 
 
 We offered an appropriate equation of state (EOS) with nuclear-fog interpolation (Eq.~\ref{Eq:5} with Eq.~\ref{Eq:6}, Sec.~\ref{s:4-1-1}).
 Using this EOS, we derived the explicit criterion of instability (via Eq.~\ref{Eq:9}, Sec.~\ref{s:4-2-1}). 
 
 We discussed the effects of deceleration and compression/decompression (Sec.~\ref{s:3-2} and Sec.~\ref{s:5-1}). 
 
We constructed the evolution equations for the nucleogenesis (Eq.~\ref{Eq:1}-Eq.~\ref{Eq:3}, Sec.~\ref{s:3-3-1}), and 
discussed  
  the necessary inputs (Sec.~\ref{s:3-3-2}) and 
the difficulties in formulation of the initial conditions for the system (Sec.~\ref{s:3-3-1} and Sec.~\ref{s:3-3-2}). 

We constructed the nucleogenetic cascades (Sec.~\ref{s:3-3-3}, Figs.~\ref{Fig:14} - \ref{Fig:17}). 

We discussed the significance of the obtained results (Sec.~\ref{s:5-3}) for  
the problem of $p$-elements, the \guillemotleft solar modeling problem\guillemotright{}, meteoritic data and anomalies, 
the 
\guillemotleft meter-size barrier\guillemotright {}  problem of the accretion model, the enrichment of gaseous giants, and the bimodal structure of the solar system. 
Comparison with multi-planetary exosystems is provided (Sec.~\ref{s:5-4}).

For the event of nuclear-fission-driven nucleogenesis in the inner part of the solar system 
to be realistic---to not contradict the laws of physics and the common sense---
a feasible scenario and an implementation mechanism are required.  
We proposed a {\em "small" fragment}---of a super-dense compact object torn apart and catapulted by the galactic supermassive black hole---as a possible carrier of the (fission-capable) neutron-rich nuclear matter. 
Upon the encounter with some random stellar system (which we now call the Solar System), 
the by-then quasi-stable nuclear matter exploded because "sufficient" deceleration-induced 
internal (localized) decompression created the state of {\em nuclear fog}, 
reducing density locally below $\rho_{drip}$
thus allowing beta-decays to start, leading to fragmentation, fission, and other cascades of nuclear transformations. 
This idea may seem bizarre to those unfamiliar with recent advances
in the fields dealing with super-dense matter and super-heavy nuclei, but the idea is grounded in
and incorporates well-established multi-disciplinary facts, both observed and experimental, which
are extensively cited throughout the article.

The creation of {\em nuclear fog}---a mixture of two phases of nuclear matter, either liquid droplets surrounded by gas of nucleons, or generally homogeneous neutron-liquid with neutron-gas bubbles --
is the key to successful realization of the scenario. 
The state of nuclear-fog is experimentally well established.  
In this paper, we presented a model of a compact super-dense fission-capable stellar fragment 
--  a liquid macroscopic nuclear-drop---
and examined its stability/instability near the boundary of $T(\rho)$-phase transition 
specifically near the nuclear-fog domain (spinodal zone). 
We established a quantitative criterion for the fragment's instability  
and a quantitative characterization for the fragment's  "small size". 
We conclude that the described process is indeed physically feasible and scientifically sound. 
The abundant references cited in this paper offer substantial support to each aspect of the proposed hypothesis. 

Obviously, many aspects of the proposed hypothesis require further detailed exploration. 
Unfortunately, the following stumbling blocks limit the progress at present: 
(1) Tighter constraints on the models of super-dense nuclear matter must come from direct experimental investigations.
For that, obviously, significant financial investments in the existing and new facilities and their operations are paramount; 
(2) Understanding of {\em fragmentation and fission} of giant-nuclei (nuclear-fog droplets) would greatly benefit from 
the  progress in experimental studies of super-heavy nuclei, which could be used to advance theoretical 
 models of giant-nuclei ($ln A \gg 1$); 
 (3) Once the reliable (experimentally-supported) understanding of the fragmentation/fission of giant-/mega-/super-nuclei is achieved, detailed numerical studies of fission cascades would become realizable.  However, since the phenomenon of fission is probabilistic, numerical simulations of nucleogenetic cascades (involving a much expanded nuclear network) will require substantially increased computing power. 
 In other words, significant funding and systematic, milestone by milestone, advancement in  astrophysical, cosmochemical, and obviously nuclear physical sub-fields, are necessary.
 
 Naturally, the implications of the proposed hypothesis significantly affect the current understanding of the evolution of the solar system, and therefore, we discussed the relevant experimental data and insights from the existing models 
from various planetary and astrophysical sub-fields. 
Thus, this discussion is of interest to the broad scientific community spanning many disciplines.
The so-far aggregated evidence reveal that a broad range of data appear to be {\em consistent} with the proposed scenario of {\em two-stage planetary formation} and  {\em "local" enrichment} of the solar system with exotic nuclei.
Even better, many of the outstanding puzzles may find their resolutions 
within the proposed "expanded" paradigm of the solar system evolution.


\vspace{24pt}
\authorcontributions{Conceptualization and Writing, E. P. T. and V. I. P.}

\funding{This research was funded by a private donor who wishes to remain anonymous.}


\conflictsofinterest{The authors declare that there is no conflict of interests regarding the publication of this article.}

\newpage

\addcontentsline{toc}{section}{List of Figures}
\listoffigures
\label{s:LOF}

\newpage
\reftitle{References}
%

\end{document}